\documentclass[
preprint%
twocolumn, secnumarabic%
,tightenlines%
,amssymb,
amsmath,
nobibnotes,
nofootinbib, aps, prl, showpacs,showkeys]{revtex4}

\usepackage{subfigure}
\usepackage{tikz}
\usepackage{bm}%
\usepackage{graphicx} %
\usepackage{rotating} %
\usepackage{hyperref} %
\usepackage{url} %
\usepackage{srcltx} %
\usepackage{float}
\usepackage{caption} 
\usepackage{listings}
\usepackage{simplewick}
\usepackage{comment}
\usepackage{caption}
\usepackage{natbib} 
\usepackage{ulem} 

\usepackage{bbm, dsfont,mdframed,slashed}
\expandafter \ifx \csname package@font\endcsname\relax\else
\expandafter \expandafter \expandafter
\usepackage
\expandafter \expandafter
\expandafter{\csname package@font\endcsname}%
\fi
\usepackage{epstopdf}
\usepackage{slashed}
\usepackage{ulem}
\usepackage{srcltx}
\usepackage{soul}
\usepackage{amsmath}
\usepackage{natbib} 
\usepackage{ulem} 

\usepackage{float} 
\extrafloats{200}


\usepackage{epstopdf}

\usepackage{enumerate}

\begin{document}

\title{Spin asymmetries in quasielastic charged current neutrino-nucleon scattering}

\author{Krzysztof M. Graczyk}
\email{krzysztof.graczyk@uwr.edu.pl}

\author{Beata E. Kowal}
\email{beata.kowal@uwr.edu.pl}

\affiliation{Institute of Theoretical Physics, University of Wroc\l aw, plac Maxa Borna 9,
50-204, Wroc\l aw, Poland}

\date{\today}%

\begin{abstract} 
The work concerns the quasielastic charged current neutrino-neutron and antineutrino-proton interactions. Single, double, and triple spin asymmetries are computed and analyzed.  The spin asymmetries are sensitive to the axial form factor of the nucleon. 
In particular, the target-recoil double spin asymmetry and the lepton-target-recoil triple spin asymmetry depend strongly on the axial form factor of the nucleon. Indeed, the sign and shape of these components depend on the axial mass parameter. All the asymmetries, except the lepton polarization, are  observables well suited to study the nonstandard interactions described by the second class current contribution.
    \end{abstract}

\maketitle

  \section{Introduction}

  The spin observables, in particles collisions, have been intensively studied for many years \cite{Bourrely:1980mr,leader_2001,Fraas:1970je,Alguard:1976bk}. 
They contain information about the structure of the scattering amplitude and the structure of the particles participating in the interaction. 
  The meson photoproduction is a good example of the process for which the spin observables are utilized to constrain the form of the scattering amplitude~\cite{Fasano:1992es,Nakayama:2019cys}. Another example is the elastic $ep$ scattering. Indeed, from the measurements of the spin asymmetries the form factor ratio: $G_E^p/G_M^p$  is extracted~\cite{Akhiezer:1968ek,Akhiezer:1974em,Arnold:1980zj,Donnelly:1985ry,Afanasev:2017gsk,Arenhovel:2019pvb}, $G_{E/M}^p$ is electric/magnetic form factor of the proton, respectively.  

  In this work, we consider the polarization observables in the quasielastic charged current (CCQE) neutrino (antineutrino) - neutron (proton) scattering.
  Progress in studies of the fundamental properties of the neutrinos requires more precise knowledge of the neutrino-nucleon and neutrino-nucleus cross-sections~ \cite{doi:10.1146/annurev-nucl-102115-044720,Alvarez-Ruso:2017oui}.  The CCQE $\nu_\mu$-nucleon scattering is the process that is very important for the experimental investigations of the neutrino oscillation phenomenon.  Indeed, in the long-baseline experiments, such as T2K \cite{sanchez_federico_2018_1295707} or No$\nu$a \cite{sanchez_mayly_2018_1286758}, the CCQE $\nu_\mu$-nucleus events  are analyzed to obtain the neutrino oscillation parameters and $CP$-violation phase. 
  On the other hand,  the theoretical description of the CCQE neutrino-nucleon interaction is the input for the nuclear models for the $\nu$-nucleus scattering.  
  
 The CCQE $\nu_\mu n$ interaction is described by a simple model. Indeed, the scattering amplitude is parametrized by four form factors, two vector and two axial.  The axial contribution dominates the CCQE cross-section. 
  After some simplifications, it is parametrized by the axial form factor $F_A$~\cite{Bernard:2001rs}. 
  The accurate estimate of the axial nucleon form factor $F_A$ is necessary to reduce the uncertainty of the predictions of the neutrino-nucleus scattering cross-sections, which is desired to lower the systematic uncertainty of the measurement of the oscillation parameters and the $CP$-violation phase, in the lepton sector.

   The axial nucleon form factor is extracted from the neutrino-nucleus scattering data. However, it is a difficult task because of the complexity of the nuclear effects \cite{Katori:2016yel}. So far the most accurate information about $F_A$ is  obtained from the analysis of the neutrino-hydrogen and the neutrino-deuteron scattering data~\cite{Mann:1973pr,Barish:1977qk,Miller:1982qi,Baker:1981su,Kitagaki:1986ct,Kitagaki:1990vs,Kitagaki:1983px}.  
   Usually, to simplify the analysis, it is assumed that $F_A$ has a dipole form. However,  an effort has been made to search for deviations from the dipole shape~\cite{Bodek:2007ym,Amaro:2015lga,Meyer:2016oeg,Alvarez-Ruso:2018rdx,Megias:2019qdv}. Unfortunately, to get clear evidence for a non-dipole dependence, it is necessary to collect more informative data. 
   
   The so-called second class current (SCC)  introduced in 1958 by Weinberg \cite{Weinberg:1958} describes not allowed by the standard model contribution to the electroweak nucleon vertex. In last several decades the SCC has been studied  theoretically  (for recent review see Fatima \textit{et. al} \cite{Fatima:2018tzs}) and experimentally \cite{Ahrens:1988rr,Baker:1981su,Belikov:1983kg,Wilkinson:2000gx}. However, no evidence for the SCC contribution has been obtained so far.

  The studies of the spin properties in the CCQE $\nu$-nucleon interactions should extend our knowledge about the elementary electroweak nucleon vertex. Indeed, we showed \cite{Graczyk:2017rti,Graczyk:2017ngi,Graczyk:2019blt} that the spin observables contain nontrivial information about the nonresonant background contribution in the single pion production (SPP) processes induced by interactions of neutrinos with nucleons. In this work, we aim to study the axial and the SCC contributions to the spin observables in the CCQE scattering.  In particular, we focus on the investigation of the sensitivity of the spin observables to the axial and SCC form factors. 
  
 We consider seven spin observables. Among three single spin asymmetries: the recoil polarization asymmetry, the lepton asymmetry, and the polarized target asymmetry, the last one has been not discussed for the CCQE scattering yet. But it was studied for the SPP in the neutrino-nucleon scattering~\cite{Graczyk:2019blt}.
 Already in the sixties, Adler \cite{Adler1963}, Pais \cite{Pais:1971er} and Block \cite{Block:1965zol} investigated the properties of the polarization of the recoil nucleon and the charged lepton in the CCQE neutrino-nucleon interactions. Report by Llewellyn-Smith \cite{LlewellynSmith:1971uhs} summarizes these works.
Recently, the polarization properties of the $\tau$-lepton, produced in the CCQE $\nu_\tau$-nucleon scattering, have been investigated \cite{Hagiwara:2003di,Kuzmin:2004yb,Kuzmin:2004ke}. Bilenky and Christova studied the axial contribution to the nucleon polarization in the CCQE $\nu_\mu$-nucleon scattering \cite{Bilenky:2013iua,Bilenky:2013fra}. Whereas in Ref.~\textit{Fatima et al.} \cite{Fatima:2018tzs},  the SCC contribution to the lepton and the recoil nucleon polarizations is estimated. 
   Eventually,  the polarization properties of the final lepton and the knock-out nucleon, in the CCQE neutrino-nucleus scattering, are discussed in  Refs. \cite{Graczyk:2004uy,Valverde:2006yi,Sobczyk:2019urm} and \cite{Jachowicz:2004we}, respectively.
  
  Besides the polarization asymmetries, we consider the lepton-recoil, target-lepton, target-recoil double spin asymmetries as well as target-lepton-recoil triple spin asymmetry. All these observables have been not studied for the CCQE scattering yet.

  To discuss the physical properties of the asymmetries, we calculate their components in spin basis (they are given in the Appendixes). Eventually, we provide with the numerical analysis of the results. 
  We show that the spin observables contain the information about the axial structure of the nucleon, which is complementary to the spin averaged cross-section measurements. They are also sensitive to possible signals from the nonstandard interactions, described by the second class currents (SCC). In particular, the target-recoil asymmetry is  sensitive to the axial form factor. 
  Indeed, the sign and the shape of transverse-transverse and normal-normal components of the target-recoil asymmetry depend on the value of the axial mass $M_A$. Similar dependence is obtained for the components of the lepton-target-recoil asymmetry (longitudinal-normal-normal, longitudinal-transverse-transverse, and longitudinal-transverse-normal). 
   
   The paper is organized as follows: in Section~\ref{Section:Formalism}, the necessary formalism is introduced. Section~\ref{Section:Spin} contains the definition of the spin observables, Section~\ref{Section:Discussion} presents the discussion of the numerical results.  
   Appendix~\ref{Appendix_convention} presents convention and normalization used in the paper, whereas Appendix~\ref{Appendix_vector_and_tensors} contains analytic formulas for the spin asymmetries in the vector and the tensor form. Finally, Appendix~\ref{Appendix:Components} presents the components of the asymmetries calculated in spin basis.

\section{Formalism}  
\label{Section:Formalism}
    \subsection{Kinematics}
    Let us consider two CCQE  scattering processes:
   \begin{equation}
    \label{Eq:CCQE_neutrino}
    \quad \nu_\mu(k) + n(p)  \to  \mu^{-}(k')  + p(p'),   \quad
    \quad \overline{\nu}_\mu(k) + p(p)  \to  \mu^+(k')  + n(p'),  
    \end{equation}
    where $N=p$ (proton) or $N=n$ (neutron). 
    
    The momenta of neutrino, lepton, target nucleon and recoil nucleon are denoted as it is given below 
    \begin{equation}
    k^\alpha     =  (E, \mathbf{k}),\quad
    k'^{\alpha}  =  (E_{k'},\mathbf{k'}),\quad
    p^\alpha     =  (E_p,\mathbf{p}),\quad
    p'^\alpha = (E_{p'},\mathbf{p'}),
    \end{equation} 
    where $E_k=\sqrt{\mathbf{k}^2+m^2}$.
    In the next part of the text, $m$ and $M$ stand for the lepton and the averaged nucleon mass, respectively. 
    
    The four-momentum transfer is defined by:
    \begin{equation}
    q^\alpha = k^\alpha -{k'}^\alpha = (\omega, \mathbf{q}),
    \end{equation}
    where $\omega$ and $\mathbf{q}$ denote the transfer of an energy and a momentum, respectively.   
    
\begin{figure}
    \includegraphics[width=0.7\textwidth]{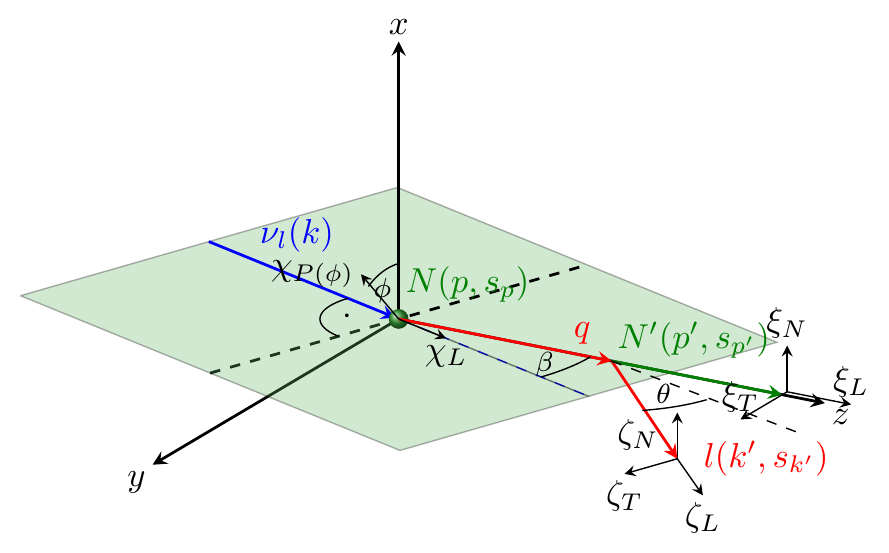}
    \caption{  The polarization basis vectors for the fermions in the CCQE $\nu$-nucleon scattering.  The full (green) ball denotes the target. The blue, red and green arrows are the momenta of the neutrino, the charged lepton, and the recoil nucleon, respectively. The vectors $\chi$, $\xi$ and $\zeta$ are the elements of the basis for the spins of target, recoil nucleon, and lepton, respectively. \label{Fig_diagramQE}
    }
\end{figure}

    The components of the asymmetry tensors are calculated in the coordinate system so that: $z$-axis goes along the momentum $\mathbf{q}$, hence
    $q^\mu = (\omega, 0,0,|\mathbf{q}|)$,
    the axis $z$ and $y$ lie in the scattering plane spanned by $\textbf{k}'$ and $\textbf{k}$ vectors, whereas  $x$ is normal to the scattering plane, see Fig.~\ref{Fig_diagramQE}.

The spin asymmetries are expressed as functions of the Mandelstam variables:
    \begin{equation}
    t = q^{\mu}q_{\mu}  \equiv q\cdot q = q^2 = - Q^2,\quad
    s = (p+k)^2,\quad
    u = (p-k')^2.
    \end{equation}
The analytic formulas take simpler form when the scalar 
\begin{equation}
    s_{u} \equiv  s-u,
\end{equation}
is used.

\subsection{Cross-section}

The differential cross-section, in the laboratory frame,  is given by the formula
    \begin{equation}
    \frac{d^6\sigma}{d^3\mathbf{k'} d^3 \mathbf{p'}}=\frac{|\mathcal{M}|^2}{64 \pi^2 M E   E_{k'} E_{p'}}   
     \delta^{(4)} (p'+k'-k-p),
    \end{equation}
    where     $\mathcal{M}$ is the scattering matrix.  
    In the one-boson exchange approximation it  has the form
    \begin{equation}
    i\mathcal{M}
    =  \cos\theta_C \frac{G_F}{\sqrt{2}}  
    j_\mu (\mathbf{k}',s^l;\mathbf{k}) J^\mu (\mathbf{p}',s^{N'};\mathbf{p},s^N),
    \end{equation}
where $G_F=1.16639\cdot 10^{-5}$~GeV$^{-2}$ is a Fermi constant; $\theta_C$ is the Cabibbo angle, $\cos \theta_C =0.9737$; by $s^l$, $s^N$ and $s^{N'}$ the charged lepton, the target nucleon and the recoil nucleon spins are denoted, respectively. The spin of the neutrino is omitted as it is a state of definite helicity. 
    
The $\mathcal{M}$ matrix is given by the contraction of
    lepton current, 
    \begin{equation}
    j_\alpha(\mathbf{k}',s^l;\mathbf{k}) =  \bar{u}(\mathbf{k}',s^l)\gamma_{\alpha} (1-\gamma_5)  u(\mathbf{k})
    \end{equation}
    with hadron current, 
    \begin{equation}
        J^\mu (\mathbf{p}',s^{N'};\mathbf{p},s^N)= \overline{u}(\mathbf{p'},s^{N'}) \Gamma_{+}^\mu(q) u(\mathbf{p},s^N),
    \end{equation}
    where $\Gamma_{+}^\mu(q)$ is the electroweak nucleon vertex. Its analytic form is given in the next section.

After several simplifications the differential cross section reads
\begin{equation}
\label{Eq:cross-section}
\frac{d\sigma_0}{d t} \equiv     \frac{d\sigma}{d t}(s^{N'}=s^l=s^N=0) 
=\frac{G_F^2  \cos^2\theta_C}{128 \pi M^2 E^2 }   
       \mathcal{I},
\end{equation}
where
\begin{equation}
    \mathcal{I} \equiv \mathrm{L}_{\mu\nu}(s^l\to 0) \, \mathrm{H}^{\mu\nu}(s^N\to 0,s^{N'}\to 0).
\end{equation}
\begin{equation}
\mathrm{H}^{\mu\nu}(s^{N'},s^N)    =  
\mathrm{Tr}\left(\Lambda(p',s^{N'},M) \Gamma_{+}^{\mu}\Lambda(p,s^N,M) \gamma_0 {\Gamma^{\nu}}^{\dagger} \gamma_0 \right), \end{equation}
is the hadronic tensor, whereas
\begin{equation}
\mathrm{L}_{\mu\nu}        
= 
 \mathrm{Tr}\left( \Lambda( k',s^l,  m)\gamma_{\mu}(1- x \gamma_5)\slashed{k}\gamma_{\nu}(1-  x \gamma_5) \right),
\end{equation}
is the lepton tensor, $x=\pm 1$ for neutrino/antineutrino, $\slashed{k}=\gamma_\mu k^\mu$.
In the above expressions the projection operator 
\begin{eqnarray}
\Lambda(p,s,M) & \equiv & \frac{1}{2}( \slashed{p} + M)\left( 1 + \gamma_5 \slashed{s}\right),
\end{eqnarray} 
is introduced. Notice that for fully polarized state $s_\mu s^{\mu} =-1 $.    The differential cross-section, averaged over initial spins and summed over final spins,  reads:

${d\overline{\sigma}}/{dt}
=
4 {d\sigma_0}/{dt}$.

\subsection{Hadronic current}

The hadronic current has the vector-axial structure. Indeed the electroweak nucleon vertex reads \cite{LlewellynSmith:1971uhs,donoghue_golowich_holstein_2014}
\begin{equation}
\label{Eq:Gamma-full}
    \Gamma_{+}^\mu (q)=   
    \gamma_{\mu}F_1^V(t) +i\sigma^{\mu \nu} q_{\nu}  \frac{F_2^V(t)}{2M } +q^{\mu} \frac{F_3^V(t)}{M }  
     -  \left(  \gamma_{\mu} F_A(t) + q_{\mu} \frac{F_P(t) }{2M } +i\sigma^{\mu \nu} q_{\nu}  \frac{F_3^A(t)}{M }  \right)\gamma_{5}.
\end{equation}
The functions $F_i^V$ and $F_i^A$ ($i=1,2,3$) refer to vector and axial form factors of the nucleon, respectively. If one assumes time reversal invariance (standard recruitment), then all form factors are real. 

It is convenient to characterize the elements of the weak current according to the $G$-parity transformation \cite{Weinberg:1958}. The vector current contributions, described by $F_1^V$ and $F_2^V$ form factors, transform similarly as their analogs in the Standard Model.  Similarly, the axial current induced by the $F_A$ and $F_P$ has the same $G$-parity property as their analog in the Standard Model.  Hence the standard part of the electroweak vertex reads 
\begin{equation}
    {\Gamma_{+}^\mu}_{(1)} (q) =   \gamma_{\mu} F_1^V (t)+  \frac{i\sigma^{\mu \nu} q_{\nu}}{2M }F_2^V(t)  
    -\left( \gamma_{\mu} F_A(t)  + \frac{q_{\mu} }{2M } F_P(t) \right)\gamma_{5}.
    \end{equation}
Then  $\overline{u}(\mathbf{p'}){\Gamma_{+}^\mu}_{(1)} u(\mathbf{p}) $  is called the first class current. 

There are many phenomenological parametrizations of the vector form factors, see e.g. \cite{Alberico:2008sz}. But without losing the generality of the discussion we consider the dipole parametrization~\cite{Thomas_book}:
    \begin{equation}
        F^V_1(t)=\frac{(1 -({t}/{4 M^2}) (\mu_p-\mu_n))}{1 - {t}/{4 M^2} } G_D(t),\quad
        F^V_2(t) = \frac{\mu_p-\mu_n-1}{1 - {t}/{4 M^2}} G_D(t),
        \end{equation}
        where $\mu_p=2.793$ and $\mu_n=-1.913$ are  proton and neutron magnetic moments expressed in nuclear magneton unit, whereas
    \begin{equation}
        G_D(t) = \left(1 - \frac{t}{M_V^2}\right)^{-2}, \quad M_V = 0.84~\mathrm{GeV}.
    \end{equation}

The parametrization of the axial form factor, used in the analysis, is given in Sec. \ref{Section:Discussion}.  The  pseudoscalar axial form factor, $F_P$, is related to $F_A$ as follows \cite{Bernard:2001rs}
\begin{equation}
F_P(t) = \frac{4M^2 }{(m_{\pi}^2 - t )} F_A(t),
\end{equation}
where $m_\pi$ is the pion mass.    

If the $G$-parity is weakly violated, then the form factors $F_3^V$ and $F_3^A$ do not vanish, and contribute to the so-called second class current:
\begin{eqnarray}
    {\Gamma_{+}^\mu}_{(2)} (q)&=&  q^{\mu} \frac{F_3^V(t)}{M }   -  i\sigma^{\mu \nu} q_{\nu} \gamma_{5}  \frac{F_3^A(t)}{M }. 
\end{eqnarray}

 We impose the conserved vector current theorem to constrain the SCC. Then   $F_3^V=0$, however, there is no such constraint for the axial $F_3^A$ form factor.  Therefore $F_3^A$ is the function that parametrizes the SCC contribution. 
 
To distinguish between the first and the second class  contributions we introduce the notation: 
\begin{equation}
\mathrm{H}^{\mu \nu}=\mathrm{H}^{\mu \nu}_{(1)} + \mathrm{H}^{\mu \nu}_{(12)} +  \mathrm{H}^{\mu \nu}_{(2)}, 
\end{equation} 
where
\begin{eqnarray}
    \mathrm{H}^{\mu \nu}_{(1)} &\equiv& \mathrm{Tr}\left(\Lambda(p',s^{N'},M) \Gamma_{+,(1)}^{\mu}\Lambda(p,s^{N},M) \overline{\Gamma}^\nu_{+,(1)} \right)
    \\
    \mathrm{H}^{\mu \nu}_{(2)} &\equiv&
    \mathrm{Tr}\left(\Lambda(p',s^{N'},M) \Gamma_{+,(2)}^{\mu}\Lambda(p,s^N,M) \overline{\Gamma}^\nu_{+,(2)} \right)
    \\
    \mathrm{H}^{\mu \nu}_{(12)} &\equiv&
    \mathrm{Tr}\left(\Lambda(p',s^{N'},M) \Gamma_{+,(1)}^{\mu}\Lambda(p,s^{N},M) \overline{\Gamma}^\nu_{+,(2)} \right)
    \nonumber \\
    & & +
    \mathrm{Tr}\left(\Lambda(p',s^{N'},M) \Gamma_{+,(2)}^{\mu}\Lambda(p,s^N,M) \overline{\Gamma}^\nu_{+,(1)}. \right)
    \end{eqnarray}
In the above $\mathrm{H}^{\mu \nu}_{(1)}$ refers to the standard hadronic tensor (the first-class contribution only), whereas $\mathrm{H}^{\mu \nu}_{(12)}$ and $\mathrm{H}^{\mu \nu}_{(2)}$ describe the second class current contributions to the hadronic tensor.

The spin averaged cross-section is easily calculated. Indeed, assuming the full form of (\ref{Eq:Gamma-full}) and reality of the form factors one gets:
\begin{equation}
    \mathcal{I} =  \mathcal{I}_{(1)}+\mathcal{I}_{(2)}+\mathcal{I}_{(12)},
\end{equation}
where
\begin{equation}
\mathcal{I}_{(a)} = \mathrm{L}_{0,\mu\nu}  \mathrm{H}^{\mu\nu}_{(a)}(s^N\to 0,s^{N'}\to 0), \quad a=1,\, 12,\, 2.
\end{equation}
and
\begin{eqnarray}
\mathcal{I}_{(1)}&=&
\frac{1}{4M^2}\left[\right.
8 F_A F_P m^2 M^2 (t-m^2)  
  + F_P^2 m^2 t (t-m^2) 
  + 4 F_A^2 M^2 (m^2 (4 M^2-m^2) + s_u^2 - 4 M^2 t + t^2) \nonumber\\ 
  & & + 4 {F_1^V}^2 M^2 (t^2 - m^2 (m^2 + 4 M^2) + s_u^2 + 4 M^2 t )  
  + 8 F_1^V F_2^V M^2 (t-m^2)(2 t + m^2) \nonumber \\ 
  & & + {F_2^V}^2 (t^3 -4 m^4 M^2 - s_u^2 t + (4 M^2 -m^2) t^2 ) 
  - 16 (F_1^V  +  F_2^V )F_A M^2 {s_u}  t  x 
\left. \right]
\\
\mathcal{I}_{(2)}
&=& \frac{1}{2M^2}\left[ 
{m^2}{F_3^V}^2   (5 m^2 M^2 - (2 m^2 + 5 M^2) t + 2 t^2)\right. 
\left. + 2t {F_3^A}^2 (4 m^2 M^2 - {s_u} ^2 - (m^2 + 4 M^2) t + t^2)\right] 
\\
\mathcal{I}_{(12)} &=&
{- 4 m^2}{s_u}   
\left[
(  F_1^V  + \frac{t}{4  M^2} F_2^V ) F_3^V + (  F_A  + \frac{t}{4  M^2} F_P )  F_3^A \right] .
\end{eqnarray}
Above results are consistent with~\cite{LlewellynSmith:1971uhs}.

\section{Spin observables}

\label{Section:Spin}
In the most general case the differential cross-section formula has the form:
\begin{equation}
    d\sigma 
    = 
    d\sigma_0   \left(1 + \mathcal{P}_l^\mu s_{\mu}^l + \mathcal{T}_{N}^\mu s_{\mu}^N  + \mathcal{P}_{N'}^\mu s_{\mu}^{N'} 
        +s_{\mu}^l s_{\nu}^{N'} \mathcal{A}^{\mu\nu}_{lN'}
    +s_{\mu}^l s_{\nu}^N \mathcal{B}^{\mu\nu}_{lN}
    +
    s_{\mu}^{N} s_{\nu}^{N'} \mathcal{C}^{\mu\nu}_{NN'}  +
     s_{\mu}^{l} s_{\nu}^N s_{\alpha}^{N'} \mathcal{D}^{\mu\nu\alpha}_{lNN'}
    \right). 
\end{equation} 
 The above expression contains seven spin  observables~\cite{Fasano:1992es}:
\begin{enumerate}[(i)]
        \item \label{type_A1}
        recoil polarization asymmetry   $\mathcal{P}_{N'}^\mu$;
        
        \item \label{type_A2}
        lepton polarization asymmetry   $\mathcal{P}_l^\mu$;
      
        \item \label{type_B} polarized target asymmetry $\mathcal{T}^\mu_N$;

        \item \label{type_C} lepton-recoil  asymmetry $A_{lN'}^{\mu\nu}$;

        \item \label{type_D1}  target-lepton asymmetry $B^{\mu\nu}_{lN}$;
        
        \item \label{type_D2} target-recoil asymmetry  $\mathcal{C}^{\mu\nu}_{NN'}$;
        
       \item \label{type_E} target-lepton-recoil asymmetry $\mathcal{D}^{\mu\nu\alpha}_{lNN'}$.
    \end{enumerate}

We calculated the spin asymmetry vectors and tensors using the full form of the current (\ref{Eq:Gamma-full}) and assuming that the form factors are complex-value functions. The analytic results are given in Appendix~\ref{Appendix_vector_and_tensors}.
\begin{figure}
	\centering{\includegraphics[width=0.9\textwidth]{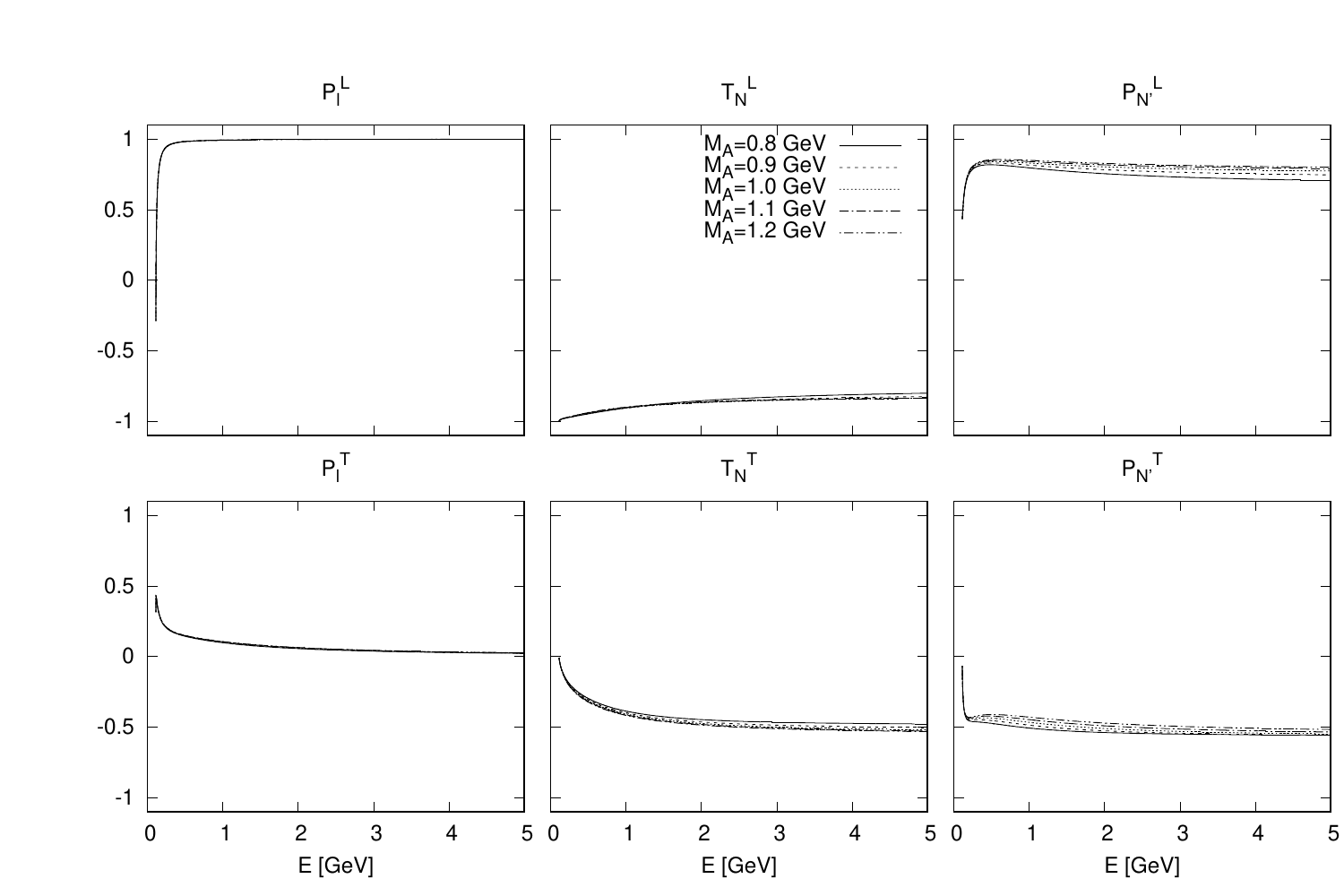}}
	\caption{Energy dependence of the  components of $\mathcal{P}_l $ (left column), $\mathcal{T}_{N}$ (middle column)  and  $\mathcal{P}_{N'}$ (right column). Results obtained  for the CCQE $\nu_\mu n$  scattering, the axial mass  $M_A=0.8$, $0.9$, $1.0$, $1.1$, $1.2$~GeV, and $F_3^A=0$. The longitudinal/transverse components are shown in top/bottom row. \label{Fig:Polarizations_nu_mu_total_MA}}
\end{figure}
\begin{figure}
	\centering{\includegraphics[width=0.9\textwidth]{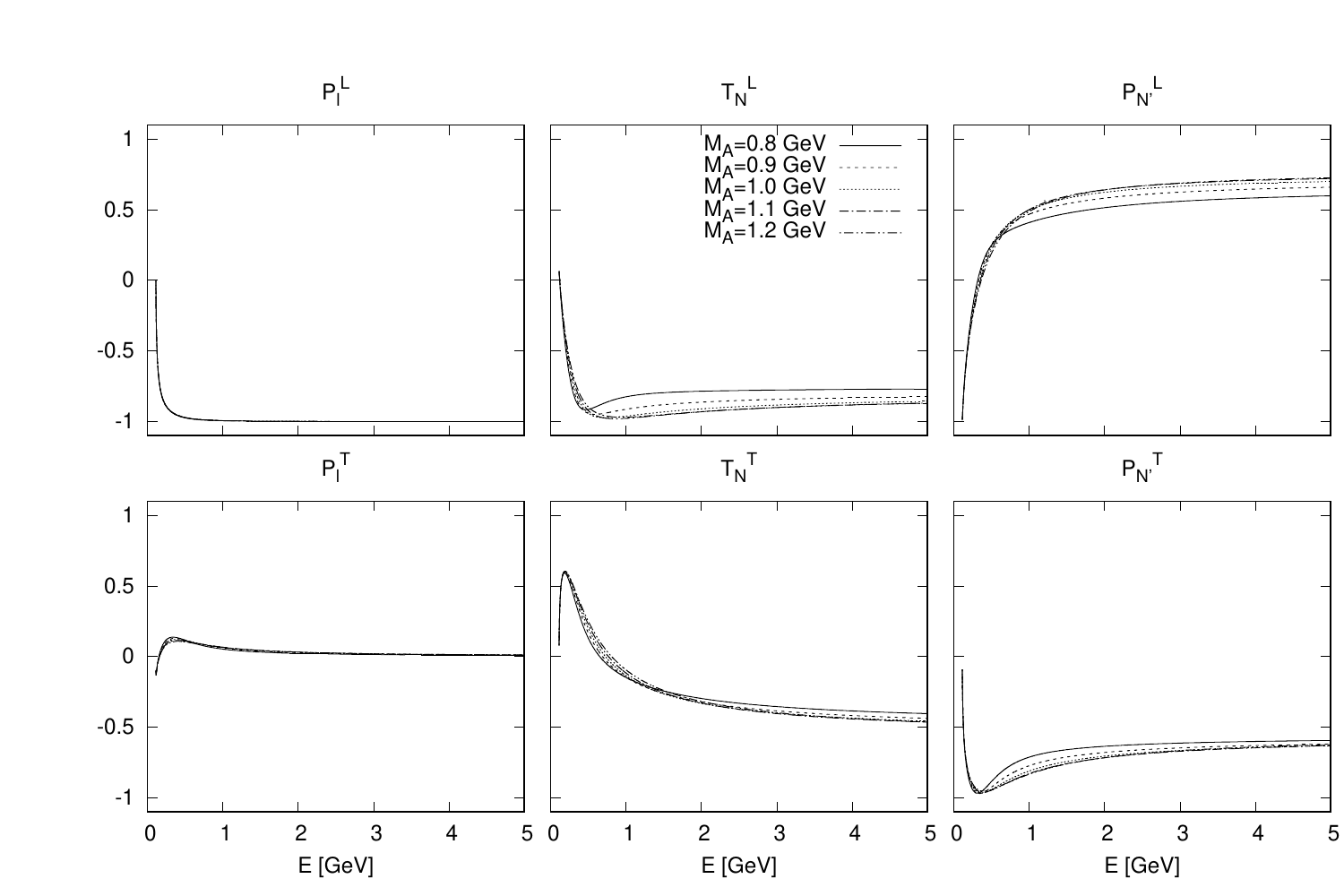}}
	\caption{Caption the same as in Fig. \ref{Fig:Polarizations_nu_mu_total_MA}   but for the CCQE $\overline{\nu}_\mu p$  scattering.
	\label{Fig:Polarizations_anti_nu_mu_total_MA}
	}
\end{figure}
\begin{figure} 
	\centering{\includegraphics[width=0.9\textwidth]{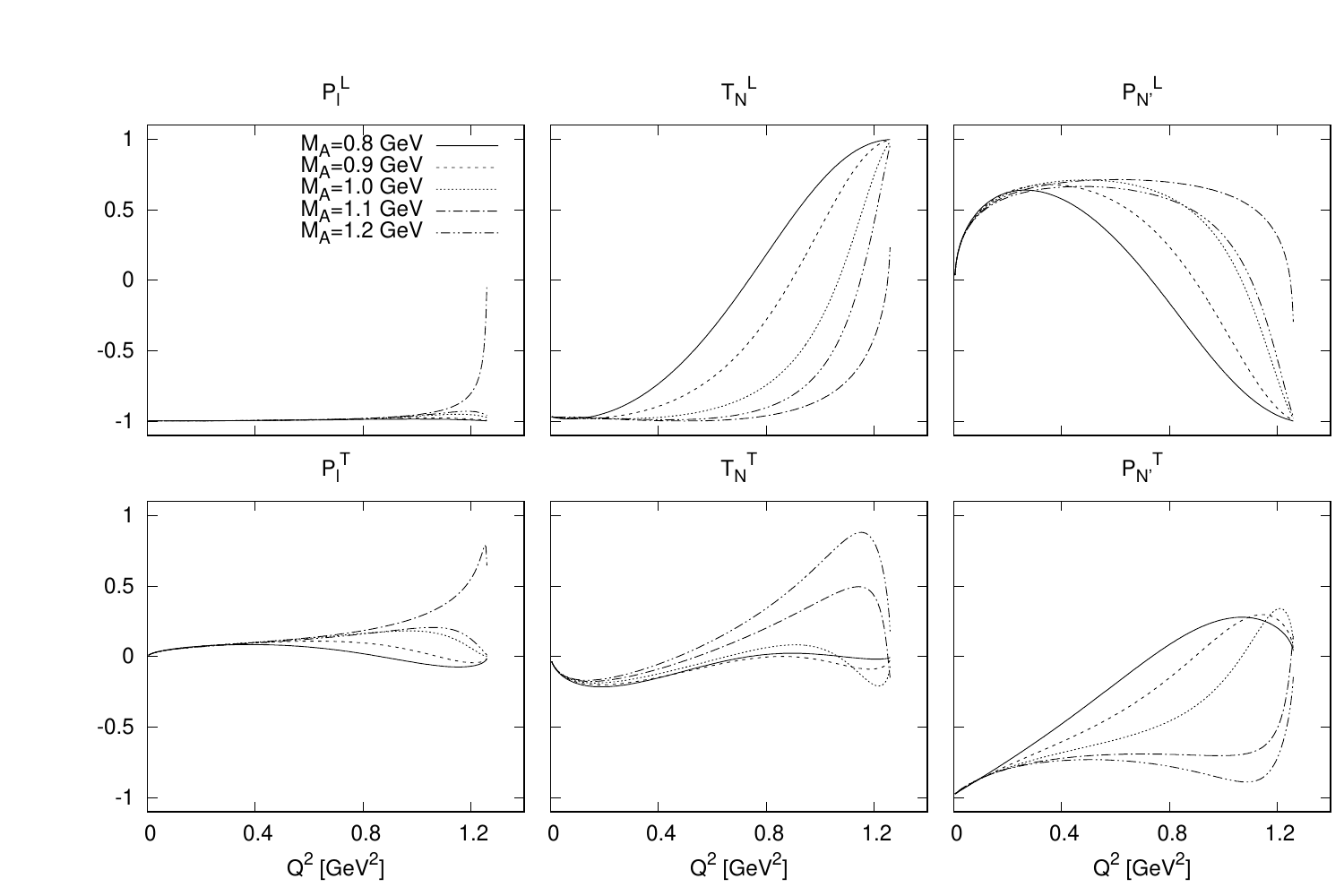}}
	\caption{Dependence of the  components of $\mathcal{P}_l $ (left column), $\mathcal{T}_{N}$ (middle column)  and  $\mathcal{P}_{N'}$ (right column) on $Q^2$. Results obtained  for the CCQE $\overline{\nu}_\mu p$ scattering,  $M_A=0.8$, $0.9$, $1.0$, $1.1$, $1.2$~GeV, $F_3^A=0$ and for energy $E=1$~GeV. In the top/bottom row longitudinal/transverse components are plotted.  \label{Fig:Polarizations_anti_nu_mu_diff_Q2_MA}}
\end{figure}
\begin{figure}
	\includegraphics[width=0.9\textwidth]{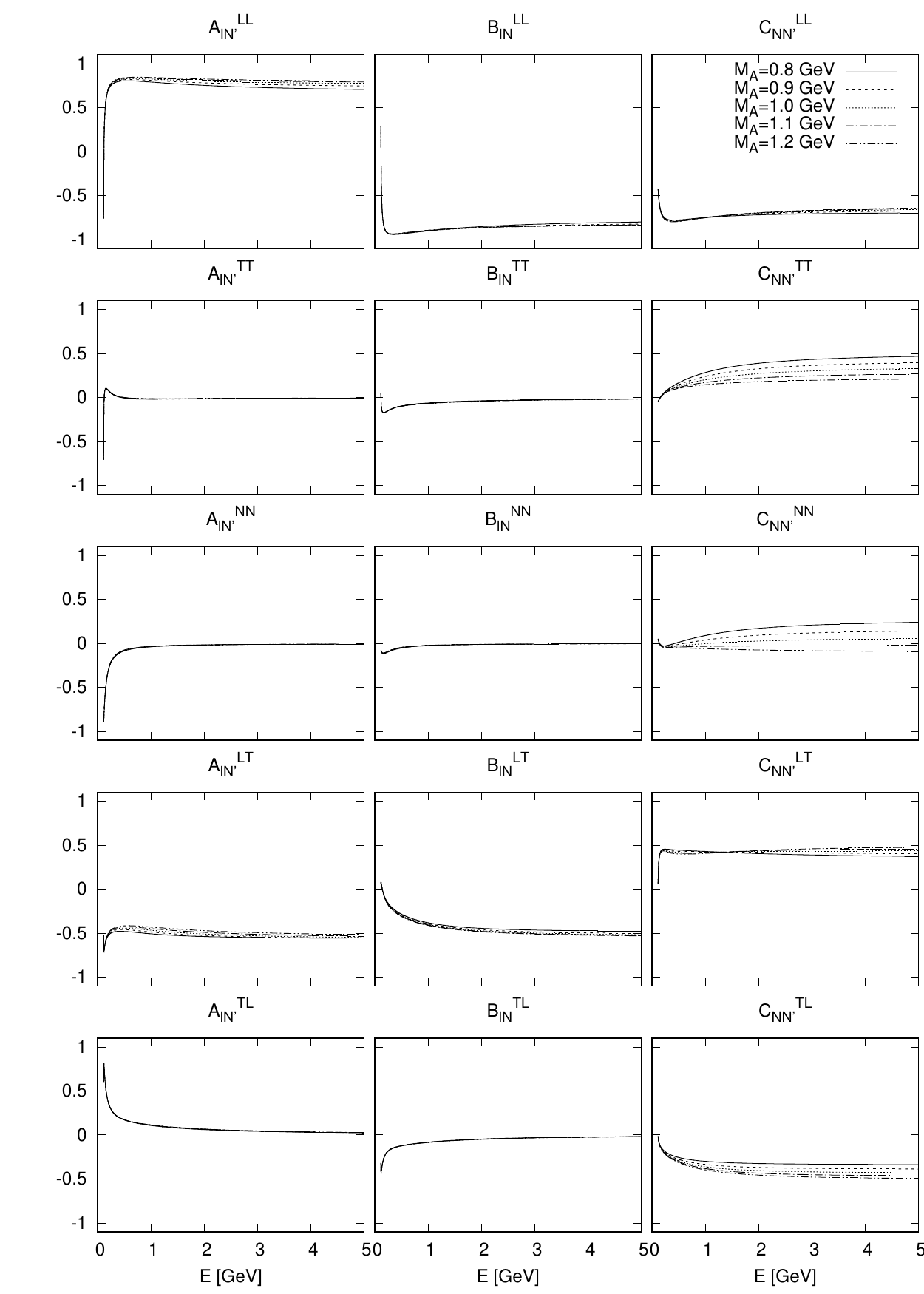}
	\caption{ Energy dependence of the components of  double spin asymmetries: $\mathcal{A}_{lN'}$ (left column),  $\mathcal{B}_{lN'}$ (middle column),  $\mathcal{C}_{NN'}$ (right column)    calculated for the CCQE $\nu_\mu n$  scattering for $M_A=0.8$, $0.9$, $1.0$, $1.1$, $1.2$~GeV and $F_3^A=0$.  
	\label{Fig:double_spin_total_nu_mu_MA}
	}
\end{figure}
\begin{figure}
	\includegraphics[width=0.9\textwidth]{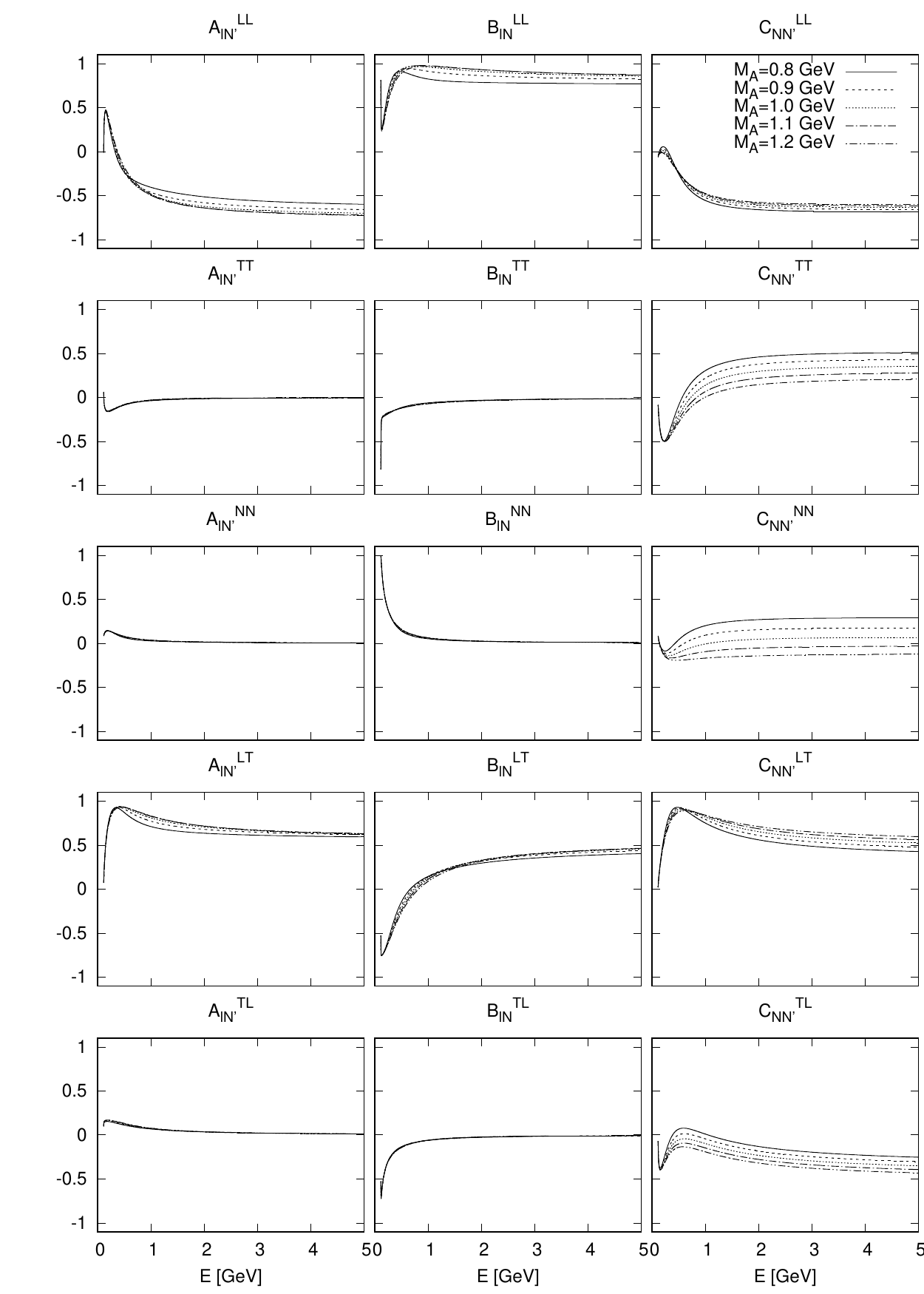}
	\caption{Caption the same as in Fig.~\ref{Fig:double_spin_total_nu_mu_MA} but for the CCQE $\overline{\nu}_\mu p$ scattering. \label{Fig:double_spin_total_antinu_mu_MA}
	}
\end{figure}

\subsection{Vector and tensor components of spin asymmetries}

We introduce  the spin basis to discuss the physical properties of the spin asymmetries:
\begin{itemize}
    \item basis for the outgoing lepton (see Fig. \ref{Fig_diagramQE})
    \begin{equation}
    \label{basis_lepton}
\zeta_L^\mu    = \frac{1}{m } \left(|\textbf{k}'| ,  \frac{ E_{k'}\textbf{k}'}{ |\textbf{k}'|} \right) \quad
\zeta_T^\mu        =  \left(0,\frac{\textbf{k}'\times(\textbf{k}\times \textbf{q})}{|\textbf{k}'\times(\textbf{k}\times \textbf{q})|}\right) 
\quad
\zeta_N^\mu        = \left(0,\frac{\textbf{k}\times \textbf{q}}{|\textbf{k}\times \textbf{q}|} \right),
\end{equation}
where $L$, $T$ and $N$ denote: longitudinal to the lepton momentum, transverse to the lepton momentum and normal to the scattering plane components, respectively;
\item basis  for the recoil nucleon:
\begin{equation}
\label{basis_recoil}
\xi_L^\mu        = \frac{1}{M} \left(|\textbf{q}|,\frac{ E_{p'} \textbf{q} }{|\textbf{q}|}\right),
\quad
\xi_T^\mu        =  \left(0,\frac{\textbf{q}\times(\textbf{k}\times \textbf{q})}{|\textbf{q}\times(\textbf{k}\times \textbf{q})|}\right), 
\quad
\xi_N^\mu        = \left(0,\frac{\textbf{k}\times \textbf{q}}{|\textbf{k}\times \textbf{q}|} \right); 
\end{equation}
\item basis for the target nucleon:
\begin{equation} 
\label{basis_target}
\chi_L^\mu    = \frac{1}{E} \left(0,\textbf{k} \right), \quad
\chi_T^\mu        = \left(0,\frac{\textbf{k}\times(\textbf{k}\times \textbf{q})}{|\textbf{k}\times(\textbf{k}\times \textbf{q})|}\right), \quad
\chi_N^\mu    =\left(0,\frac{\textbf{k}\times \textbf{q}}{|\textbf{k}\times \textbf{q}|}  \right). 
\end{equation}
\end{itemize}
Notice that, we have made the standard choice for  the basis vectors for the lepton (\ref{basis_lepton}) and recoil nucleon (\ref{basis_recoil}). For the target nucleon, the longitudinal vector is parallel to the neutrino momentum, and $\chi_N^\mu$ is normal to the scattering plane.   

Having the spin basis we calculate the longitudinal (L), transverse (T), and normal (N) components of the polarization asymmetry for the recoil nucleon and the outgoing lepton 
\begin{equation}
\mathcal{P}_{N'}^X \equiv    \xi^X_\mu  \mathcal{P}^{\mu}_{N'} \quad \mathrm{and} \quad
\mathcal{P}_{l}^X \equiv    \zeta^X_\mu  \mathcal{P}^{\mu}_{l},\quad X=L, T, N, 
\end{equation}
respectively. 

If all the form factors are real then 
\begin{equation}
    \mathcal{P}_{N'}^N = \mathcal{P}_{l}^N =0.
\end{equation}
Three independent components of the polarized target asymmetry are given by:
\begin{equation}
\mathcal{T}_{N}^X \equiv    \chi^X_\mu  \mathcal{T}^{\mu}_{N}, \quad X=L, T, N. 
\end{equation}
For real form factors $\mathcal{T}_{N}^N=0 $. 

The components of the double spin asymmetry tensor  are defined by
\begin{equation}
\mathcal{B}^{XY}_{lN} \equiv    \zeta^X_\mu \chi_\nu^Y \mathcal{B}^{\mu\nu}_{lN}, 
\quad
\mathcal{A}^{XY}_{lN'}  =  \zeta_\mu^X\xi_\nu^Y\mathcal{A}^{\mu\nu}_{lN'}, \quad \mathcal{C}^{XY}_{NN'}  =  \chi_\mu^X\xi_\nu^Y\mathcal{C}^{\mu\nu}_{NN'}.
\end{equation}
For real form factors 
\begin{equation}
0 =    \mathcal{A}^{y N}_{lN'} = \mathcal{B}^{y N}_{lN} = \mathcal{C}^{y N}_{NN'} = \mathcal{A}^{N y}_{lN'} = \mathcal{B}^{N y}_{lN} = \mathcal{C}^{N y}_{NN'}, y = L\; \mathrm{or} \; T.
\end{equation}
Then each tensor has five non-vanishing components. 

Eventually  the components of the triple spin asymmetry are defined by
\begin{equation}
\mathcal{D}_{lNN'}^{XYZ} \equiv  \zeta_\mu^X   \chi_\nu^Y \xi_\alpha^Z \mathcal{D}_{lNN'}^{\mu\nu\alpha}.
\end{equation}
When the form factors are real the components given by the contraction of $\mathcal{D}_{lNN'}^{\mu\nu\alpha}$ tensor with odd number of normal basis vector vanish e.g. $\mathcal{D}_{lNN'}^{N L T}=0$ or $\mathcal{D}_{lNN'}^{N NN}=0$, hence, there are $14$ independent components.  

Tensor components of the spin asymmetries, calculated assuming that form factors are real and $F_3^V=0$,  are given in Appendix~\ref{Appendix:Components}.

\begin{figure}
	\includegraphics[width=0.9\textwidth]{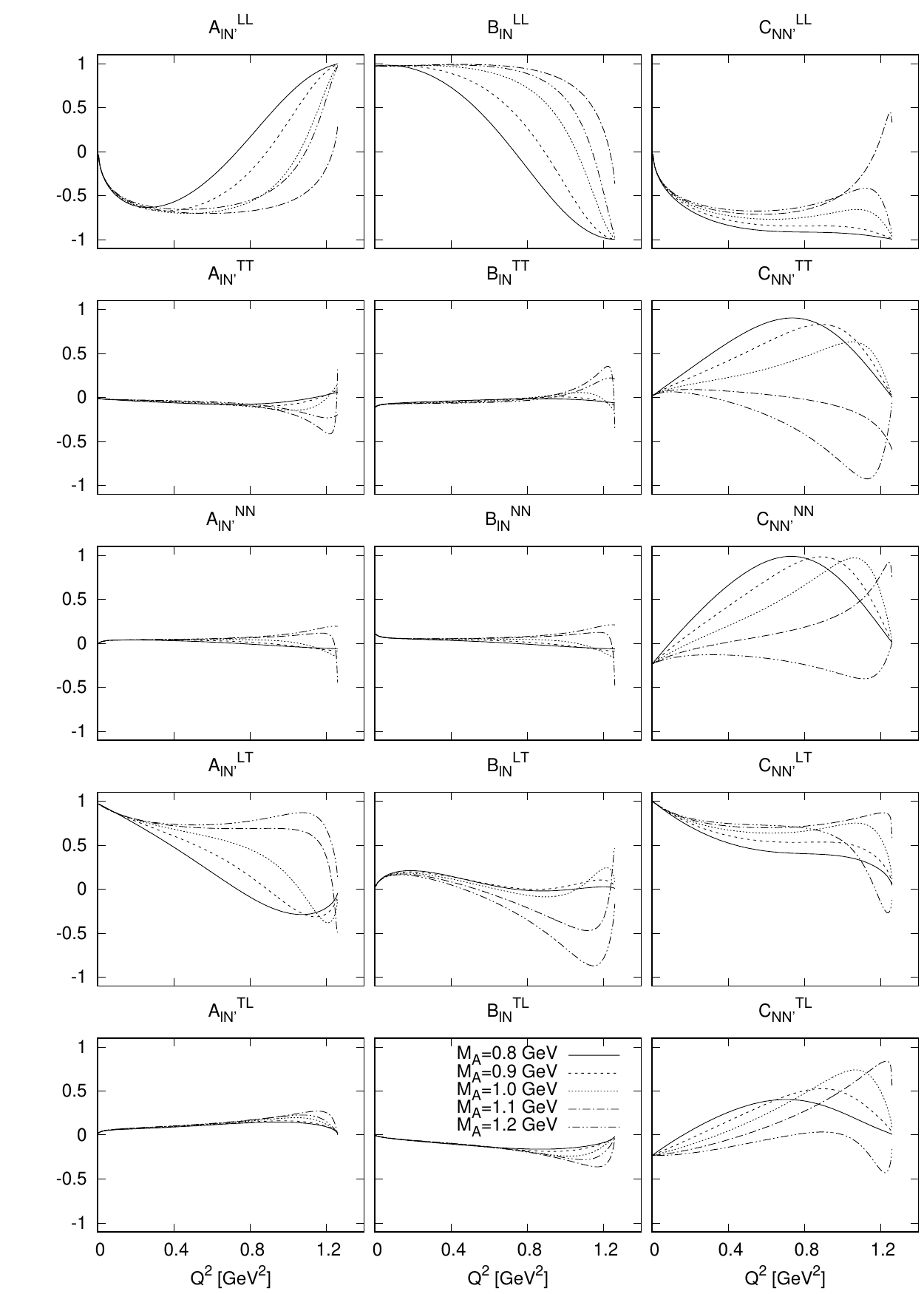}
	\caption{Dependence of the  components of $\mathcal{A}_{lN'} $ (left column), $\mathcal{B}_{lN}$ (middle column)  and  $\mathcal{C}_{NN'}$ (right column) on $Q^2$. Results obtained  for the CCQE $\overline{\nu}_\mu p$ scattering,  $M_A=0.8$, $0.9$, $1.0$, $1.1$, $1.2$~GeV, $F_3^A=0$ and energy $E=1$~GeV. In  \label{Fig:double_spin_Q2_antinu_mu_MA}
	}
\end{figure}
\begin{figure}
	\centering\includegraphics[width=0.55 \textwidth]{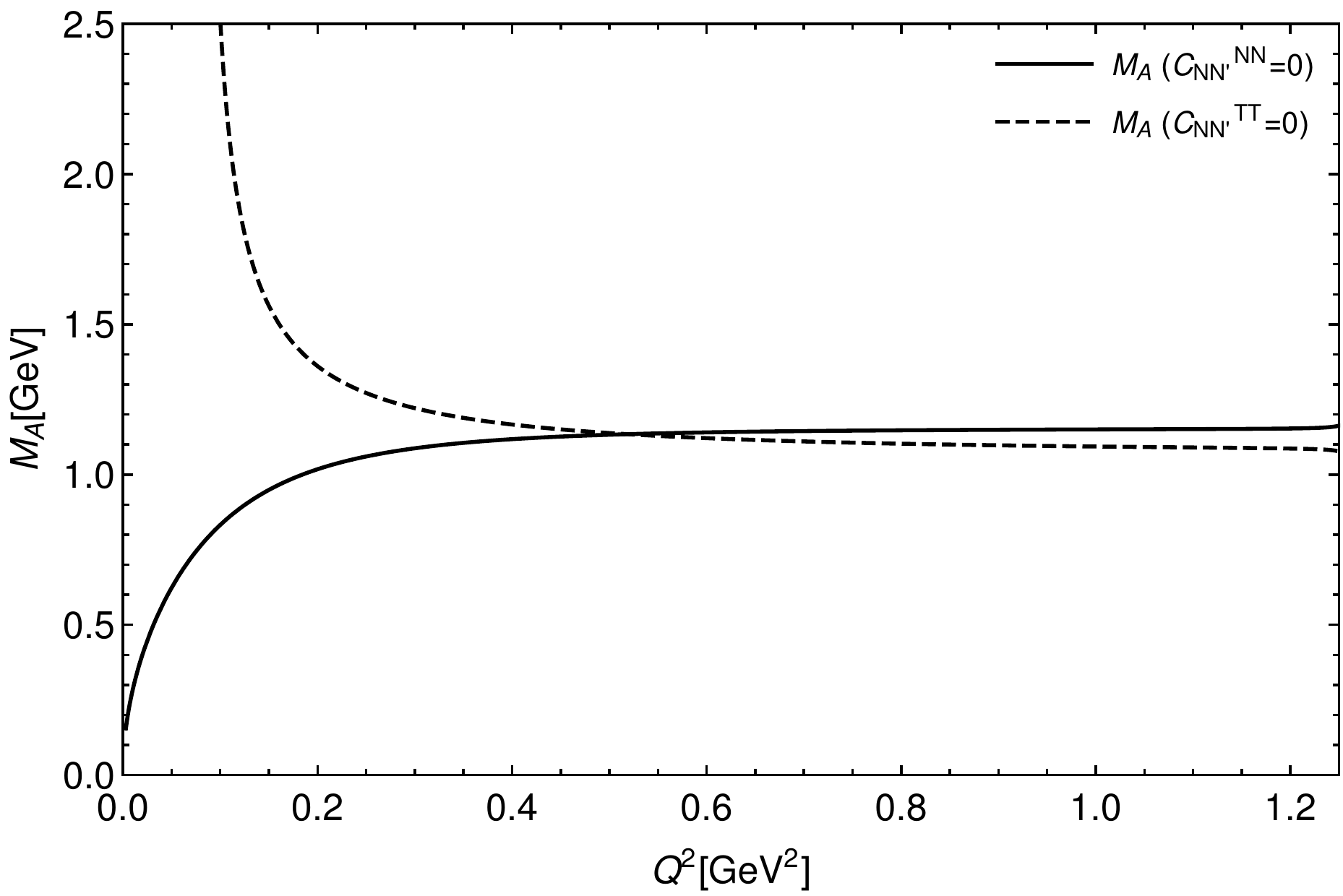}
	\caption{Axial mass dependence of the roots of the equations:  $C_{NN'}^{NN}(E=1~\mathrm{GeV},Q^2)=0$ (solid line) and $C_{NN'}^{TT}(E=1~\mathrm{GeV},Q^2)=0$ (dashed line) obtained for the CCQE $\overline{\nu}_\mu p$ scattering.   \label{Fig:MA}
	}
\end{figure}
\begin{figure}
	\includegraphics[width=\textwidth]{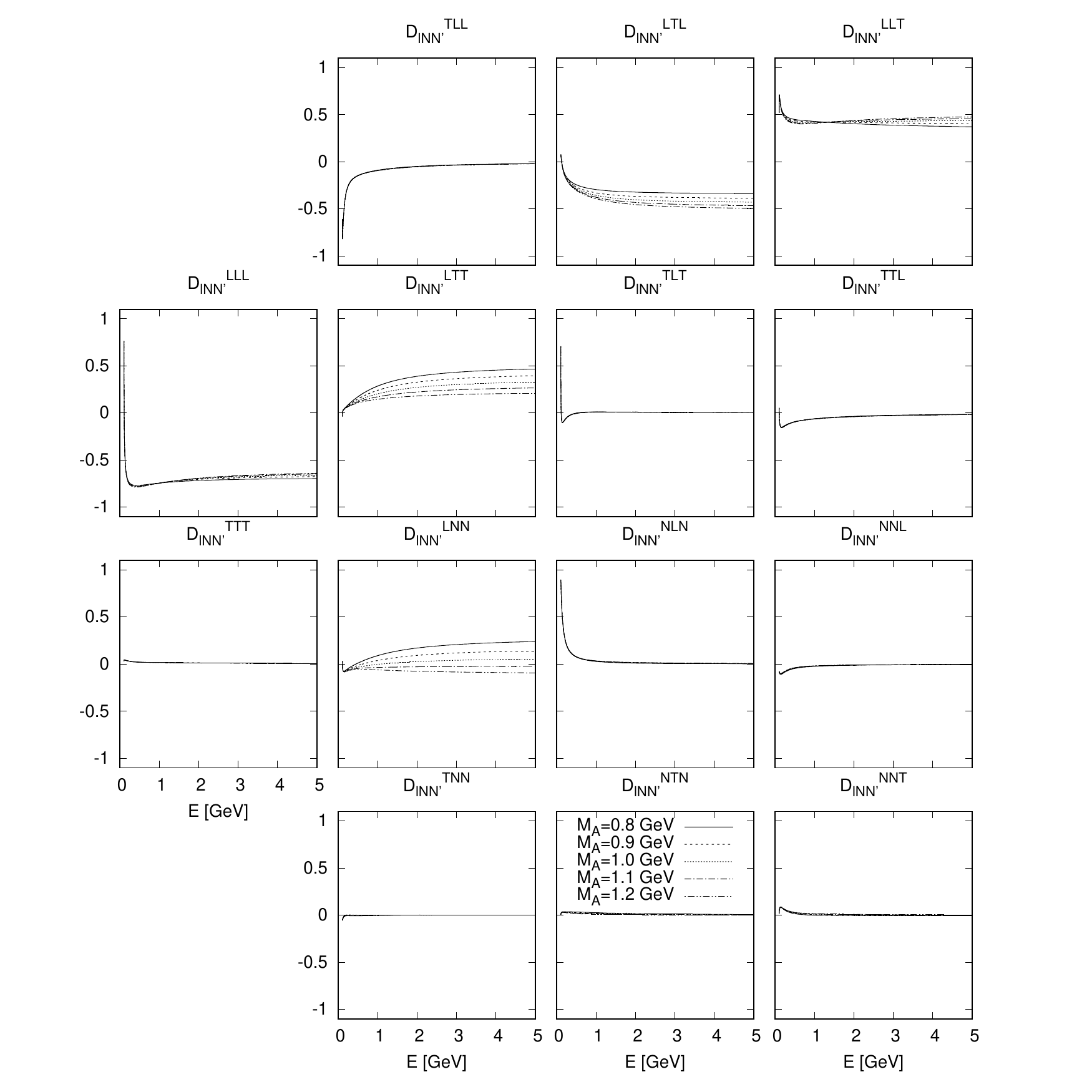}
	\caption{Energy dependence of the target-recoil-lepton spin asymmetry $\mathcal{D}_{lNN'}$   calculated for the CCQE $\nu_\mu n$ scattering, for $M_A=0.8$, $0.9$, $1.0$, $1.1$, $1.2$~GeV, $F_3^A=0$. 
	\label{Fig:triple_spin_total_nu_mu_MA}
	}
\end{figure}
\begin{figure}
	\includegraphics[width=\textwidth]{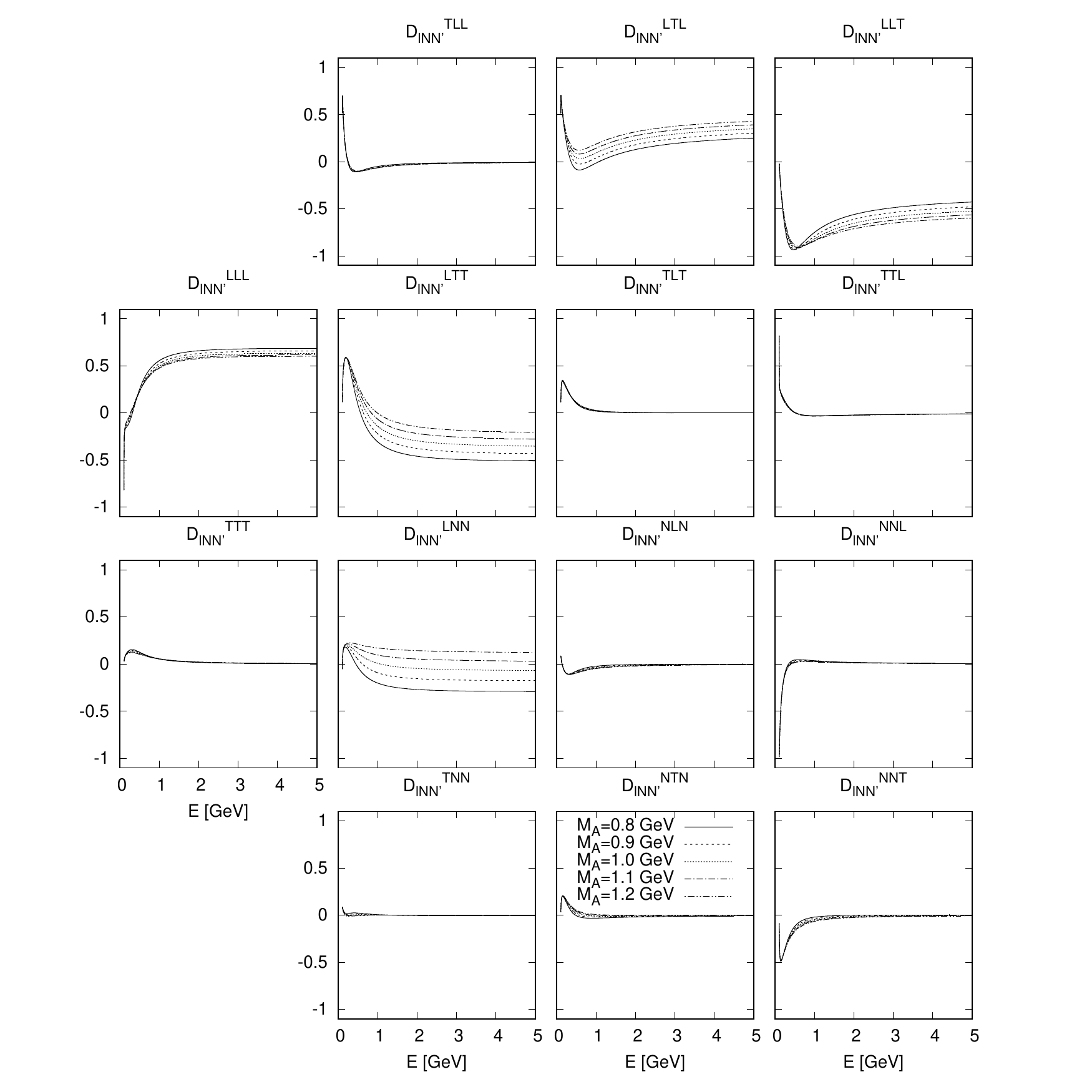}
	\caption{     Caption the same as in Fig.~\ref{Fig:triple_spin_total_nu_mu_MA} but for the CCQE $\overline{\nu}_\mu p$ scattering. 
	\label{Fig:triple_spin_total_antinu_mu_MA}
	}
\end{figure}
\begin{figure}
	\includegraphics[width=\textwidth]{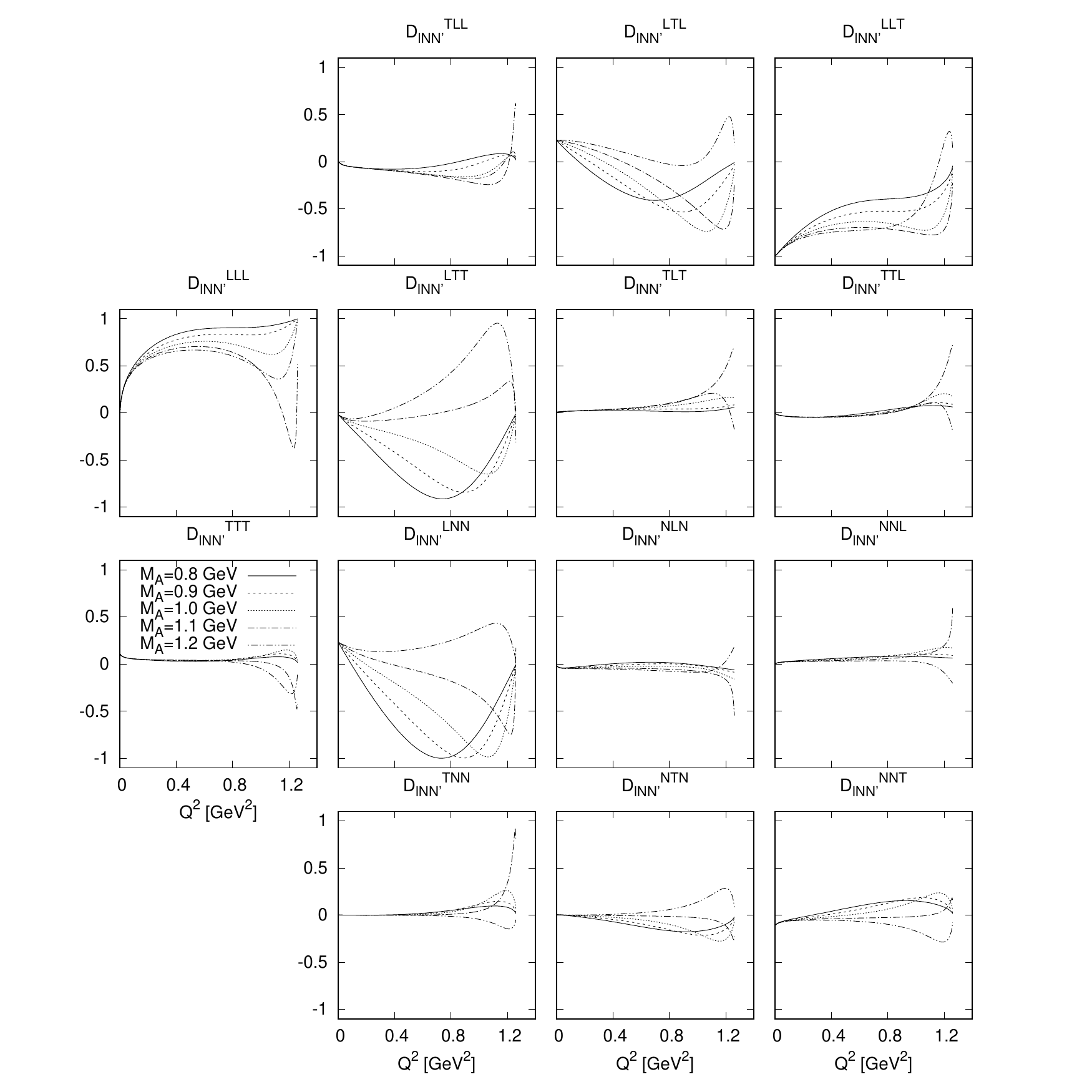}
	\caption{  The dependence of   the target-recoil-lepton asymmetry $\mathcal{D}_{lNN'}$  on $Q^2$, obtained for the CCQE $\bar{\nu}_\mu p$  scattering for $M_A=0.8$, $0.9$, $1.0$, $1.1$, $1.2$~GeV and $F_3^A=0$.   
	\label{Fig:triple_spin_Q2_antinu_mu_MA}
	}
\end{figure}
\begin{figure}
	\centering{\includegraphics[width=\textwidth]{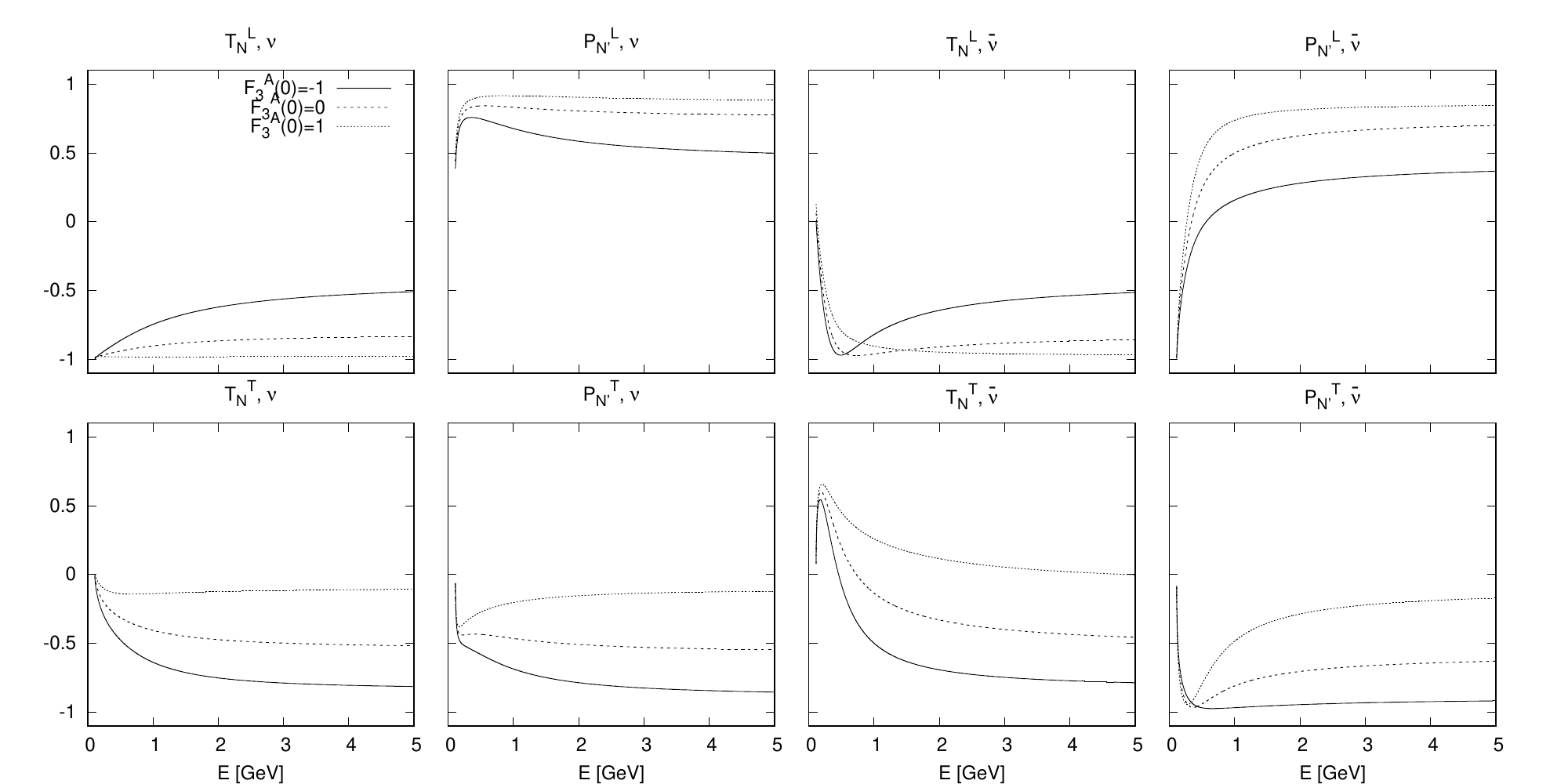}}
	\caption{  Dependence of the longitudinal (top row) and transverse (bottom row) components of the polarized target and recoil nucleon asymmetry, on  $F_3^A(0)$. The results obtained for the CCQE $\nu_\mu n$ (the first and the second columns) and $\overline{\nu}_\mu p$ (the third and fourth columns) interactions, $M_A=1.0$~GeV and $F_3^A(0)=-1$, $0$, $1$.  
	\label{Fig:Polarizations_nu_mu_and_anti_nu_mu_total_FA3}
	}
\end{figure}
\begin{figure}
	\includegraphics[width=\textwidth]{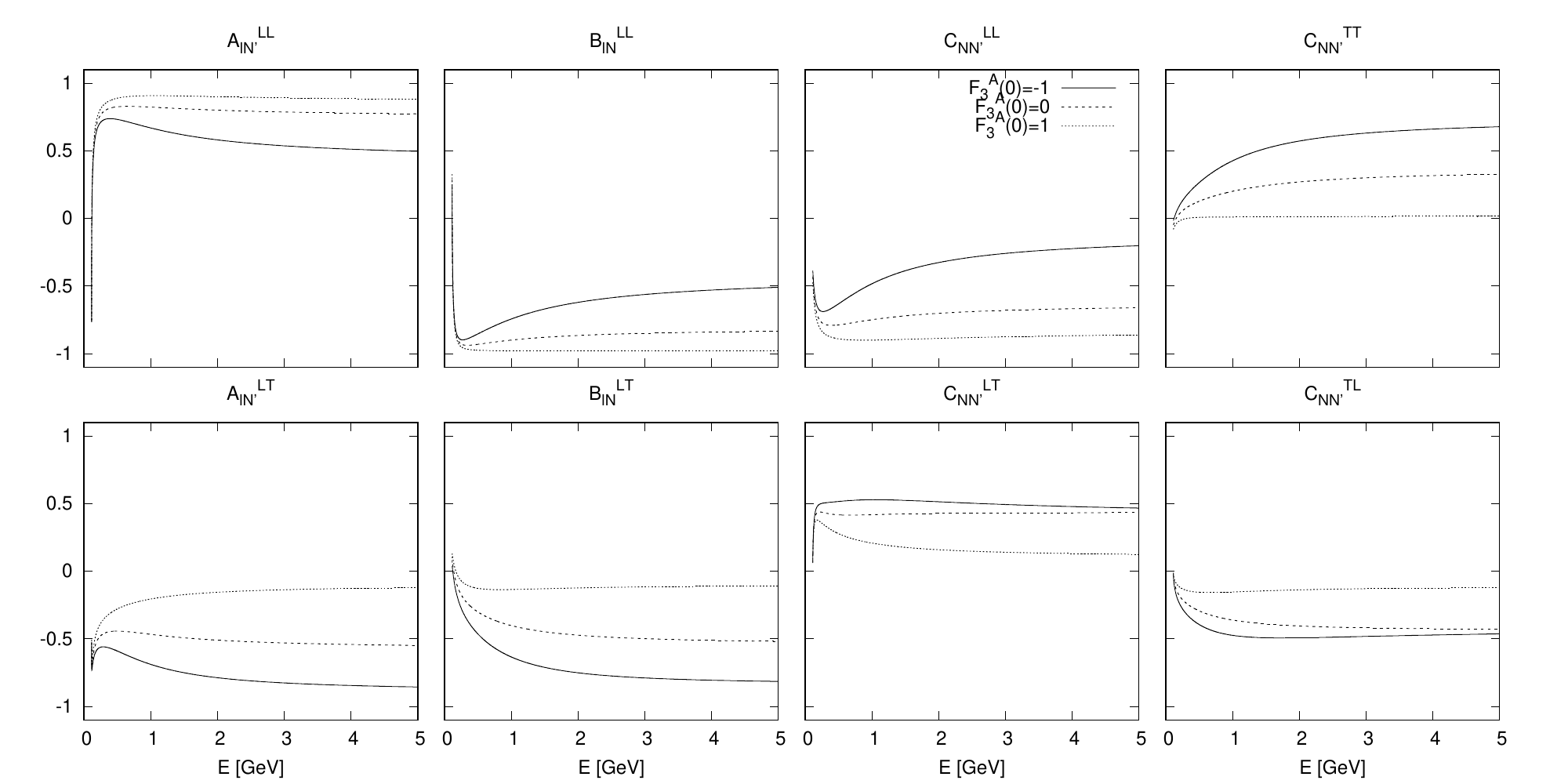}
	\caption{ Energy dependence of the components of  double spin asymmetries: $\mathcal{A}_{lN'}$ ,  $\mathcal{B}_{lN'}$ ,  $\mathcal{C}_{NN'}$    calculated for the CCQE $\nu_\mu n$  scattering with $M_A=1.0$~GeV and for  $F_3^A(0)=-1$, $0$ or $1$.
	\label{Fig:double_spin_total_nu_mu_F3A}
	}
\end{figure}
\begin{figure}
	\includegraphics[width=\textwidth]{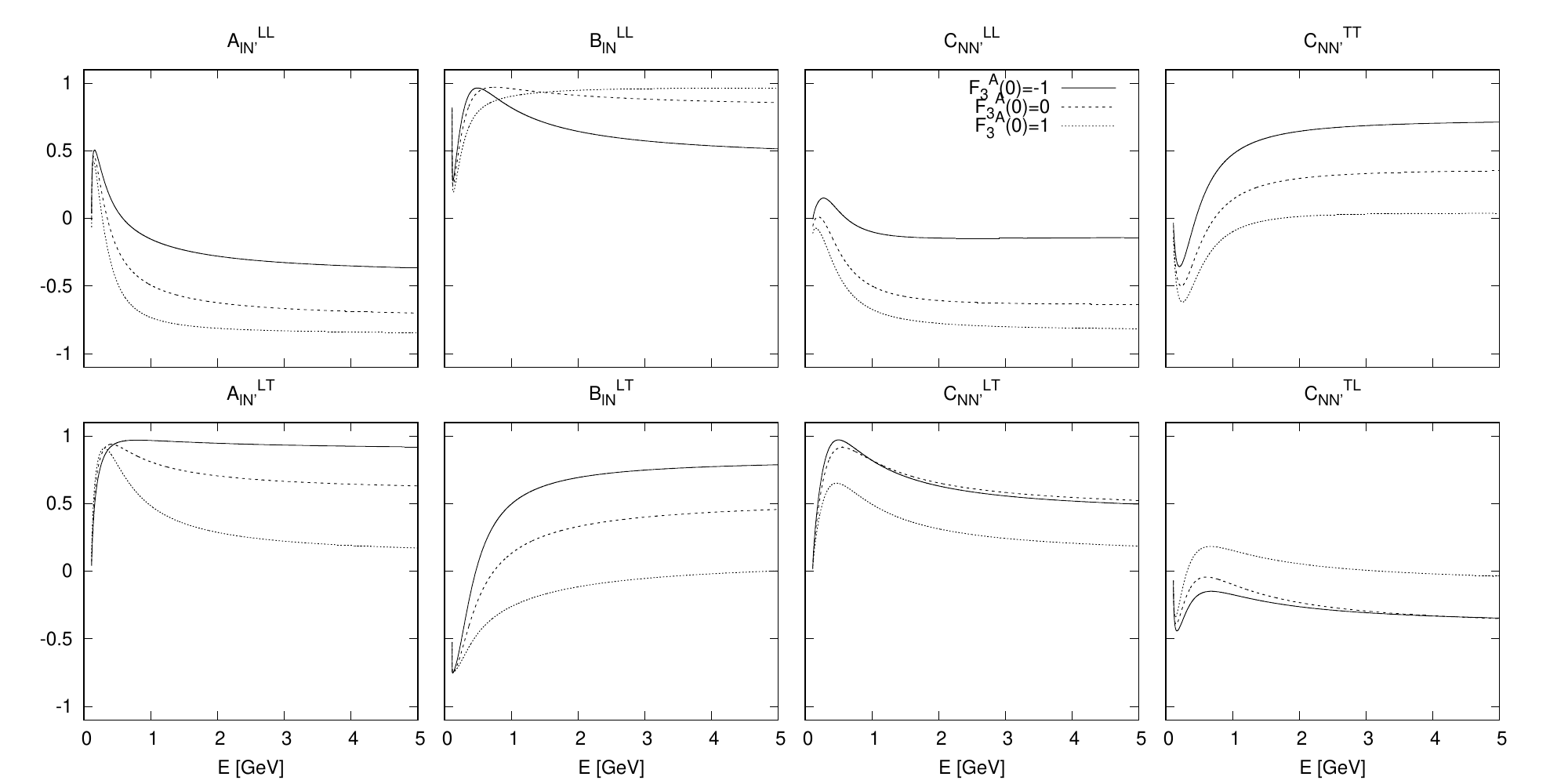}
	\caption{ Caption the same as in Fig.~\ref{Fig:double_spin_total_nu_mu_F3A} but for the CCQE $\overline{\nu}_\mu p$ scattering. \label{Fig:double_spin_total_antinu_mu_F3A}
	}
\end{figure}
\begin{figure}
	\includegraphics[width=\textwidth]{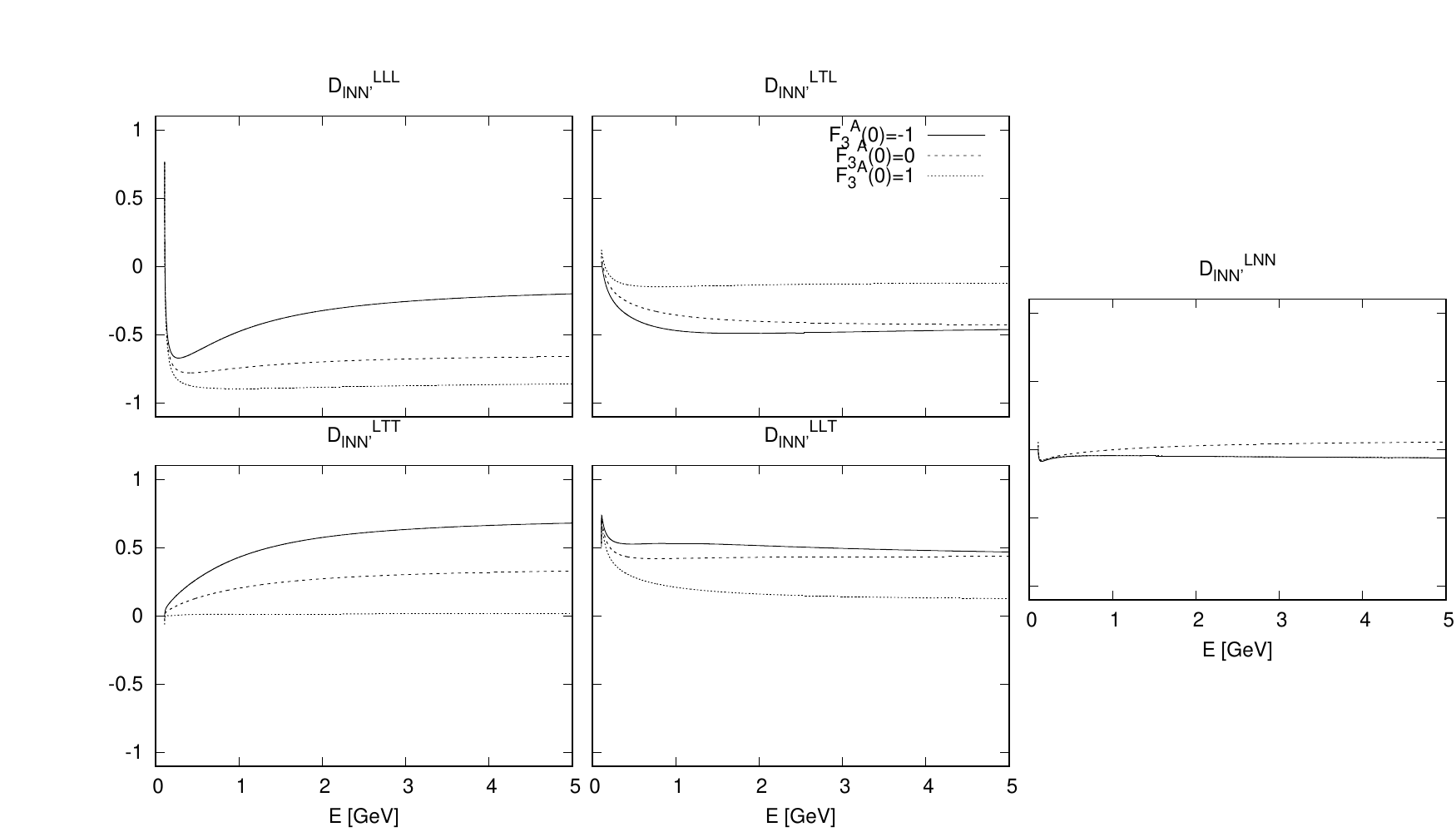}
	\caption{ Energy dependence of the components of  triple spin asymmetry $\mathcal{D}_{lNN'}$ calculated for the CCQE $\nu_\mu n$  scattering with $M_A=1.0$~GeV and for  $F_3^A(0)=-1$, $0$ or $1$.
	\label{Fig:triple_spin_total_nu_mu_F3A}
	}
\end{figure}
\begin{figure}
	\includegraphics[width=\textwidth]{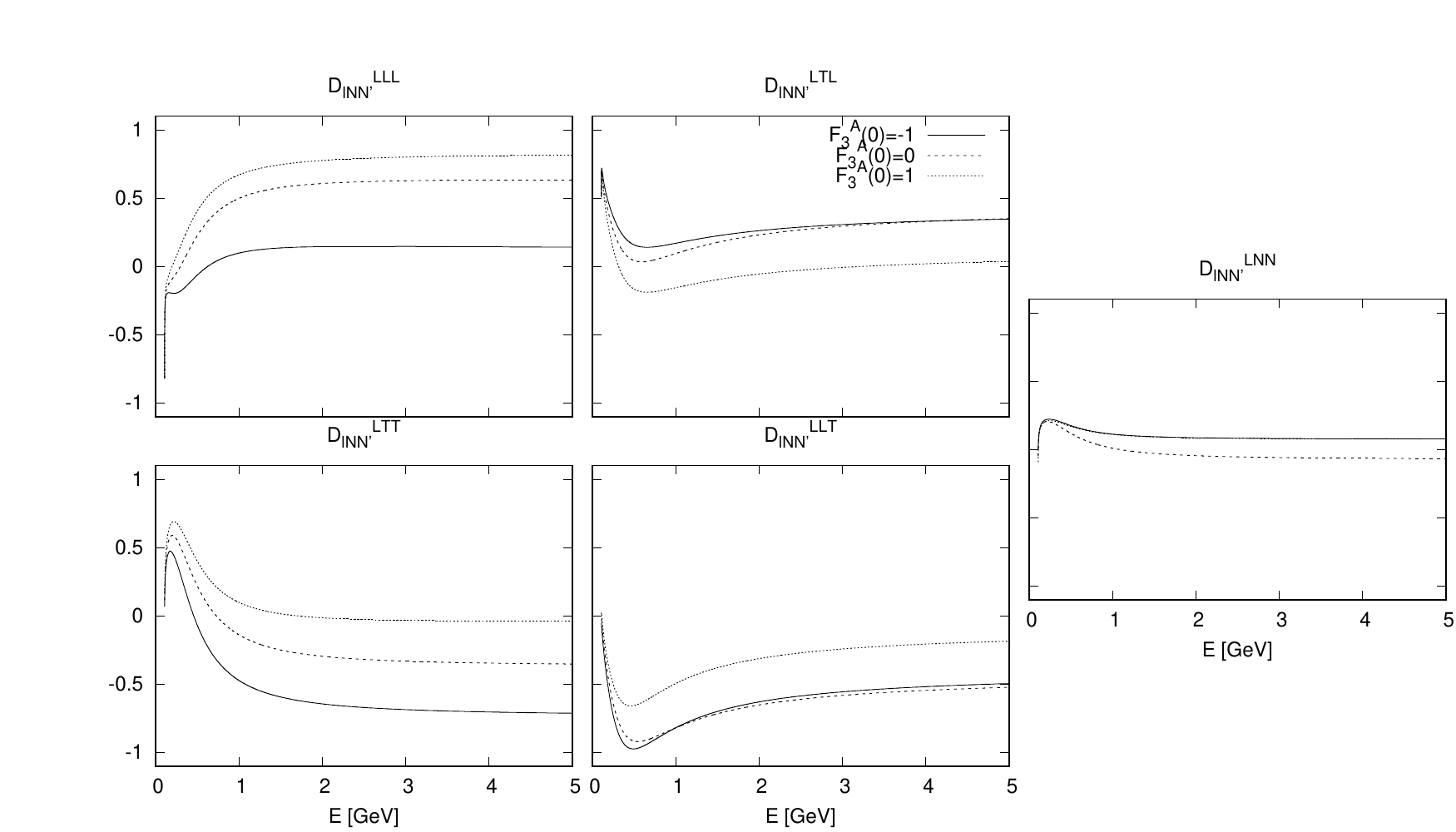}
	\caption{Caption the same as in \ref{Fig:triple_spin_total_nu_mu_F3A} but for the CCQE $\overline{\nu}_\mu p$ scattering.
	\label{Fig:triple_spin_total_antinu_mu_F3A}
	}
\end{figure}

\section{Numerical results and discussion}
\label{Section:Discussion}

We start the discussion of the results from the analyses of the sensitivity of spin asymmetries to the axial contribution. The numerical calculations are made assuming that: 
\begin{itemize}
\item All form factors are real;
\item  the SCC vanishes, hence  $F_3^V=0$ and $F_3^A=0$;
\item the axial form factor is given by the dipole parametrization: 
\begin{equation}
\label{Eq:axial_dipole}
F_A    (t)= \frac{g_A }{\displaystyle\left(1 -\frac{t }{M_A^2} \right)^2},
\end{equation}
where $M_A$ is the axial mass, which for the default value takes $1$~GeV, and $g_A = 1.2723 \pm 0.0023$~\cite{Patrignani:2016xqp}.
\end{itemize}
In the pre-analyses,  we considered also the non-dipole parametrization of $F_A$ defined by the sum of two poles \cite{Amaro:2015lga}.  But it turned out that effectively this parametrization works very similarly as the dipole with $M_A=1.4$~GeV. Therefore, in this work, we present the results only for $F_A$ given by Eq.~\ref{Eq:axial_dipole}. To mimic different shapes of the $F_A$ the value of  $M_A$ is varied, namely, $M_A=0.8$, $0.9$, $1.0$, $1.1$ and $1.2$~GeV.

In Figs. \ref{Fig:Polarizations_nu_mu_total_MA} and \ref{Fig:Polarizations_anti_nu_mu_total_MA} we  present the dependence of the components of the single-spin asymmetries on the energy. Both the CCQE processes are taken into consideration, namely $\nu_\mu n$ and $\overline{\nu}_\mu p$ interactions. The impact of the axial contribution on the recoil nucleon and the lepton polarization is discussed in Refs.~\cite{Fatima:2018tzs,Bilenky:2013iua,Fatima:2018tzs}. Our results confirm the conclusions of previous investigations, namely, the outgoing muon polarization, for both types of processes, weakly depends on the shape of the axial form factor. But the recoil polarization, for the $\overline{\nu}_\mu p$ scattering, is sensitive to $F_A$.    

As has been mentioned in the introduction,  the polarized target asymmetry has been not studied yet. Similarly as for  $\mathcal{P}_{N'}$ the longitudinal and transverse components of the $\mathcal{T}_N$, calculated for  $\overline{\nu}_\mu p$ scattering, are sensitive to the axial contribution. The change of the functional $Q^2$-dependence of $F_A$ results in distortion of the shape of the components of $\mathcal{T}_N$. This effect is the strongest for the antineutrino scattering for large $Q^2$ values,  see  Fig.~\ref{Fig:Polarizations_anti_nu_mu_diff_Q2_MA}.

The double spin asymmetries, as a function of energy, are plotted in Figs.~\ref{Fig:double_spin_total_nu_mu_MA} and \ref{Fig:double_spin_total_antinu_mu_MA}. 
Among three tensors $\mathcal{A}_{lN'}$, $\mathcal{B}_{lN}$ and $\mathcal{C}_{NN'}$, the last one - the target-recoil double spin asymmetry is the most sensitive to the axial contribution. In particular, this effect is visible for the components $C^{TT}_{NN'}$, and $C^{NN}_{NN'}$.
Indeed, a reduction of the axial mass leads to an increasing of the amplitude of the components $C^{TT}_{NN'}$, and $C^{NN}_{NN'}$ and a change of their signs from positive to negative. 
Indeed, for instance for $\overline{\nu}_\mu p$ scattering at fixed energy ($E=1$~GeV) for large values of  $M_A$ the observables $C^{TT}_{NN'}$ and $C^{NN}_{NN'}$ are negative whereas for large $M_A$ they take the positive values in full $Q^2$ range, see Fig.~\ref{Fig:double_spin_Q2_antinu_mu_MA}. Hence, the asymmetries $C^{TT}_{NN'}$ and  $C^{NN}_{NN'}$ are well suited  observables for estimation of the value of the $M_A$ and for studies of the $Q^2$-dependence of $F_A$. For an illustration of this property let us present the numerical solutions of  two equations, obtained for the CCQE $\overline{\nu}_\mu p$ scattering: 
 \begin{eqnarray}
     \label{Eq:MA_roots1}
     0&=&\mathcal{C}_{NN'}^{TT}(E=1~\mathrm{GeV},Q^2), \\
     \label{Eq:MA_roots2}
     0 &=& \mathcal{C}_{NN'}^{NN}(E=1~\mathrm{GeV},Q^2).
 \end{eqnarray} 
 The roots of the above equations depend on the value of $M_A$, see  Fig.~\ref{Fig:MA}. Notice that for too small values of the axial mass the first equation has no solution, whereas for too large values of $M_A$ the second equation has no roots. Moreover, for the value of $M_A\approx 1.0$~GeV suggested by the recent neutrino scattering data analyses,    the roots of Eq.~\ref{Eq:MA_roots2}  monotonically depend on $M_A$.

The triple spin asymmetry, $\mathcal{D}_{lNN'}$ is shown in Figs.~\ref{Fig:triple_spin_total_nu_mu_MA} and \ref{Fig:triple_spin_total_antinu_mu_MA}. Similarly as for the tensor $\mathcal{C}_{NN'}$ the spin asymmetry components calculated for $\overline{\nu}_\mu p$ scattering are more sensitive to  the axial form factor shape than the components calculated for neutrino scattering. In Fig.~\ref{Fig:triple_spin_Q2_antinu_mu_MA} we plot the $Q^2$-dependence of the components of the $\mathcal{D}_{lNN'}$ tensor.  The most sensitive to the shape of the axial form factor are the symmetries obtained from the contraction of the longitudinal component of the lepton spin vector with the tensor $\mathcal{D}_{lNN'}^{\mu\nu\alpha}$, namely, the components $\mathcal{D}_{lNN'}^{LNN}$, $\mathcal{D}_{lNN'}^{LTT}$ and $\mathcal{D}_{lNN'}^{LTN}$. Analogically as for the double spin asymmetry case, the sign of these components depends on the value of $M_A$. 

The  SCC contribution in the lepton and the recoil spin asymmetries was studied in Ref. \cite{Fatima:2018tzs}.  In our analysis, we additionally consider polarized target asymmetry, three double spin asymmetries, and triple spin asymmetry. In this part of the work, the numerical results are obtained assuming that:
\begin{itemize}
\item All form factors are real;
\item 
$F_3^A$ form factor is given by~\cite{Fatima:2018tzs}:
\begin{equation}
\label{Eq:dipoleF3A}
F_A^3(t) = \frac{F_A^3(0)}{(1 -t/M_A^2 )},  
\end{equation}
where $F_A^{(0)}$ is unknown parameter and for numerical calculations $F_A^3(0)=-1$, $0$ and $1$;
\item $F_3^V=0$;
\item $M_A=1.0$~ GeV in (\ref{Eq:axial_dipole}) and (\ref{Eq:dipoleF3A}).
\end{itemize}
To reduce the number of figures, we show only the components of the spin asymmetries which are sensitive to change of $F_A^3$. In Fig. \ref{Fig:Polarizations_nu_mu_and_anti_nu_mu_total_FA3} the single spin asymmetries are plotted whereas in Figs.~\ref{Fig:double_spin_total_nu_mu_F3A} and \ref{Fig:double_spin_total_antinu_mu_F3A} we show the double spin asymmetries, as a function of energy, calculated for $\nu_\mu n$ and $\overline{\nu}_\mu p$ interactions, respectively. The triple spin asymmetries are plotted in Figs.~\ref{Fig:triple_spin_total_nu_mu_F3A} and~\ref{Fig:triple_spin_total_antinu_mu_F3A}. Concluding this part, we state that outgoing muon spin asymmetry is insensitive to the SCC contribution, whereas other single, double and triple spin asymmetries are good observables for studying the signal from the non-standard interactions.

\section{Summary}

The spin observables in the CCQE $\nu_\mu n$ and $\overline{\nu}_\mu p$ interactions have been studied. We considered seven spin asymmetries. Five of them, namely,  the polarized target asymmetry, double and triple asymmetries have been not discussed yet.  All asymmetries were calculated analytically, they are given, in the covariant form, in  Appendix~\ref{Appendix_vector_and_tensors}.  To discuss the physical properties of the asymmetries, we introduced the spin basis vectors and the physical components of the asymmetries were computed and given in Appendix~\ref{Appendix:Components}. Eventually, we studied the dependence of the spin asymmetries on the axial and SCC contributions. 

We showed that the target-recoil asymmetry, $\mathcal{C}_{NN'}$, as well as the lepton-target-recoil asymmetry, $\mathcal{D}_{lNN'}$, are  well suited observables for studying the  the axial nucleon form factor. Indeed, the sign and magnitude of the components $\mathcal{C_{NN'}^{NN}}$, $\mathcal{C_{NN'}^{TT}}$ depend strongly on the axial mass value. Similar effect is found for  the components $\mathcal{D}_{lNN'}^{LNN}$, $\mathcal{D}_{lNN'}^{LTT}$ and $\mathcal{D}_{lNN'}^{LTN}$.
Eventually, we showed that all the spin asymmetries, except the lepton polarization asymmetry, are promising observables for investigation of the non-standard interactions in the neutrino scattering processes.

\section*{Acknowledgments}

 \begin{acknowledgments}
 The calculations have been carried out in Wroclaw Centre for Networking and Supercomputing (\url{http://www.wcss.wroc.pl}), grant No. 268.

A part of the algebraic calculations presented in this talk has been performed using FORM language \cite{Vermaseren:2000nd} and FeynCalc package \cite{Mertig:1990an,Shtabovenko:2016sxi}.
  \end{acknowledgments}

\appendix

\section{Convention, normalization and kinematics}
\label{Appendix_convention}

We work with the metric tensor  
\begin{equation}
g_{\mu\nu}=\mathrm{diag} (1,-1,-1,-1).
\end{equation}
For the Levi-Civita tensor we keep the normalization so that
\begin{equation}
\epsilon_{0123} = 1.
\end{equation}
The Dirac field of $1/2$-particle of mass $M$, momentum $p$ and spin $s$ is normalized so that:
\begin{equation}
\overline{u}(p,s) u(p,s') = 2 M \delta_{ss'},
\quad
u(p,s) \overline{u}(p,s) = \frac{1}{2}(1+\gamma_5 \slashed{s})(\slashed{p}+M).
\end{equation}

\section{Spin asymmetry vectors and tensors}

\label{Appendix_vector_and_tensors}

The results are obtained for full form of the charged current (\ref{Eq:Gamma-full}) assuming that the form factors are complex.
This appendix contains first class current result denoted by subscript $(1)$, interference between first and second class current contributions $(12)$ and pure second class current contribution denote by $(2)$.

\subsubsection{Recoil polarization asymmetry}
\begin{footnotesize}
\begin{eqnarray}
 \mathcal{P} _{N'(1)}^{\mu} &=&\frac{1}{M\mathcal{I}}
\left[ \right.
 4( Re(F_1^V) Re(F_A) + Im(F_1^V) Im(F_A)) M^2 (m^2 (2 k^{\mu} + p^{\mu}) + (2 k^{\mu} + p^{\mu}) {s_u}  - p^{\mu} t) \nonumber\\
+&&
4 \epsilon(k,p,q,{\mu}) (Re(F_P) Im(F_A)  -  Re(F_A) Im(F_P) )m^2 
+ 4 \epsilon(k,p,q,{\mu}) (Re(F_2^V) Im(F_1^V)  - Re(F_1^V) Im(F_2^V) ){s_u} \nonumber \\
&+& (Re(F_2^V) Re(F_P)  + Im(F_2^V) Im(F_P) )m^2 (-(m^2 p^{\mu}) + (2 k^{\mu} + p^{\mu}) t) \nonumber\\
&+&  (Re(F_1^V) Re(F_P)  + Im(F_1^V) Im(F_P) )m^2 (-(m^2 p^{\mu}) + p^{\mu} {s_u}  + (2 k^{\mu} + p^{\mu}) t) \nonumber\\
&+& (Re(F_1^V) Re(F_A) + Im(F_1^V) Im(F_A)) (p^{\mu} {s_u} ^2 + 4 M^2 (m^2 (2 k^{\mu} + p^{\mu}) - p^{\mu} t) + {s_u}  (-(m^2 p^{\mu}) + (2 k^{\mu} + p^{\mu}) t)) 
\nonumber\\
&+& 16 \epsilon(k,p,q,{\mu}) (Re(F_A) Im(F_1^V) - Re(F_1^V) Im(F_A)) M^2  x  \nonumber\\
&+& 4 |F_A|^2 M^2 p^{\mu} {s_u}   x   
 + 4 \epsilon(k,p,q,{\mu})( Re(F_A) Im(F_2^V)  -  Re(F_2^V) Im(F_A)) t  x \nonumber \\
&+& 4 |F_1^V|^2 M^2 (m^2 p^{\mu} + (-2 k^{\mu} - p^{\mu}) t)  x    
+ |F_2^V|^2 t (m^2 p^{\mu} - p^{\mu} {s_u}  + (-2 k^{\mu} - p^{\mu}) t)  x  \nonumber\\
&+& (Re(F_1^V) Re(F_2^V) +Im(F_1^V) Im(F_2^V) )(4 m^2 M^2 p^{\mu} + (m^2 p^{\mu} - 4 M^2 (2 k^{\mu} + p^{\mu})) t - p^{\mu} {s_u}  t + (-2 k^{\mu} - p^{\mu}) t^2)  x 
\left. \right] 
\\
\mathcal{P} _{N'(2)}^{\mu} &=&\frac{4}{M\mathcal{I}}
\left[ \right. 
-Re(F_3^A) Re(F_3^V) m^2 p^{\mu} {s_u}
\left. \right]
\\
 \mathcal{P} _{N'(12)}^{\mu}& =&\frac{2}{M \mathcal{I}}
\left[ \right. 
4 \epsilon(k,p,q,{\mu}) Re(F_3^V) (Im(F_1^V) +Im(F_2^V)) m^2 + 4 \epsilon(k,p,q,{\mu}) Re(F_3^A) (Im(F_A)  - Im(F_3^A) ){s_u} \nonumber \\
&+& Re(F_3^V) Re(F_P) m^2 p^{\mu} (-m^2 + t) + Re(F_3^V) Re(F_A) m^2 (-8 k^{\mu} M^2 - m^2 p^{\mu} + p^{\mu} {s_u}  + (2 k^{\mu} + p^{\mu}) t) \nonumber\\
&+& (Re(F_1^V) Re(F_3^A)+Im(F_1^V) Im(F_3^A)) (p^{\mu} {s_u} ^2 + 4 M^2 p^{\mu} (-m^2 + t) + {s_u}  (-(m^2 p^{\mu}) + (2 k^{\mu} + p^{\mu}) t)) \nonumber\\
&+&( Re(F_2^V) Re(F_3^A)+Im(F_2^V) Im(F_3^A)) (p^{\mu} t (-m^2 + t) + {s_u}  (-(m^2 p^{\mu}) + (2 k^{\mu} + p^{\mu}) t)) \nonumber\\
&-& 4 \epsilon(k,p,q,{\mu}) Re(F_3^A) (Im(F_1^V) + Im(F_2^V)) t  x + 4 \epsilon(k,p,q,{\mu}) (Re(F_1^V) +Re(F_2^V)) Im(F_3^A) t  x  \nonumber \\
&+& (Re(F_3^A) Re(F_A) +Im(F_3^A) Im(F_A)) (-4 m^2 M^2 p^{\mu} + (m^2 p^{\mu} + 4 M^2 (2 k^{\mu} + p^{\mu})) t - p^{\mu} {s_u}  t - (2 k^{\mu} + p^{\mu}) t^2)  x \left. \right] 
\end{eqnarray}
\end{footnotesize}

\subsubsection{Polarization asymmetry of the charged lepton}

\begin{footnotesize} 
 \begin{eqnarray}
\mathcal{P} _{l(1)}^{\lambda} &=&\frac{m}{2M^2\mathcal{I}}
\left[ \right.
8 (Re(F_1^V) Re(F_2^V) + Im(F_1^V) Im(F_2^V)) k^{\lambda}  M^2 (m^2 - 2 t)  x  + |F_P|^2 k^{\lambda} m^2 t  x  \nonumber \\
&+& |F_2^V|^2  ((-k^{\lambda} - 2 p^{\lambda}) {s_u}  t + k^{\lambda} (4 m^2 M^2 - 4 M^2 t - t^2))  x \nonumber \\
&+& 8 ((Re(F_1^V)  + Re(F_2^V) )Re(F_A) + (Im(F_1^V)  + Im(F_2^V)) Im(F_A))  M^2 (m^2 (k^{\lambda} + 2 p^{\lambda}) + k^{\lambda} {s_u}  + (-k^{\lambda} - 2 p^{\lambda}) t) \nonumber \\
&+& 8 (Re(F_A) Re(F_P)+ Im(F_A) Im(F_P)) k^{\lambda} m^2 M^2  x  + 4( |F_1^V|^2 +|F_A|^2 )  M^2 ((k^{\lambda} + 2 p^{\lambda}) {s_u}  + k^{\lambda} (m^2 + 4 M^2 - t))  x  
\left. \right]
 \end{eqnarray}
 \begin{eqnarray}
\mathcal{P} _{l(2)}^{\lambda}&=&\frac{m}{M^2\mathcal{I}}
\left[ \right.
(Im(F_3^V)^2 - Re(F_3^V)^2)k^{\lambda} m^2 (5 M^2 - 2 t)  x  
+ 2|F_3^A|^2  ((-k^{\lambda} - 2 p^{\lambda}) {s_u}  + k^{\lambda} (4 M^2 - t)) t  x  
\left. \right]
 \\
\mathcal{P} _{l(12)}^{\lambda} &=&\frac{m}{M^2\mathcal{I}}
\left[ \right.
16 \epsilon(k,p,q,{\lambda}) (Re(F_3^V) Im(F_1^V)+ Re(F_A) Im(F_3^A) - Re(F_3^A) Im(F_A))  M^2 \nonumber \\
&+& 4 \epsilon(k,p,q,{\lambda})( Re(F_3^V) Im(F_2^V) + Re(F_P) Im(F_3^A) - Re(F_3^A) Im(F_P) )  t \nonumber \\
&+& 4 (Re(F_1^V) Re(F_3^V) + Re(F_3^A) Re(F_A) +Im(F_3^A) Im(F_A))   M^2 (m^2 (-k^{\lambda} - 2 p^{\lambda}) + k^{\lambda} {s_u}  + (k^{\lambda} + 2 p^{\lambda}) t)  x  \nonumber \\
&+& (Re(F_2^V) Re(F_3^V) +Re(F_3^A) Re(F_P) + Im(F_3^A) Im(F_P))  t (m^2 (-k^{\lambda} - 2 p^{\lambda}) + k^{\lambda} {s_u}  + (k^{\lambda} + 2 p^{\lambda}) t)  x 
\left. \right]
\end{eqnarray}
\end{footnotesize}

\subsubsection{Polarized target asymmetry}

\begin{footnotesize}
\begin{eqnarray}
\mathcal{T} _{N(1)}^{\mu} &=&\frac{1}{M\mathcal{I}}
\left[ \right.4 (Re(F_1^V) Re(F_A) + Im(F_1^V) Im(F_A)) M^2 (m^2 (-2 k^{\nu} + q^{\nu}) + (2 k^{\nu} - q^{\nu}) {s_u}  - q^{\nu} t) \nonumber \\
&-& 
4 \epsilon(k,p,q,{\nu}) (Re(F_P) Im(F_A) - Re(F_A) Im(F_P)) m^2\nonumber \\
&+& 4 \epsilon(k,p,q,{\nu}) (Re(F_2^V) Im(F_1^V) -Re(F_1^V) Im(F_2^V)) {s_u}  \nonumber \\
&+& 
(Re(F_2^V) Re(F_P) + Im(F_2^V) Im(F_P)) m^2 (-(m^2 q^{\nu}) + (-2 k^{\nu} + q^{\nu}) t) \nonumber \\
&+& (Re(F_1^V) Re(F_P) + Im(F_1^V) Im(F_P)) m^2 (-(m^2 q^{\nu}) - q^{\nu} {s_u}  + (-2 k^{\nu} + q^{\nu}) t) \nonumber \\
&+& (Re(F_2^V) Re(F_A) + Im(F_2^V) Im(F_A)) (q^{\nu} {s_u} ^2 + {s_u}  (m^2 q^{\nu} + (2 k^{\nu} - q^{\nu}) t) + 4 M^2 (m^2 (-2 k^{\nu} + q^{\nu}) - q^{\nu} t)) \nonumber \\
&+& 16 \epsilon(k,p,q,{\nu}) (Re(F_A) Im(F_1^V) -Re(F_1^V) Im(F_A)) M^2  x\nonumber \\
&+&4 \epsilon(k,p,q,{\nu}) (Re(F_A) Im(F_2^V) - Re(F_2^V) Im(F_A)) t  x  \nonumber \\
&+& 4 |F_A|^2 M^2 q^{\nu} {s_u}   x  + 4 |F_1^V|^2 M^2 (-(m^2 q^{\nu}) + (-2 k^{\nu} + q^{\nu}) t)  x  + |F_2^V|^2 t (-(m^2 q^{\nu}) - q^{\nu} {s_u}  + (-2 k^{\nu} + q^{\nu}) t)  x  \nonumber \\
&+& (Re(F_1^V) Re(F_2^V) + Im(F_1^V) Im(F_2^V))    (-4 m^2 M^2 q^{\nu} + (-(m^2 q^{\nu}) + 4 M^2 (-2 k^{\nu} + q^{\nu})) t - q^{\nu} {s_u}  t + (-2 k^{\nu} + q^{\nu}) t^2)  x 
\left. \right]
\end{eqnarray}
\begin{eqnarray}
\mathcal{T} _{N(2)}^{\mu} &=&\frac{4}{M\mathcal{I}}Re(F_3^A) Re(F_3^V) m^2 q^{\nu} {s_u} 
\end{eqnarray}
\begin{eqnarray}
\mathcal{T} _{N(12)}^{\mu} &=&\frac{2}{M\mathcal{I}}
\left[ \right.
Re(F_3^V) Re(F_P) m^2 q^{\nu} (m^2 - t) + Re(F_3^V) Re(F_A) m^2 (-8 k^{\nu} M^2 + m^2 q^{\nu} + q^{\nu} {s_u}  + (2 k^{\nu} - q^{\nu}) t) \nonumber \\
&+& 
4 \epsilon(k,p,q,{\nu}) (Re(F_3^V) Im(F_1^V) +Re(F_3^V) Im(F_2^V)) m^2 
+ 4 \epsilon(k,p,q,{\nu}) (Re(F_A) Im(F_3^A) - Re(F_3^A) Im(F_A)) {s_u}  \nonumber \\
&+& 
(Re(F_1^V) Re(F_3^A) + Im(F_1^V) Im(F_3^A)) (-(q^{\nu} {s_u} ^2) + 4 M^2 q^{\nu} (m^2 - t) + {s_u}  (-(m^2 q^{\nu}) + (-2 k^{\nu} + q^{\nu}) t)) \nonumber \\
&+& (Re(F_2^V) Re(F_3^A) + Im(F_2^V) Im(F_3^A)) (q^{\nu} (m^2 - t) t + {s_u}  (-(m^2 q^{\nu}) + (-2 k^{\nu} + q^{\nu}) t)) \nonumber \\
&+& 4 \epsilon(k,p,q,{\nu}) (Re(F_3^A) Im(F_1^V) +Re(F_3^A) Im(F_2^V) - Re(F_1^V) Im(F_3^A) - Re(F_2^V) Im(F_3^A)) t  x  \nonumber \\
&+& (Re(F_3^A) Re(F_A) +Im(F_3^A) Im(F_A)) (-4 m^2 M^2 q^{\nu} + (m^2 q^{\nu} + 4 M^2 (-2 k^{\nu} + q^{\nu})) t + q^{\nu} {s_u}  t + (2 k^{\nu} - q^{\nu}) t^2)  x 
\left. \right]
\end{eqnarray}
\end{footnotesize}

\subsubsection{Lepton-recoil asymmetry}
\begin{footnotesize}
\begin{eqnarray}
\mathcal{A}_{lN'(1)}^{\lambda\mu} &=&\frac{m}{M\mathcal{I}}
\left[ \right.
4 \epsilon(k,p,q,{\lambda}) (-Re(F_P) Im(F_1^V) +Re(F_A) Im(F_2^V) - Re(F_2^V) Im(F_A) +Re(F_1^V) Im(F_P)) m p^{\mu} \nonumber \\
&+& 4 |F_A|^2   M^2 (2 k^{\mu} (-k^{\lambda} - 2 p^{\lambda}) +  g^{\lambda \mu}  {s_u} ) + 4 |F_1^V|^2   M^2 (2 k^{\lambda} k^{\mu} + m^2  g^{\lambda \mu}  -  g^{\lambda \mu}  t) \nonumber \\
&+& |F_2^V|^2  (m^2 (k^{\lambda} + 2 p^{\lambda}) p^{\mu} + k^{\lambda} p^{\mu} {s_u}  + (2 k^{\lambda} k^{\mu} - (k^{\lambda} + 2 p^{\lambda}) p^{\mu} + m^2  g^{\lambda \mu} ) t -  g^{\lambda \mu}  t^2) \nonumber \\
&+& (Re(F_1^V) Re(F_2^V)+Im(F_1^V) Im(F_2^V) )   (k^{\lambda} (8 k^{\mu} M^2 + m^2 p^{\mu}) + 2 m^2 (p^{\lambda} p^{\mu} + 2 M^2  g^{\lambda \mu} ) + k^{\lambda} p^{\mu} {s_u}  \nonumber \\
&&+ (2 k^{\lambda} k^{\mu} - (k^{\lambda} + 2 p^{\lambda}) p^{\mu} + (m^2 - 4 M^2)  g^{\lambda \mu} ) t -  g^{\lambda \mu}  t^2) \nonumber \\
&+& (Re(F_A) Re(F_P) + Im(F_A) Im(F_P))  (m^2 (k^{\lambda} + 2 p^{\lambda}) p^{\mu} - (k^{\lambda} + 2 p^{\lambda}) (2 k^{\mu} + p^{\mu}) t + {s_u}  (k^{\lambda} p^{\mu} +  g^{\lambda \mu}  t)) \nonumber \\
&+& 4 \epsilon(k,p,q,{\mu})(- Re(F_P) Im(F_A) + Re(F_A) Im(F_P)) k^{\lambda}   x  \nonumber \\
&+& 4 \epsilon(k,p,q,{\mu}) (Re(F_1^V) Im(F_2^V) +Re(F_2^V) Im(F_1^V))   (k^{\lambda} + 2 p^{\lambda})  x  \nonumber \\
&+& 4 (Re(F_1^V) Re(F_A) +Im(F_1^V) Im(F_A)) m M^2 (4 k^{\mu} p^{\lambda} + m^2  g^{\lambda \mu}  +  g^{\lambda \mu}  {s_u}  -  g^{\lambda \mu}  t)  x  \nonumber \\
&+& (Re(F_1^V) Re(F_P) + Im(F_1^V) Im(F_P))   (m^2 (3 k^{\lambda} + 2 p^{\lambda}) p^{\mu} - k^{\lambda} p^{\mu} {s_u}  \nonumber \\
&&+ (-2 k^{\lambda} k^{\mu} - 3 k^{\lambda} p^{\mu} - 2 p^{\lambda} p^{\mu} + m^2  g^{\lambda \mu} ) t -  g^{\lambda \mu}  t^2)  x  \nonumber \\
&+& (Re(F_2^V) Re(F_P) +Im(F_2^V) Im(F_P))  (2 k^{\lambda} m^2 p^{\mu} + (-2 k^{\lambda} (k^{\mu} + p^{\mu}) + m^2  g^{\lambda \mu} ) t -  g^{\lambda \mu}  t^2)  x  \nonumber \\
&+& (Re(F_2^V) Re(F_A) + Im(F_2^V) Im(F_A))   (-(k^{\lambda} (8 k^{\mu} M^2 + m^2 p^{\mu})) - 2 m^2 (p^{\lambda} p^{\mu} - 2 M^2  g^{\lambda \mu} ) \nonumber \\
&&+ ((k^{\lambda} + 2 p^{\lambda}) (2 k^{\mu} + p^{\mu}) - 4 M^2  g^{\lambda \mu} ) t + {s_u}  ((3 k^{\lambda} + 4 p^{\lambda}) p^{\mu} +  g^{\lambda \mu}  t))  x 
\left. \right]
\end{eqnarray}
\begin{eqnarray}
\mathcal{A}_{lN'(2)}^{\lambda\mu} &=&\frac{4m}{M\mathcal{I}}
\left[ \right.
 Re(F_3^A) Re(F_3^V)   p^{\mu} (m^2 (-k^{\lambda} - 2 p^{\lambda}) + k^{\lambda} {s_u}  + (k^{\lambda} + 2 p^{\lambda}) t)  x 
+ 4 \epsilon(k,p,q,{\lambda}) Re(F_3^V) Im(F_3^A)  p^{\mu}
\left. \right]
\end{eqnarray}

\begin{eqnarray}
\mathcal{A}_{lN'(12)}^{\lambda\mu} &=&\frac{2m}{M\mathcal{I}}
\left[ \right.
 (Re(F_3^A) Re(F_A)+ Im(F_3^A) Im(F_A))   (k^{\lambda} (-8 k^{\mu} M^2 + m^2 p^{\mu}) + 2 m^2 (p^{\lambda} p^{\mu} - 2 M^2  g^{\lambda \mu} ) + k^{\lambda} p^{\mu} {s_u}  \nonumber \\
&&+ (2 k^{\lambda} k^{\mu} - (k^{\lambda} + 2 p^{\lambda}) p^{\mu} + (m^2 + 4 M^2)  g^{\lambda \mu} ) t -  g^{\lambda \mu}  t^2) +
 2 \epsilon(q,k,\mu,\lambda) Im(F_3^V) Im(F_A) (4 M^2 - t) \nonumber \\
 &+& 4 \epsilon(k,p,q,{\lambda}) (- Re(F_3^A) (Im(F_1^V) + Im(F_2^V)) + (Re(F_1^V) + Re(F_2^V)) Im(F_3^A) - Re(F_3^V) Im(F_A))   p^{\mu} \nonumber \\
&+& (Re(F_1^V) Re(F_3^V) + Re(F_2^V) Re(F_3^V))   (m^2 (k^{\lambda} + 2 p^{\lambda}) p^{\mu} - (k^{\lambda} + 2 p^{\lambda}) (2 k^{\mu} + p^{\mu}) t + {s_u}  (k^{\lambda} p^{\mu} +  g^{\lambda \mu}  t))\nonumber \\
&-& 4 \epsilon(k,p,q,{\mu}) (Re(F_3^V) Im(F_1^V) + Re(F_3^V) Im(F_2^V)) k^{\lambda}    x  \nonumber \\
&+& 4 \epsilon(k,p,q,{\mu}) (-Re(F_A) Im(F_3^A)+Re(F_3^A) Im(F_A)) m (k^{\lambda} + 2 p^{\lambda})  x  + 2 Re(F_3^V) Re(F_P) k^{\lambda} m^2 p^{\mu}  x  \nonumber \\
&+& 2 \epsilon(p,q,{\mu},{\lambda}) (Im(F_1^V) Im(F_3^V) +Im(F_2^V) Im(F_3^V))  (-m^2 + t)  x  \nonumber \\
&+& Re(F_3^V) Re(F_A)   (k^{\lambda} (3 m^2 p^{\mu} + 8 M^2 (k^{\mu} + p^{\mu})) + 2 m^2 (p^{\lambda} p^{\mu} - 2 M^2  g^{\lambda \mu} ) - k^{\lambda} p^{\mu} {s_u}  \nonumber \\
&&+ (-2 k^{\lambda} k^{\mu} - 3 k^{\lambda} p^{\mu} - 2 p^{\lambda} p^{\mu} + m^2  g^{\lambda \mu}  + 4 M^2  g^{\lambda \mu} ) t -  g^{\lambda \mu}  t^2)  x  \nonumber \\
&+& (Re(F_2^V) Re(F_3^A)+ Im(F_2^V) Im(F_3^A) )  (m^2 (-k^{\lambda} - 2 p^{\lambda}) p^{\mu} + (2 k^{\lambda} k^{\mu} + 4 k^{\mu} p^{\lambda} - k^{\lambda} p^{\mu} + 2 p^{\lambda} p^{\mu}) t \nonumber + {s_u}  (k^{\lambda} p^{\mu} +  g^{\lambda \mu}  t))  x  \nonumber \\
&+& (Re(F_1^V) Re(F_3^A) + Im(F_1^V) Im(F_3^A))   ((-(k^{\lambda} (m^2 + 8 M^2)) - 2 m^2 p^{\lambda}) p^{\mu} + (k^{\lambda} + 2 p^{\lambda}) (2 k^{\mu} + p^{\mu}) t \nonumber \\
&&+ {s_u}  ((3 k^{\lambda} + 4 p^{\lambda}) p^{\mu} +  g^{\lambda \mu}  t))  x 
\left. \right]
 \end{eqnarray}
\end{footnotesize}

\subsubsection{Target-lepton asymmetry}

\begin{footnotesize}
 \begin{eqnarray}
\mathcal{B}_{lN(1)}^{\lambda\nu} &=&\frac{m}{M\mathcal{I}}
\left[ \right.
4 \epsilon(k,p,q,{\lambda})( Re(F_P) Im(F_1^V) - Re(F_A) Im(F_2^V) + Re(F_2^V) Im(F_A) -Re(F_1^V) Im(F_P))   q^{\nu} \nonumber \\
&+& 4 |F_A|^2   M^2 (2 k^{\nu} (k^{\lambda} + 2 p^{\lambda}) -  g^{\lambda \nu}  {s_u} ) + 4 |F_1^V|^2   M^2 (2 k^{\lambda} k^{\nu} + m^2  g^{\lambda \nu}  -  g^{\lambda \nu}  t) \nonumber \\
&+& |F_2^V|^2   (m^2 (k^{\lambda} + 2 p^{\lambda}) q^{\nu} + k^{\lambda} q^{\nu} {s_u}  + (2 k^{\lambda} k^{\nu} - (k^{\lambda} + 2 p^{\lambda}) q^{\nu} + m^2  g^{\lambda \nu} ) t -  g^{\lambda \nu}  t^2) \nonumber \\
&+& (Re(F_1^V) Re(F_2^V) + Im(F_1^V) Im(F_2^V))  (k^{\lambda} (8 k^{\nu} M^2 + m^2 q^{\nu}) + 2 m^2 (p^{\lambda} q^{\nu} + 2 M^2  g^{\lambda \nu} ) + k^{\lambda} q^{\nu} {s_u}  \nonumber \\
&&+ (2 k^{\lambda} k^{\nu} - (k^{\lambda} + 2 p^{\lambda}) q^{\nu} + (m^2 - 4 M^2)  g^{\lambda \nu} ) t -  g^{\lambda \nu}  t^2) \nonumber \\
&+& (Re(F_A) Re(F_P) + Im(F_A) Im(F_P))  (m^2 (k^{\lambda} + 2 p^{\lambda}) q^{\nu} + (k^{\lambda} + 2 p^{\lambda}) (2 k^{\nu} - q^{\nu}) t + {s_u}  (k^{\lambda} q^{\nu} -  g^{\lambda \nu}  t)) \nonumber \\
&+& 4 \epsilon(k,p,q,{\nu}) (Re(F_P) Im(F_A) - Re(F_A) Im(F_P)) k^{\lambda}  x  \nonumber \\
&+& 4 \epsilon(k,p,q,{\nu}) (-Re(F_1^V) Im(F_2^V) + Re(F_2^V) Im(F_1^V))   (k^{\lambda} + 2 p^{\lambda})  x  \nonumber \\
&+& 4 (Re(F_1^V) Re(F_A) + Im(F_1^V) Im(F_A))  M^2 (4 k^{\nu} (k^{\lambda} + p^{\lambda}) - m^2  g^{\lambda \nu}  +  g^{\lambda \nu}  {s_u}  +  g^{\lambda \nu}  t)  x  \nonumber \\
&+& (Re(F_2^V) Re(F_P) + Im(F_2^V) Im(F_P))   (2 k^{\lambda} m^2 q^{\nu} + (2 k^{\lambda} (k^{\nu} - q^{\nu}) - m^2  g^{\lambda \nu} ) t +  g^{\lambda \nu}  t^2)  x  \nonumber \\
&+& (Re(F_1^V) Re(F_P) +Im(F_1^V) Im(F_P))   (m^2 (k^{\lambda} - 2 p^{\lambda}) q^{\nu} + k^{\lambda} q^{\nu} {s_u}  + (2 k^{\lambda} k^{\nu} - k^{\lambda} q^{\nu} + 2 p^{\lambda} q^{\nu} - m^2  g^{\lambda \nu} ) t +  g^{\lambda \nu}  t^2)  x  \nonumber \\
&+& (Re(F_2^V) Re(F_A) + Im(F_2^V) Im(F_A))   (k^{\lambda} (8 k^{\nu} M^2 + m^2 q^{\nu}) + 2 m^2 (p^{\lambda} q^{\nu} - 2 M^2  g^{\lambda \nu} )\nonumber \\
&&+ ((k^{\lambda} + 2 p^{\lambda}) (2 k^{\nu} - q^{\nu}) + 4 M^2  g^{\lambda \nu} ) t + {s_u}  ((k^{\lambda} + 4 p^{\lambda}) q^{\nu} +  g^{\lambda \nu}  t))  x 
\left. \right]
 \end{eqnarray}
 \begin{eqnarray}
\mathcal{B}_{lN(2)}^{\lambda\nu}& =&\frac{4m}{M\mathcal{I}}
\left[ 
-4 \epsilon(k,p,q,{\lambda}) Re(F_3^V) Im(F_3^A)  q^{\nu} 
+  Re(F_3^A) Re(F_3^V)   q^{\nu} (m^2 (k^{\lambda} + 2 p^{\lambda}) - k^{\lambda} {s_u}  + (-k^{\lambda} - 2 p^{\lambda}) t)  x 
 \right] \\
\mathcal{B}_{lN,(12)}^{\lambda\nu} &=&\frac{2m}{ M\mathcal{I}}
\left[ \right. (Re(F_3^A) Re(F_A) + Im(F_3^A) Im(F_A))   (8 k^{\lambda} k^{\nu} M^2 - k^{\lambda} m^2 q^{\nu} - 2 m^2 p^{\lambda} q^{\nu} + 4 m^2 M^2  g^{\lambda \nu}  - k^{\lambda} q^{\nu} {s_u}  \nonumber \\
&+&7 (2 p^{\lambda} q^{\nu} + k^{\lambda} (-2 k^{\nu} + q^{\nu}) - (m^2 + 4 M^2)  g^{\lambda \nu} ) t +  g^{\lambda \nu}  t^2) + 2 \epsilon(q,k,{\lambda},{\nu}) Im(F_3^V) Im(F_A)  (-4 M^2 + t) \nonumber \\
&+& 4 \epsilon(k,p,q,{\lambda}) (-Re(F_3^A) (Im(F_1^V) + Im(F_2^V)) + (Re(F_1^V) + Re(F_2^V)) Im(F_3^A)- Re(F_3^V) Im(F_A))   q^{\nu} \nonumber \\
&+& (Re(F_1^V) Re(F_3^V) + Re(F_2^V) Re(F_3^V))   (m^2 (-k^{\lambda} - 2 p^{\lambda}) q^{\nu} - (k^{\lambda} + 2 p^{\lambda}) (2 k^{\nu} - q^{\nu}) t + {s_u}  (-(k^{\lambda} q^{\nu}) +  g^{\lambda \nu}  t)) \nonumber \\
&-& 4 \epsilon(k,p,q,{\nu}) (Re(F_3^V) Im(F_1^V) + Re(F_3^V) Im(F_2^V)) k^{\lambda}   x  \nonumber \\
&+& 4 \epsilon(k,p,q,{\nu}) Re(F_3^A) Im(F_A)   (-k^{\lambda} - 2 p^{\lambda})  x  + 4 \epsilon(k,p,q,{\nu}) Re(F_A) Im(F_3^A) m (k^{\lambda} + 2 p^{\lambda})  x  \nonumber \\
&-& 2 Re(F_3^V) Re(F_P) k^{\lambda} m^2 q^{\nu}  x  + 2 \epsilon(p,q,{\lambda},{\nu}) (Im(F_1^V) Im(F_3^V) +Im(F_2^V) Im(F_3^V))   (m^2 - t)  x  \nonumber \\
&+& Re(F_3^V) Re(F_A)   (8 k^{\lambda} k^{\nu} M^2 - k^{\lambda} (m^2 + 8 M^2) q^{\nu} + 2 m^2 (p^{\lambda} q^{\nu} - 2 M^2  g^{\lambda \nu} ) - k^{\lambda} q^{\nu} {s_u}  \nonumber \\
&&+ (-2 p^{\lambda} q^{\nu} + k^{\lambda} (-2 k^{\nu} + q^{\nu}) + (m^2 + 4 M^2)  g^{\lambda \nu} ) t -  g^{\lambda \nu}  t^2)  x  \nonumber \\
&+& (Re(F_2^V) Re(F_3^A) + Im(F_2^V) Im(F_3^A))   (m^2 (-k^{\lambda} - 2 p^{\lambda}) q^{\nu} + (-2 k^{\lambda} k^{\nu} - 4 k^{\nu} p^{\lambda} + 3 k^{\lambda} q^{\nu} + 2 p^{\lambda} q^{\nu}) t + {s_u}  (k^{\lambda} q^{\nu} -  g^{\lambda \nu}  t))  x  \nonumber \\
&+& (Re(F_1^V) Re(F_3^A) + Im(F_1^V) Im(F_3^A))   ((-(k^{\lambda} m^2) + 8 k^{\lambda} M^2 - 2 m^2 p^{\lambda}) q^{\nu} - (k^{\lambda} + 2 p^{\lambda}) (2 k^{\nu} - q^{\nu}) t\nonumber \\
&&+ {s_u}  ((-k^{\lambda} - 4 p^{\lambda}) q^{\nu} -  g^{\lambda \nu}  t))  x
\left. \right]
\end{eqnarray}
\end{footnotesize}
\subsubsection{Target-recoil nucleon asymmetry}
\begin{footnotesize}
 \begin{eqnarray}
\mathcal{C}_{NN'(1)}^{\nu\mu} &=&\frac{1}{8M^2\mathcal{I}}
\left[ \right.
4 Re(F_2^V) Im(F_P) m^2 (3 \epsilon(k,p,q,{\nu}) p^{\mu} - 4 \epsilon(k,p,q,{\mu}) q^{\nu}) \nonumber \\
&+& 64 (Re(F_1^V)  + Re(F_2^V)) Im(F_A) M^2 (2 \epsilon(k,p,q,{\mu}) k^{\nu} + 2 \epsilon(k,p,q,{\nu}) k^{\mu} + \epsilon(k,p,q,{\nu}) p^{\mu} - \epsilon(k,p,q,{\mu}) q^{\nu}) \nonumber \\
&+& 4 Re(F_P) Im(F_2^V) m^2 (-3 \epsilon(k,p,q,{\nu}) p^{\mu} + 4 \epsilon(k,p,q,{\mu}) q^{\nu}) \nonumber \\
&+& 64 (Re(F_A) Im(F_1^V) + Re(F_A) Im(F_2^V)) M^2 (-(\epsilon(k,p,q,{\nu}) (2 k^{\mu} + p^{\mu})) + \epsilon(k,p,q,{\mu}) (-2 k^{\nu} + q^{\nu})) \nonumber \\
&+& (Re(F_2^V) Re(F_P) + Im(F_2^V) Im(F_P)) m^2 (\epsilon(p,q,{\mu},{\nu}) m^2 + 4 \epsilon(q,k,{\mu},{\nu}) M^2 + \epsilon(p,q,{\mu},{\nu}) {s_u}  + (2 \epsilon(p,k,{\mu},{\nu}) - \epsilon(p,q,{\mu},{\nu})) t) \nonumber \\
&+& 16 (Re(F_A) Re(F_P)+Im(F_A) Im(F_P)) m^2 M^2 (2 k^{\nu} p^{\mu} - 2 k^{\mu} q^{\nu} - m^2 g^{\nu \mu}  + g^{\nu \mu}  t) \nonumber \\
&+& 16 (Re(F_1^V) Re(F_2^V) +Im(F_1^V) Im(F_2^V)) M^2 (2 (k^{\nu} p^{\mu} + k^{\mu} q^{\nu}) {s_u}  + m^2 (-4 k^{\nu} p^{\mu} + 4 k^{\mu} q^{\nu} + 2 p^{\mu} q^{\nu} + m^2 g^{\nu \mu} ) \nonumber \\
&&+ (2 (2 k^{\mu} + p^{\mu}) (2 k^{\nu} - q^{\nu}) - m^2 g^{\nu \mu} ) t) \nonumber \\
&+& 8 |F_A|^2 M^2 (4 k^{\nu} (m^2 p^{\mu} + 4 M^2 (2 k^{\mu} + p^{\mu})) + 4 (-(k^{\nu} p^{\mu}) - k^{\mu} q^{\nu}) {s_u}  + {s_u} ^2 g^{\nu \mu} - (m^2 + 4 M^2) (4 k^{\mu} q^{\nu} + m^2 g^{\nu \mu} ) \nonumber \\
&&+ 2 (-2 k^{\nu} (2 k^{\mu} + p^{\mu}) + 2 k^{\mu} q^{\nu} + (m^2 + 2 M^2) g^{\nu \mu} ) t - g^{\nu \mu}  t^2) 
+ 2 |F_P|^2 m^2 (-2 m^2 p^{\mu} q^{\nu} + (2 p^{\mu} q^{\nu} - m^2 g^{\nu \mu} ) t + g^{\nu \mu}  t^2) \nonumber \\
&+& 8 |F_1^V|^2 M^2 (4 (k^{\nu} p^{\mu} + k^{\mu} q^{\nu}) {s_u}  - {s_u} ^2 g^{\nu \mu}  + m^2 (-4 k^{\nu} p^{\mu} + 4 k^{\mu} q^{\nu} + (m^2 + 4 M^2) g^{\nu \mu} ) \nonumber \\
&+& 2 (2 k^{\nu} (2 k^{\mu} + p^{\mu}) - 2 k^{\mu} q^{\nu} - (m^2 + 2 M^2) g^{\nu \mu} ) t + g^{\nu \mu}  t^2) \nonumber \\
&+& 2 |F_2^V|^2 (4 m^2 M^2 (-4 k^{\nu} p^{\mu} + 4 k^{\mu} q^{\nu} + 2 p^{\mu} q^{\nu} + m^2 g^{\nu \mu} ) + 2 (8 k^{\nu} M^2 (2 k^{\mu} + p^{\mu}) - 8 k^{\mu} M^2 q^{\nu} \nonumber \\
&&+ (m^2 - 4 M^2) p^{\mu} q^{\nu} - 4 m^2 M^2 g^{\nu \mu} ) t + (-2 p^{\mu} q^{\nu} + (m^2 + 4 M^2) g^{\nu \mu} ) t^2 - g^{\nu \mu}  t^3 + {s_u} ^2 (2 p^{\mu} q^{\nu} + g^{\nu \mu}  t)) \nonumber \\
&+& 64 (Re(F_2^V) Im(F_1^V) - Re(F_1^V) Im(F_2^V))M^2 (\epsilon(k,p,q,{\nu}) p^{\mu} + \epsilon(k,p,q,{\mu}) q^{\nu})  x \nonumber \\
&+& 128 (Re(F_1^V) Re(F_A) +Im(F_1^V) Im(F_A)) M^4 (k^{\nu} p^{\mu} + k^{\mu} q^{\nu})  x 
- 4 \epsilon(k,p,q,{\nu}) Re(F_3^A) Im(F_3^V) m^2 p^{\mu} 
\nonumber \\
&+& 8 (Re(F_2^V) Re(F_A) + Im(F_2^V) Im(F_A)) M^2 ((2 \epsilon(p,k,{\mu},{\nu}) + \epsilon(p,q,{\mu},{\nu}) + \epsilon(q,k,{\mu},{\nu})) m^2 \nonumber \\
&&+ {s_u}  (\epsilon(q,k,{\mu},{\nu}) + 4 p^{\mu} q^{\nu}  x ) + t (-\epsilon(p,q,{\mu},{\nu}) + 4 (k^{\nu} p^{\mu} + k^{\mu} q^{\nu})  x ))
\left. \right]
\end{eqnarray}
\begin{eqnarray}
\mathcal{C}_{NN'(12)}^{\nu\mu} &=&\frac{1}{M^2\mathcal{I}}
\left[ \right.
4 (Re(F_1^V) Im(F_3^A) - Re(F_3^A) Im(F_1^V))M^2 (\epsilon(k,p,q,{\nu}) (2 k^{\mu} + p^{\mu}) 
+ \epsilon(k,p,q,{\mu}) (-2 k^{\nu} + q^{\nu})) \nonumber \\
&+& 2 (Re(F_2^V) Im(F_3^A) - Re(F_3^A) Im(F_2^V))(m^2 (\epsilon(k,p,q,{\nu}) p^{\mu} + \epsilon(k,p,q,{\mu}) q^{\nu}) + (3 \epsilon(k,p,q,{\nu}) p^{\mu} - \epsilon(k,p,q,{\mu}) q^{\nu}) {s_u} ) \nonumber \\
&+& 4 (Re(F_3^A) Re(F_A) + Im(F_3^A) Im(F_A)) M^2 {s_u}  (2 k^{\nu} p^{\mu} - 2 (k^{\mu} + p^{\mu}) q^{\nu} - m^2 g^{\nu \mu} ) \nonumber \\
&+& (Re(F_3^A) Re(F_P) + Im(F_3^A) Im(F_P)) m^2 {s_u}  (-2 p^{\mu} q^{\nu} - g^{\nu \mu}  t) \nonumber \\
&+& 16 (Re(F_A) Im(F_3^A) - Re(F_3^A) Im(F_A)) M^2 (\epsilon(k,p,q,{\nu}) p^{\mu} - \epsilon(k,p,q,{\mu}) q^{\nu})  x  \nonumber \\
&+& 8 (Re(F_1^V) Re(F_3^A) + Re(F_2^V) Re(F_3^A) + Im(F_1^V) Im(F_3^A) + Im(F_2^V) Im(F_3^A) ) \times\nonumber \\
&& \times M^2 (-(m^2 p^{\mu} q^{\nu}) + (-(k^{\nu} p^{\mu}) + (k^{\mu} + p^{\mu}) q^{\nu}) t)  x \left. \right]
\end{eqnarray}
\end{footnotesize}

\subsubsection{Target-recoil-nucleon-charged lepton asymmetry}

\begin{footnotesize}

 \begin{eqnarray}
\mathcal{D}_{lNN'(1)}^{\lambda\nu\mu} &=&\frac{m}{2M^2\mathcal{I}}
\left[ \right.
32 (Re(F_1^V) Re(F_A) + Im(F_1^V) Im(F_A))   M^4 (-(k^{\mu}  g^{\lambda \nu} ) + k^{\nu}  g^{\lambda \mu} ) \nonumber \\
&+& 4 (Re(F_1^V) Re(F_P) +Im(F_1^V) Im(F_P))  M^2 (2 k^{\lambda} (k^{\nu} p^{\mu} + k^{\mu} q^{\nu}) + m^2 (p^{\mu}  g^{\lambda \nu}  + q^{\nu}  g^{\lambda \mu} ) \nonumber \\
&&+ (-2 k^{\mu}  g^{\lambda \nu}  - p^{\mu}  g^{\lambda \nu}  + 2 k^{\nu}  g^{\lambda \mu}  - q^{\nu}  g^{\lambda \mu} ) t) \nonumber \\
&+& (Re(F_P) Im(F_2^V) - Re(F_2^V) Im(F_P))k^{\lambda}   (3 \epsilon(k,p,q,{\nu}) p^{\mu} - 4 \epsilon(k,p,q,{\mu}) q^{\nu})  x  \nonumber \\
&+& 16 (Re(F_1^V) Im(F_A) +Re(F_2^V) Im(F_A) - Re(F_A) Im(F_1^V) - Re(F_A) Im(F_2^V)) \times\nonumber \\
&&\times  M^2 (\epsilon(k,p,q,{\mu})  g^{\lambda \nu}  + \epsilon(k,p,q,{\nu})  g^{\lambda \mu} )  x  \nonumber \\
&+& 4 (Re(F_A) Re(F_P) +Im(F_A) Im(F_P) )  M^2 (m^2 (p^{\mu}  g^{\lambda \nu}  - q^{\nu}  g^{\lambda \mu} ) + 2 k^{\lambda} (-(k^{\nu} p^{\mu}) + k^{\mu} q^{\nu} + 2 p^{\mu} q^{\nu} + m^2 g^{\nu \mu} ) \nonumber \\
&&+ (-(p^{\mu}  g^{\lambda \nu} ) + q^{\nu}  g^{\lambda \mu} ) t)  x  + |F_P|^2 k^{\lambda} m^2 (2 p^{\mu} q^{\nu} + g^{\nu \mu}  t)  x  \nonumber \\
&+& 4 |F_1^V|^2   M^2 (4 k^{\nu} p^{\lambda} p^{\mu} + 4 k^{\mu} p^{\lambda} q^{\nu} - m^2 p^{\mu}  g^{\lambda \nu}  + m^2 q^{\nu}  g^{\lambda \mu}  + k^{\lambda} (4 k^{\nu} p^{\mu} - 4 p^{\mu} q^{\nu} - m^2 g^{\nu \mu}  + 4 M^2 g^{\nu \mu} ) \nonumber \\
&&+ {s_u}  (p^{\mu}  g^{\lambda \nu}  + q^{\nu}  g^{\lambda \mu}  - (k^{\lambda} + 2 p^{\lambda}) g^{\nu \mu} ) + (2 k^{\mu}  g^{\lambda \nu}  + p^{\mu}  g^{\lambda \nu}  + 2 k^{\nu}  g^{\lambda \mu}  - q^{\nu}  g^{\lambda \mu}  - k^{\lambda} g^{\nu \mu} ) t)  x  \nonumber \\
&+& 4 |F_A|^2   M^2 (-4 k^{\nu} p^{\lambda} p^{\mu} - 4 k^{\mu} p^{\lambda} q^{\nu} + 8 k^{\mu} M^2  g^{\lambda \nu}  + m^2 p^{\mu}  g^{\lambda \nu}  + 8 k^{\nu} M^2  g^{\lambda \mu}  - m^2 q^{\nu}  g^{\lambda \mu}  \nonumber \\
&&+ k^{\lambda} (-4 k^{\nu} p^{\mu} + 4 p^{\mu} q^{\nu} + m^2 g^{\nu \mu}  - 4 M^2 g^{\nu \mu} ) + {s_u}  (-(p^{\mu}  g^{\lambda \nu} ) - q^{\nu}  g^{\lambda \mu}  + (k^{\lambda} + 2 p^{\lambda}) g^{\nu \mu} ) \nonumber \\
&&+ (-((2 k^{\mu} + p^{\mu})  g^{\lambda \nu} ) - 2 k^{\nu}  g^{\lambda \mu}  + q^{\nu}  g^{\lambda \mu}  + k^{\lambda} g^{\nu \mu} ) t)  x  \nonumber \\
&+& 4 (Re(F_1^V) Re(F_2^V) + Im(F_1^V) Im(F_2^V) )  M^2 ((p^{\mu}  g^{\lambda \nu}  + q^{\nu}  g^{\lambda \mu} ) {s_u}  + 2 (2 k^{\nu} p^{\lambda} p^{\mu} + 2 k^{\mu} p^{\lambda} q^{\nu} - m^2 p^{\mu}  g^{\lambda \nu}  \nonumber \\
&&+ m^2 q^{\nu}  g^{\lambda \mu}  + k^{\lambda} (3 k^{\nu} p^{\mu} - (k^{\mu} + 2 p^{\mu}) q^{\nu} - m^2 g^{\nu \mu} ) + (2 k^{\mu}  g^{\lambda \nu}  + p^{\mu}  g^{\lambda \nu}  + 2 k^{\nu}  g^{\lambda \mu}  - q^{\nu}  g^{\lambda \mu} ) t))  x  \nonumber \\
&+& |F_2^V|^2   (4 M^2 (m^2 (-(p^{\mu}  g^{\lambda \nu} ) + q^{\nu}  g^{\lambda \mu} ) + k^{\lambda} (2 k^{\nu} p^{\mu} - 2 (k^{\mu} + p^{\mu}) q^{\nu} - m^2 g^{\nu \mu} )) \nonumber \\
&&+ 2 (2 M^2 (2 k^{\mu}  g^{\lambda \nu}  + p^{\mu}  g^{\lambda \nu}  + 2 k^{\nu}  g^{\lambda \mu}  - q^{\nu}  g^{\lambda \mu} ) + k^{\lambda} (p^{\mu} q^{\nu} - 2 M^2 g^{\nu \mu} )) t + k^{\lambda} g^{\nu \mu}  t^2 \nonumber \\
&&+ {s_u}  (2 (k^{\lambda} + 2 p^{\lambda}) p^{\mu} q^{\nu} + (k^{\lambda} + 2 p^{\lambda}) g^{\nu \mu}  t))  x  \nonumber \\
&+& (Re(F_2^V) Re(F_A) + Im(F_2^V) Im(F_A))  M^2 (8 (k^{\lambda} + 2 p^{\lambda}) (k^{\nu} p^{\mu} - k^{\mu} q^{\nu}) - (4 \epsilon(p,k,{\mu},{\nu}) k^{\lambda} - 8 \epsilon(p,q,{\mu},{\nu}) k^{\lambda} \nonumber \\
&&+ 4 \epsilon(p,q,{\mu},{\lambda}) k^{\nu} + 4 \epsilon(p,q,{\lambda},{\nu}) k^{\mu} + (2 \epsilon(p,{\nu},{\mu},{\lambda}) + \epsilon(q,{\nu},{\mu},{\lambda})) m^2 - 4 \epsilon(q,k,{\mu},{\nu}) p^{\lambda})  x  \nonumber \\
&&+ {s_u}  (-4 p^{\mu}  g^{\lambda \nu}  + 4 q^{\nu}  g^{\lambda \mu}  - \epsilon(q,{\nu},{\mu},{\lambda})  x ) + t (-8 k^{\mu}  g^{\lambda \nu}  + 8 k^{\nu}  g^{\lambda \mu}  + (2 \epsilon(p,{\nu},{\mu},{\lambda}) + \epsilon(q,{\nu},{\mu},{\lambda}))  x )) \nonumber \\
&+& \frac{Re(F_2^V) Re(F_P) + Im(F_2^V) Im(F_P) }{4}  (8 m^2 (k^{\lambda} + 2 p^{\lambda}) p^{\mu} q^{\nu} - (4 \epsilon(q,k,{\mu},{\nu}) k^{\lambda} M^2 + \epsilon(p,q,{\mu},{\nu}) m^2 (k^{\lambda} - 2 p^{\lambda}) \nonumber \\
&&+ 2 m^2 (\epsilon(q,{\nu},{\mu},{\lambda}) M^2 + \epsilon(p,q,{\lambda},{\nu}) p^{\mu}))  x  + t^2 (-4 (2 k^{\mu} + p^{\mu})  g^{\lambda \nu}  + 8 k^{\nu}  g^{\lambda \mu}  - 4 q^{\nu}  g^{\lambda \mu}  + \epsilon(p,{\nu},{\mu},{\lambda})  x ) \nonumber \\
&&+ t (4 (4 k^{\lambda} k^{\nu} p^{\mu} + 4 k^{\nu} p^{\lambda} p^{\mu} - 4 k^{\mu} p^{\lambda} q^{\nu} - 2 k^{\lambda} p^{\mu} q^{\nu} - 4 p^{\lambda} p^{\mu} q^{\nu} + m^2 p^{\mu}  g^{\lambda \nu}  + m^2 q^{\nu}  g^{\lambda \mu} ) \nonumber \\
&&+ (-2 \epsilon(p,k,{\mu},{\nu}) k^{\lambda} - \epsilon(p,{\nu},{\mu},{\lambda}) m^2 + 2 \epsilon(q,{\nu},{\mu},{\lambda}) M^2 + \epsilon(p,q,{\mu},{\nu}) (k^{\lambda} - 2 p^{\lambda}) + 2 \epsilon(p,q,{\lambda},{\nu}) p^{\mu})  x ) \nonumber \\
&&+ {s_u}  (4 (-(p^{\mu}  g^{\lambda \nu} ) + q^{\nu}  g^{\lambda \mu} ) t + k^{\lambda} (8 p^{\mu} q^{\nu} - \epsilon(p,q,{\mu},{\nu})  x )))
\left. \right]
\end{eqnarray}
 \begin{equation}
\mathcal{D}_{lNN'(2)}^{\lambda\nu\mu}= \frac{m}{2M^2\mathcal{I}}
\left[ \right.
 4 |F_3^A|^2   ({s_u}  (2 (-k^{\lambda} - 2 p^{\lambda}) p^{\mu} q^{\nu} + (-k^{\lambda} - 2 p^{\lambda}) g^{\nu \mu}  t) + k^{\lambda} (8 M^2 p^{\mu} q^{\nu} + 2 (-(p^{\mu} q^{\nu}) + 2 M^2 g^{\nu \mu} ) t - g^{\nu \mu}  t^2))  x  
\left. \right]
\end{equation}
 \begin{eqnarray}
\mathcal{D}_{lNN'(12)}^{\lambda\nu\mu}&=&\frac{m}{ M^2\mathcal{I}}
\left[ \right.
 16 \epsilon(k,p,q,{\lambda}) (Re(F_A) Im(F_3^A)  - Re(F_3^A) Im(F_A))  M^2 g^{\nu \mu}    \nonumber \\
&+& 4 (Re(F_1^V) Re(F_3^A) + Re(F_2^V) Re(F_3^A)+Im(F_1^V) Im(F_3^A) + Im(F_2^V) Im(F_3^A)) \times \nonumber \\
&& \times M^2 (2 k^{\lambda} (k^{\nu} p^{\mu} - k^{\mu} q^{\nu}) + m^2 (p^{\mu}  g^{\lambda \nu}  - q^{\nu}  g^{\lambda \mu} ) + (-(p^{\mu}  g^{\lambda \nu} ) + q^{\nu}  g^{\lambda \mu} ) t)   \nonumber \\
&+&  4 \epsilon(k,p,q,{\lambda}) (Re(F_P) Im(F_3^A) -Re(F_3^A) Im(F_P))  (2 p^{\mu} q^{\nu} + g^{\nu \mu}  t)   \nonumber \\
&+& 4 (Re(F_2^V) Im(F_3^A) - Re(F_3^A) Im(F_2^V))   (-(\epsilon(k,p,q,{\nu}) (k^{\lambda} + 3 p^{\lambda}) p^{\mu}) + \epsilon(k,p,q,{\mu}) (k^{\lambda} + p^{\lambda}) q^{\nu})  x  \nonumber \\
&+& 4 (Re(F_3^A) Im(F_1^V) -  Re(F_1^V) Im(F_3^A) )  M^2 (\epsilon(k,p,q,{\mu})  g^{\lambda \nu}  - \epsilon(k,p,q,{\nu})  g^{\lambda \mu} )  x  \nonumber \\
&+& 4 (Re(F_3^A) Re(F_A) + Im(F_3^A) Im(F_A))   M^2 ({s_u}  (p^{\mu}  g^{\lambda \nu}  - q^{\nu}  g^{\lambda \mu}  + k^{\lambda} g^{\nu \mu} ) \nonumber \\
&&+ (k^{\lambda} + 2 p^{\lambda}) (2 k^{\nu} p^{\mu} - 2 k^{\mu} q^{\nu} - m^2 g^{\nu \mu} ) + (k^{\lambda} + 2 p^{\lambda}) g^{\nu \mu}  t)  x  \nonumber \\
&+& (Re(F_3^A) Re(F_P) +Im(F_3^A) Im(F_P))   (2 m^2 (-k^{\lambda} - 2 p^{\lambda}) p^{\mu} q^{\nu} + (k^{\lambda} + 2 p^{\lambda}) (2 p^{\mu} q^{\nu} - m^2 g^{\nu \mu} ) t  \nonumber \\
&&+ (k^{\lambda} + 2 p^{\lambda}) g^{\nu \mu}  t^2 + k^{\lambda} {s_u}  (2 p^{\mu} q^{\nu} + g^{\nu \mu}  t))  x
\left. \right]
\end{eqnarray}
\end{footnotesize}

\section{Components of the spin asymmetries}
\label{Appendix:Components}

Below we give all non-vanishing components of the single, double and triple asymmetries. They are calculated assuming that the form factors are real and $F_3^V=0$.

\subsubsection{Recoil polarization asymmetry}
\begin{footnotesize}
\begin{eqnarray}
  \mathcal{P}^L_{N'(1)} &=& -\frac{1}{M|\textbf{p}'|\mathcal{I}}  \left[ \right.
 F_1^V F_P m^2  {s_u}  t + F_2^V F_P \frac{m^2 {s_u}  t^2}{4 M^2} 
+   F_1^V F_A  (4 m^2 M^2(m^2+ {s_u})   + (s_u^2  - m^4) t - (4 M^2 - t) t^2) \nonumber \\
&+&  F_2^V F_A  
(4 m^4 M^2  +  (s_u^2  - m^4 +m^2 {s_u})t - (4 M^2  - t) t^2) 
  -    {s_u}  t^2  x ({F_1^V} + {F_2^V})  -  F_A^2  {s_u}  t (t-4 M^2)  x
  \left. \right]
\\
  \mathcal{P}^L_{N'(12)} &=& \frac{-2}{M|\textbf{p}'|\mathcal{I}}  
\left[ \right. 
F_3^A t \left( F_1^V  + \frac{t}{4 M^2}F_2^V \right) ({s_u} ^2 + (m^2 - t) (t-4 M^2 ))
  \left. \right]\\ 
\mathcal{P}^T_{N'(1)}
&=&  
 \frac{( s-M^2 ) \sin(\beta)}{ M^2 \mathcal{I}} \left[ \right. 
 -4 F_1^V F_A M^2 (m^2 + {s_u} ) 
- (F_1^V  + F_2^V   ) F_P m^2 t \nonumber \\
&-& F_2^V F_A (4 m^2 M^2 + {s_u}  t) 
+ 4 {F_1^V}^2 M^2 t  x  
+ {F_2^V}^2 t^2  x  
+ F_1^V F_2^V t (4 M^2 + t)  x 
\left. \right] \\
\mathcal{P}^T_{N'(12)}
 &=&  
 \frac{2( s-M^2 ) \sin(\beta)}{ M^2 \mathcal{I}}
     \left[  F_3^A t (-(F_1^V + F_2^V) {s_u}  + F_A (-4 M^2 + t)  x )\right]
\end{eqnarray}

\end{footnotesize}

\subsubsection{Lepton polarization asymmetry} 
 
 \begin{footnotesize}
\begin{eqnarray}
\mathcal{P}^L_{l(1)}
  &=& \frac{1}{16 M^3|\textbf{k}'|\mathcal{I}}
  \left[ \right. 
 16 (F_1^V+F_2^V) F_A M^2 (-s_u^2 t + {s_u}  (m^2 - t) t + 2 m^2 (-4 m^2 M^2 + (m^2 + 4 M^2) t - t^2)) \nonumber \\
&+& 8 F_A F_P m^2 M^2 (-m^4 - {s_u}  (m^2 + t) + 2 m^2 t - t^2)  x  
+ F_P^2 m^2 t (-m^4 - s_u  (m^2 + t) + 2 m^2 t - t^2)  x  \nonumber \\
&+& 4 F_A^2 M^2 (m^4 (-m^2 - 4 M^2) + {s_u} ^3 + m^2 (3 m^2 + 8 M^2) t + (-3 m^2 - 4 M^2) t^2 + t^3 + {s_u} ^2 (m^2 + t) \nonumber \\
&&+ {s_u}  (m^2 (-m^2 - 20 M^2) + 4 (m^2 - M^2) t + t^2))  x  \nonumber \\
&+& 4 {F_1^V}^2 M^2 (m^4 (-m^2 + 4 M^2) + {s_u} ^3 + m^2 (3 m^2 - 8 M^2) t + (-3 m^2 + 4 M^2) t^2 + t^3 + {s_u} ^2 (m^2 + t) \nonumber \\
&&+ {s_u}  (m^2 (-m^2 - 12 M^2) + 4 (m^2 + M^2) t + t^2))  x  \nonumber \\
&+& 8 F_1^V F_2^V M^2 (-m^6 + 4 m^4 t - 5 m^2 t^2 + 2 t^3 + {s_u}  (-m^4 + m^2 t + 2 t^2))  x  \nonumber \\
&+& {F_2^V}^2 (-4 m^6 M^2 + 12 m^4 M^2 t - {s_u} ^3 t + {s_u} ^2 (-m^2 - t) t + m^2 (m^2 - 12 M^2) t^2 + 2 (-m^2 + 2 M^2) t^3 + t^4 \nonumber \\
&&+ {s_u}  (-4 m^4 M^2 + 16 m^2 M^2 t + (-3 m^2 + 4 M^2) t^2 + t^3))  x   
 \left. \right]
\end{eqnarray}
\begin{eqnarray}
\mathcal{P}^L_{l(12)}
&=& \frac{-m^2 x}{4 M^3|\textbf{k}'|\mathcal{I}}   (4  F_3^A F_A M^2 +  F_3^A F_P t)
(8 M^2 t + ({s_u}  - 2 t) ({s_u}  + t) + m^2 (-8 M^2 + {s_u}  + 2 t))
\\
\mathcal{P}^L_{l,(2)}
  &=& \frac{1}{8 M^3|\textbf{k}'| \mathcal{I}} \left[ \right.
 2 {F_3^A}^2 t (-4 m^4 M^2 - {s_u} ^3 + {s_u} ^2 (-m^2 - t) + m^2 (m^2 + 8 M^2) t - 2 (-m^2 - 2 M^2) t^2 + t^3 \nonumber \\
& & + {s_u}  (12 m^2 M^2 - (3 m^2 + 4 M^2) t + t^2))  x 
 \left. \right]
\end{eqnarray}
\end{footnotesize}
\begin{footnotesize}
\begin{eqnarray}
  \mathcal{P}^T_{l(1)}
  &=& \frac{ m ( s - M^2 ) |\textbf{p}'| \sin(\beta)}{4M^3|\textbf{k}'|\mathcal{I}}  \left[ \right.  
  8 F_A F_P m^2 M^2  x  
+ 
 8 (F_1^V + F_2^V) F_A M^2 (m^2 + {s_u}  - t) 
+8 F_1^V F_2^V M^2 (m^2 - 2 t)  x  \nonumber \\
&+& 4 {F_1^V}^2 M^2 (m^2 - 4 M^2 + {s_u}  - t)  x  
+ 4 F_A^2 M^2 (m^2 + 4 M^2 + {s_u}  - t)  x  
 +  F_P^2 m^2 t  x  
+ {F_2^V}^2 (4 m^2 M^2 - 4 M^2 t - {s_u}  t - t^2)  x
 \left. \right]\\
\mathcal{P}^T_{l,(12)}
  &=&
\frac{m (s-M^2)   \sin (\beta) |\mathbf{p'}| x}{ 2M^3|\mathbf{k'}|\mathcal{I}}   \left[ 
4 M^2  F_3^A F_A   
  +   t  F_3^A F_P  
  \right](-m^2 + {s_u}  + t)
\\
\mathcal{P}^T_{l(2)}
&=&\frac{ m   ( s - M^2 ) |\textbf{p}'| \sin(\beta) x}{2M^3|\textbf{k}'|\mathcal{I}} 
\left[ 
2 {F_3^A}^2 (4 M^2 - {s_u}  - t) t   
  \right],
\end{eqnarray}
where
\begin{equation}
 \sin(\beta) = \sqrt{1+\frac{((s+ M^2) t - 2 m^2 M^2 )^2}{ (s-M^2)^2 t (t - 4M^2)}},
\end{equation}
$\beta$ is the angle between $\textbf{k}$ and $\textbf{q}$.
\end{footnotesize}

\subsubsection{Polarized target asymmetry}
 
 \begin{footnotesize}
\begin{eqnarray}
   \mathcal{T}^L_{N(1)}
   &=& \frac{1}{4M^2(s-M^2) \mathcal{I}}
\left[ \right.
4 F_1^V F_P m^2 M^2 (m^4 + {s_u}  (m^2 - t) - 2 m^2 t + t^2) \nonumber \\
&+& 4 F_2^V F_A M^2 (m^4 (m^2 - 4 M^2) + m^2 (-m^2 + 8 M^2) t + {s_u} ^2 t + (-m^2 - 4 M^2) t^2 + t^3 + {s_u}  (m^4 + m^2 t - 2 t^2)) \nonumber \\
&+& 4 F_1^V F_A M^2 (m^4 (m^2 - 4 M^2) - {s_u} ^3 + m^2 (-m^2 + 8 M^2) t + (-m^2 - 4 M^2) t^2 + t^3 + {s_u} ^2 (-m^2 + t) \nonumber \\
&&+ {s_u}  (m^2 (m^2 + 4 M^2) - 4 M^2 t - t^2)) \nonumber \\
&+& F_2^V F_P m^2 ({s_u} ^2 t + {s_u}  (m^2 - t) t + 4 M^2 (m^4 - 2 m^2 t + t^2)) \nonumber \\
&+& 4 F_A^2 M^2 {s_u}  (-4 m^2 M^2 + (m^2 + 4 M^2) t + {s_u}  t - t^2)  x  
+ 4 {F_2^V}^2 M^2 t (m^4 + {s_u}  (m^2 - t) - 2 m^2 t + t^2)  x  \nonumber \\
&+& 4 F_1^V F_2^V M^2 (4 m^4 M^2 + m^2 (m^2 - 8 M^2) t + {s_u} ^2 t + 2 {s_u}  (m^2 - t) t + 2 (-m^2 + 2 M^2) t^2 + t^3)  x  \nonumber \\
&+& 4 {F_1^V}^2 M^2 ({s_u} ^2 t + {s_u}  (m^2 - t) t + 4 M^2 (m^4 - 2 m^2 t + t^2))  x 
   \left. \right]
\\
\mathcal{T}^L_{N(12)}
&=& \frac{1}{2M^2(s-M^2)\mathcal{I}} 
 \left[ \right. F_2^V F_3^A ({s_u} ^3 t + {s_u} ^2 (m^2 - t) t + {s_u}  (4 m^4 M^2 - 8 m^2 M^2 t + (m^2 + 4 M^2) t^2 - t^3)\nonumber \\
&&+ t (-4 m^4 M^2 + m^2 (m^2 + 8 M^2) t + 2 (-m^2 - 2 M^2) t^2 + t^3)) \nonumber \\
&+& 4 F_1^V F_3^A M^2 (-4 m^4 M^2 + m^2 {s_u}  (m^2 - t) + {s_u} ^2 (m^2 - t) + m^2 (m^2 + 8 M^2) t + 2 (-m^2 - 2 M^2) t^2 + t^3) \nonumber \\
&+& 4 F_3^A F_A M^2 (4 m^4 M^2 + m^2 (-m^2 - 8 M^2) t + {s_u} ^2 t + 2 (m^2 + 2 M^2) t^2 - t^3)  x
 \left. \right]
\end{eqnarray}
\begin{eqnarray}
 \mathcal{T}^T_{N(1)}&=& 
  \frac{|\textbf{p}'|\sin(\beta)}{M \mathcal{I}}  \left[ \right.
  F_2^V F_P m^2 (m^2 - t) 
+ F_1^V F_P m^2 (m^2 + s_u  - t) \nonumber \\
&+& 4 F_1^V F_A M^2 (-m^2 + {s_u}  + t) 
+ F_2^V F_A (-{s_u} ^2 + (4 M^2+{s_u}) (t-m^2)    ) 
- 4 F_A^2 M^2 {s_u}   x  \nonumber \\
& + & 4 {F_1^V}^2 M^2 (m^2 - t)  x  
+ {F_2^V}^2 (m^2 + {s_u}  - t) t  x  
+ F_1^V F_2^V (4 m^2 M^2 + (m^2 - 4 M^2+{s_u}) t     - t^2)  x
   \left. \right]
\\
\mathcal{T}^T_{N(12)} 
 &=& \frac{2|\textbf{p}'|\sin(\beta)}{ M}  \left[ \right.   
  F_1^V F_3^A (s_u^2 + {s_u}  (m^2 - t) + 4 M^2 (t-m^2 )) +  F_2^V F_3^A ((m^2 - t)(s_u- t)) 
\nonumber \\
&+& F_3^A F_A (4 m^2 M^2 - (m^2 + 4 M^2) t - {s_u}  t + t^2)  x 
  \left. \right].
\end{eqnarray}
\end{footnotesize}


\subsubsection{Target-lepton asymmetry} 

\begin{footnotesize}
\begin{eqnarray}
\mathcal{B}^{LL}_{lN(1)} &=&
\frac{1}{(s-M^2) 4 M^3|\mathbf{k'}|\mathcal{I}} 
\left[ \right.
2 F_A F_P m^2 M^2 (4 m^4 M^2 + m^2 (-m^2 - 8 M^2) t + {s_u} ^2 t + 2 (m^2 + 2 M^2) t^2 - t^3)  \nonumber \\
&+& {F_2^V}^2 M^2 (8 m^6 M^2 + m^4 (-3 m^2 - 16 M^2) t + {s_u} ^2 (m^2 - t) t + m^2 (7 m^2 + 8 M^2) t^2 - 5 m^2 t^3 + t^4 + 2 m^2 {s_u}  t (-m^2 + t))  \nonumber \\
&+& F_A^2 M^2 ({s_u} ^3 t + 4 M^2 {s_u} ^2 (m^2 + t) + {s_u}  (12 m^4 M^2 + m^2 (-3 m^2 - 16 M^2) t + 4 (m^2 + M^2) t^2 - t^3)  \nonumber \\
&&+ 2 m^2 (4 m^4 M^2 + m^2 (-m^2 - 8 M^2) t + 2 (m^2 + 2 M^2) t^2 - t^3))  \nonumber \\
&+& F_1^V F_2^V M^2 (12 m^6 M^2 + 5 m^4 (-m^2 - 4 M^2) t + {s_u} ^3 t + {s_u} ^2 (m^2 - t) t + m^2 (11 m^2 + 4 M^2) t^2 + (-7 m^2 + 4 M^2) t^3  \nonumber \\
&&+ t^4 + {s_u}  (4 m^4 M^2 + m^2 (-5 m^2 - 8 M^2) t + 2 (3 m^2 + 2 M^2) t^2 - t^3))  \nonumber \\
&+& {F_1^V}^2 M^2 ({s_u} ^3 t + {s_u}  (4 m^4 M^2 + m^2 (-3 m^2 - 8 M^2) t + 4 (m^2 + M^2) t^2 - t^3)  \nonumber \\
&&+ 2 (2 m^6 M^2 + m^4 (-m^2 - 2 M^2) t + 2 m^2 (m - M) (m + M) t^2 + (-m^2 + 2 M^2) t^3))  \nonumber \\
&+& F_1^V F_P m^2 M^2 (m^4 (m^2 - 8 M^2) + 2 m^2 {s_u}  (m^2 - t) + {s_u} ^2 (m^2 - t) - m^2 (m^2 - 16 M^2) t - (m^2 + 8 M^2) t^2 + t^3)  x   \nonumber \\
&+& F_1^V F_A M^2 (m^6 (m^2 + 4 M^2) - 2 m^2 {s_u} ^3 - {s_u} ^4 + 4 m^4 (-m^2 - 3 M^2) t + 6 m^2 (m^2 + 2 M^2) t^2  + 4 (-m^2 - M^2) t^3  \nonumber \\
&&+ t^4 + 4 {s_u} ^2 (3 m^2 M^2 + (-m^2 - M^2) t) + 2 {s_u}  (m^4 (m^2 + 8 M^2)  + 4 m^2 (-m^2 - M^2) t + (3 m^2 - 4 M^2) t^2))  x   \nonumber \\
&+& F_2^V F_A M^2 (m^6 (m^2 + 4 M^2) + 4 m^4 (-m^2 - 3 M^2) t + {s_u} ^3 t + 6 m^2 (m^2 + 2 M^2) t^2 + 4 (-m^2 - M^2) t^3  \nonumber \\
&&+ t^4 + {s_u} ^2 (m^4 - 2 m^2 t - t^2) + {s_u}  (2 m^4 (m^2 + 6 M^2) + m^2 (-7 m^2 - 8 M^2) t + 2 (3 m^2 - 2 M^2) t^2 - t^3))  x   \nonumber \\
&+& \frac{F_2^V F_P}{4} m^2 ({s_u} ^3 t + 2 {s_u} ^2 (m^2 - t) t + {s_u}  (4 m^4 M^2 + m^2 (m^2 - 8 M^2) t + 4 M^2 t^2 - t^3)  \nonumber \\
&&+ 2 (2 m^6 M^2 - 10 m^4 M^2 t + m^2 (m^2 + 14 M^2) t^2 + 2 (-m^2 - 3 M^2) t^3 + t^4))  x 
 \left. \right]
\end{eqnarray}
\begin{eqnarray}
\mathcal{B}^{LL}_{lN(12)} &=&
\frac{1}{(s-M^2) 8 M^3|\mathbf{k'}|\mathcal{I}}  
 \left[ \right. 
 4 F_3^A F_A M^2 (-4 m^6 M^2 + m^4 (m^2 + 12 M^2) t + {s_u} ^3 t + 3 m^2 (-m^2 - 4 M^2) t^2  \nonumber \\
&&+ (3 m^2 + 4 M^2) t^3 - t^4 + {s_u} ^2 t (-m^2 + t) + {s_u}  (4 m^4 M^2 + m^2 (-m^2 - 8 M^2) t + 2 (m^2 + 2 M^2) t^2 - t^3))  \nonumber \\
&+& F_2^V F_3^A (-32 m^6 M^4 + 4 m^4 M^2 (3 m^2 + 16 M^2) t + 2 m^2 {s_u} ^3 t + {s_u} ^4 t + m^2 (-m^4 - 28 m^2 M^2 - 32 M^4) t^2  \nonumber \\
&&+ m^2 (3 m^2 + 20 M^2) t^3 + (-3 m^2 - 4 M^2) t^4 + t^5 + {s_u} ^2 (4 m^4 M^2 + m^2 (m^2 - 16 M^2) t  \nonumber \\
&&+ (3 m^2 + 4 M^2) t^2 - 2 t^3) + 2 m^2 {s_u}  (2 m^4 M^2 - 8 m^2 M^2 t + (m^2 + 6 M^2) t^2 - t^3))  x   \nonumber \\
&+& 4 F_1^V F_3^A M^2 (-4 m^6 M^2 + {s_u} ^3 (m^2 - t) + m^4 (m^2 + 4 M^2) t + m^2 (-m^2 + 4 M^2) t^2 - (m^2 + 4 M^2) t^3  \nonumber \\
&& + t^4 + {s_u} ^2 (2 m^4 - m^2 t - t^2) + {s_u}  (m^4 (m^2 - 12 M^2) + m^2 (m^2 + 16 M^2) t - (3 m^2 + 4 M^2) t^2 + t^3))  x 
 \left. \right]
\end{eqnarray}
\begin{eqnarray}
\mathcal{B}^{TT}_{lN(1)} &=&
\frac{m}{(s-M^2) 8 M^2|\mathbf{k'}|\mathcal{I}}
\left[ \right. 
4 F_A^2 M^2 {s_u}  (m^2 (-m^2 + 8 M^2) - 2 m^2 {s_u}  - {s_u} ^2 - 8 M^2 t + t^2)  \nonumber \\
&+& 4 {F_1^V}^2 M^2 (m^4 (m^2 - 8 M^2) + 2 m^2 {s_u}  (m^2 - t) + {s_u} ^2 (m^2 - t) + m^2 (-m^2 + 16 M^2) t + (-m^2 - 8 M^2) t^2 + t^3)  \nonumber \\
&+& F_A F_P (4 m^6 M^2 + m^4 (-m^2 - 12 M^2) t + {s_u} ^2 (-m^2 - t) t + 3 m^2 (m^2 + 4 M^2) t^2 + (-3 m^2 - 4 M^2) t^3 + t^4  \nonumber \\
&& + 2 {s_u}  (2 m^4 M^2 - m^4 t + (m^2 - 2 M^2) t^2))  \nonumber \\
&+& {F_2^V}^2 ({s_u} ^3 t + 2 {s_u} ^2 (m^2 - t) t + {s_u}  (4 m^4 M^2 + m^2 (m^2 - 8 M^2) t + 4 M^2 t^2 - t^3)  \nonumber \\
&&+ 2 (2 m^6 M^2 - 10 m^4 M^2 t + m^2 (m^2 + 14 M^2) t^2 + 2 (-m^2 - 3 M^2) t^3 + t^4))  \nonumber \\
&+& F_1^V F_2^V ({s_u} ^3 t + 2 {s_u} ^2 (2 m^2 M^2 + (m^2 - 2 M^2) t - t^2) + {s_u}  (12 m^4 M^2 + m^2 (m^2 - 16 M^2) t + 4 M^2 t^2 - t^3)  \nonumber \\
&&+ 2 (4 m^4 M^2 (m^2 - 4 M^2) + 4 m^2 M^2 (-3 m^2 + 8 M^2) t + (m^4 + 12 m^2 M^2 - 16 M^4) t^2 -2 (m^2 + 2 M^2) t^3 + t^4))  \nonumber \\
&+& F_2^V F_P (8 m^6 M^2 + m^4 (-3 m^2 - 16 M^2) t + {s_u} ^2 (m^2 - t) t + m^2 (7 m^2 + 8 M^2) t^2 - 5 m^2 t^3 + t^4  + 2 m^2 {s_u}  t (-m^2 + t))  x  \nonumber \\
&+& 4 F_1^V F_A M^2 (m^4 (-m^2 + 8 M^2) + {s_u} ^3 + m^2 (m^2 - 16 M^2) t + (m^2 + 8 M^2) t^2 - t^3 + {s_u} ^2 (m^2 + t)  \nonumber \\
&&+ {s_u}  (m^2 (-m^2 - 8 M^2) + 2 (m^2 + 4 M^2) t - t^2))  x   \nonumber \\
&+& F_2^V F_A (32 m^4 M^4 + m^2 (-m^4 - 8 m^2 M^2 - 64 M^4) t + 2 {s_u} ^3 t + (3 m^4 + 16 m^2 M^2 + 32 M^4) t^2 - (3 m^2 + 8 M^2) t^3  \nonumber \\
&&+ t^4 + {s_u} ^2 (-4 m^2 M^2 + (3 m^2 + 4 M^2) t - t^2) + 2 {s_u}  (-2 m^4 M^2 - 4 m^2 M^2 t + (m^2 + 6 M^2) t^2 - t^3))  x   \nonumber \\
&+& F_1^V F_P ({s_u} ^3 t + {s_u}  (4 m^4 M^2 - m^2 (3 m^2 + 8 M^2) t + 4 (m^2 + M^2) t^2 - t^3)  \nonumber \\
&&+ 2 (2 m^6 M^2 + m^4 (-m^2 - 2 M^2) t + 2 m^2 (m - M) (m + M) t^2 + (-m^2 + 2 M^2) t^3))  x
 \left. \right]
 \end{eqnarray}
 \begin{eqnarray}
\mathcal{B}^{TT}_{lN(12)} &=&
\frac{m}{(s-M^2) 4 M^2|\mathbf{k'}|\mathcal{I}}
 \left[ \right.  F_2^V F_3^A (-4 m^6 M^2 + m^4 (m^2 + 20 M^2) t + m^2 (-5 m^2 - 28 M^2) t^2  \nonumber \\
 &&+ (7 m^2 + 12 M^2) t^3 - 3 t^4 + 3 {s_u} ^2 t (-m^2 + t) + 2 {s_u}  (2 m^4 M^2 - m^4 t + (m^2 - 2 M^2) t^2))  x   \nonumber \\
&+&
F_3^A F_A (-({s_u} ^3 t) + 2 {s_u} ^2 (2 m^2 M^2 + (-m^2 - 2 M^2) t + t^2) + {s_u}  (4 m^4 M^2 - m^4 t - 4 M^2 t^2 + t^3)  \nonumber \\
&&+ 2 (-16 m^4 M^4 + 8 m^2 M^2 (m^2 + 4 M^2) t + (-m^4 - 16 m^2 M^2 - 16 M^4) t^2 + 2 (m^2 + 4 M^2) t^3 - t^4))  \nonumber \\
&+&  F_1^V F_3^A (4 m^4 M^2 (-m^2 + 8 M^2) + m^2 (m^4 + 4 m^2 M^2 - 64 M^4) t - 2 {s_u} ^3 t + (-3 m^4 + 4 m^2 M^2 + 32 M^4) t^2  \nonumber \\
&&+ (3 m^2 - 4 M^2) t^3 - t^4 + {s_u} ^2 t (-3 m^2 + 8 M^2 + t) + 2 {s_u}  (-2 m^4 M^2 + 8 m^2 M^2 t + (-m^2 - 6 M^2) t^2 + t^3))  x   \left. \right]
\end{eqnarray}
\begin{eqnarray}
\mathcal{B}^{NN}_{lN(1)} &=&
\frac{m}{ M \mathcal{I}} \left[ \right. 4 F_A^2 M^2 {s_u}  
+ F_A F_P {s_u}  t 
+ 4 {F_1^V}^2 M^2 (-m^2 + t) 
+ {F_2^V}^2 t (-m^2 + t) 
+ 4 F_1^V F_A M^2 (m^2 - {s_u}  - t)  x   \nonumber \\
&+&( F_1^V + F_2^V )F_P (m^2 - t) t  x  
+ F_2^V F_A (4 M^2 (m^2 - t) - {s_u}  t)  x 
+ F_1^V F_2^V (-4 m^2 M^2 + (-m^2 + 4 M^2) t + t^2)  
 \left. \right]  \\
\mathcal{B}^{NN}_{lN,(12)} &=&
\frac{2m}{M \mathcal{I}} \left[ \right. 
  F_3^A F_A (-4 m^2 M^2 + (m^2 + 4 M^2) t - t^2) 
+   (F_1^V + F_2^V)F_3^A {s_u}  t  x   
 \left. \right]
\end{eqnarray}
\begin{eqnarray}
\mathcal{B}^{LT}_{lN(1)}  &=&
\frac{  |\mathbf{q}|\sin(\beta)}{ 4 M^2|\mathbf{k'}|\mathcal{I}} 
\left[ \right.
4 F_A^2 M^2 {s_u}  (-m^2 - {s_u}  - t) 
+ 4 {F_1^V}^2 M^2 (m^4 + {s_u}  (m^2 - t) - t^2)   \nonumber \\
&+& 2 F_A F_P m^2 (4 m^2 M^2 + (-m^2 - 4 M^2) t - {s_u}  t + t^2) 
+ {F_2^V}^2 (8 m^4 M^2 + m^2 (-m^2 - 8 M^2) t + {s_u} ^2 t + 2 m^2 t^2 - t^3)   \nonumber \\
&+& F_1^V F_2^V (12 m^4 M^2 + 4 M^2 {s_u}  (m^2 - t) + m^2 (-m^2 - 8 M^2) t + {s_u} ^2 t + 2 (m^2 - 2 M^2) t^2 - t^3)   \nonumber \\
&+& F_1^V F_P m^2 (m^2 (m^2 - 8 M^2) + {s_u} ^2 + 2 {s_u}  (m^2 - t) + 2 (-m^2 + 4 M^2) t + t^2)  x    \nonumber \\
&+& 4 F_1^V F_A M^2 (-m^4 + {s_u} ^2 + 2 {s_u}  t + t^2)  x  + F_2^V F_P m^2 (m^4 + {s_u}  (m^2 - t) - 4 m^2 t + 3 t^2)  x    \nonumber \\
&+& F_2^V F_A (-2 m^2 {s_u} ^2 - {s_u} ^3 + {s_u}  (m^2 (-m^2 + 12 M^2) + 2 (-m^2 + 2 M^2) t + t^2)   \nonumber \\
&&+ 2 (2 m^4 M^2 + m^2 (-m^2 - 4 M^2) t + (m^2 + 2 M^2) t^2))  x \left. \right]
\\
\mathcal{B}^{LT}_{lN(12)} &=&
\frac{ |\mathbf{q}| \sin(\beta)}{ 2 M^2|\mathbf{k'}|\mathcal{I}} 
 \left[ \right. 
  F_3^A F_A (-4 m^4 M^2 + 4 M^2 {s_u}  (m^2 - t) + m^2 (m^2 + 8 M^2) t - {s_u} ^2 t + 2 (-m^2 - 2 M^2) t^2 + t^3)   \nonumber \\
&+&  F_2^V F_3^A (-8 m^4 M^2 + m^2 {s_u}  (m^2 - t) + {s_u} ^2 (m^2 - t) + m^2 (3 m^2 + 8 M^2) t - 4 m^2 t^2 + t^3)  x    \nonumber \\
&+& F_1^V F_3^A (2 m^2 {s_u} ^2 + {s_u} ^3 + {s_u}  (m^2 (m^2 - 12 M^2) + 2 (m^2 + 2 M^2) t - t^2)  + 2 (-2 m^4 M^2 + m^4 t + (-m^2 + 2 M^2) t^2))  x 
 \left. \right]
\end{eqnarray}
\begin{eqnarray}
\mathcal{B}^{TL}_{lN(1)} &=&
\frac{m |\mathbf{q}| \sin(\beta)}{ 2M^3|\mathbf{k'}|\mathcal{I}} 
\left[ \right.
 {F_1^V}^2 M^2 (m^2 (-m^2 + 8 M^2) - {s_u} ^2 + 2 (m^2 - 4 M^2) t - t^2 + 2 {s_u}  (-m^2 + t)) \nonumber \\
&+& F_A F_P M^2 (-m^4 + {s_u}  (-m^2 - t) + 2 m^2 t - t^2) 
+ {F_2^V}^2 M^2 (-m^4 + 4 m^2 t - 3 t^2 + {s_u}  (-m^2 + t)) \nonumber \\
&+&  
F_A^2 M^2 (-m^4 - {s_u} ^2 + 2 m^2 t - t^2 + 2 {s_u}  (-m^2 - 4 M^2 + t)) \nonumber \\
&+&  F_1^V F_2^V M^2 (-{s_u} ^2 + 3 {s_u}  (-m^2 + t) + 2 (m^2 (-m^2 + 4 M^2) + (3 m^2 - 4 M^2) t - 2 t^2)) \nonumber \\
&+& \frac{F_2^V F_P}{4} (-8 m^4 M^2 + m^2 (m^2 + 8 M^2) t - {s_u} ^2 t - 2 m^2 t^2 + t^3)  x 
+  F_1^V F_P M^2 (-m^4 + t^2 + {s_u}  (-m^2 + t))  x  \nonumber \\
&+& 2 F_1^V F_A M^2 (m^2 (-m^2 - 4 M^2) - {s_u} ^2 + 2 (m^2 + 2 M^2) t - t^2 + 2 {s_u}  (-m^2 + 2 M^2 + t))  x  \nonumber \\
&+&  F_2^V F_A M^2 (-{s_u} ^2 + {s_u}  (-3 m^2 + 5 t) + 2 (m^2 (-m^2 - 4 M^2) + 2 (m^2 + 2 M^2) t - t^2))  x 
 \left. \right]
\end{eqnarray}
\begin{eqnarray}
\mathcal{B}^{TL}_{lN(12)}&=&
\frac{m  |\mathbf{q}| \sin(\beta)}{ M^3|\mathbf{k'}|\mathcal{I}} 
 \left[ \right. 
  F_3^A F_A M^2 (-{s_u} ^2 + {s_u}  (-m^2 + t) + 2 (4 m^2 M^2 + (-m^2 - 4 M^2) t + t^2))  \nonumber \\
&+&  F_1^V F_3^A M^2 (m^2 (m^2 - 8 M^2) + {s_u}  (m^2 - t) + 8 M^2 t - t^2)  x   \nonumber \\
&+&\frac{ F_2^V F_3^A}{2} (2 m^4 M^2 - 8 m^2 M^2 t + {s_u} ^2 t + (m^2 + 6 M^2) t^2 - t^3 + {s_u}  (-2 m^2 M^2 + (m^2 - 2 M^2) t))  x
 \left. \right]
\end{eqnarray}

\end{footnotesize}

\subsubsection{Lepton-recoil asymmetry}

\begin{footnotesize}

\begin{eqnarray}
 \mathcal{A}^{LL}_{lN'(1)} &=&
\frac{1}{ 4 M^4 |\mathbf{k}'||\mathbf{q}|\mathcal{I}} 
\left[ \right.  
({F_1^V}+{F_2^V})^2 M^2 t 
({s_u} ^2 t + {s_u}  t (-m^2 + t) + 2 m^2 (4 m^2 M^2 + (-m^2 - 4 M^2) t + t^2)) \nonumber \\
&+& F_A^2 M^2 ({s_u} ^2 t (-4 M^2 + t) + {s_u}  t (4 m^2 M^2 - (m^2 + 4 M^2) t + t^2) + 2 m^2 (-16 m^2 M^4 + 8 M^2 (m^2 + 2 M^2) t - (m^2 + 8 M^2) t^2 + t^3)) \nonumber \\
&+& \frac{F_P}{4} (F_2^V  t + 4 M^2 F_1^V ) m^2 t  (-{s_u} ^2 + {s_u}  (-m^2 + t) + 2 (4 m^2 M^2 + (-m^2 - 4 M^2) t + t^2))  x  \nonumber \\
&+& F_1^V F_A M^2 (4 m^4 M^2 (-m^2 + 8 M^2) + m^2 (m^4 + 4 m^2 M^2 - 32 M^4) t - {s_u} ^3 t - m^2 (3 m^2 + 4 M^2) t^2 - t^4 + (3 m^2 + 4 M^2) t^3 \nonumber \\
&&- {s_u} ^2 (4 m^2 M^2 + m^2 t + t^2)
+ {s_u}  (-8 m^4 M^2 + m^2 (m^2 + 20 M^2) t - 4 (m^2 - M^2) t^2 - t^3))  x  \nonumber \\
&+& F_2^V F_A M^2 (-4 m^6 M^2 + m^4 (m^2 + 20 M^2) t - {s_u} ^3 t + {s_u} ^2 (-2 m^2 - t) t + 5 m^2 (-m^2 - 4 M^2) t^2 + (5 m^2 + 4 M^2) t^3 - t^4 \nonumber \\
&&+ {s_u}  (-4 m^4 M^2 + 16 m^2 M^2 t + (-3 m^2 + 4 M^2) t^2 - t^3))  x 
 \left. \right]
 \end{eqnarray}
 \begin{eqnarray}
\mathcal{A}^{LL}_{lN'(12)} &=&
\frac{1}{ 8 M^4 |\mathbf{k}'||\mathbf{q}|\mathcal{I}}  
 \left[ \right. 
 4 F_1^V F_3^A M^2 t (-4 m^4 M^2 - {s_u} ^3 + {s_u} ^2 (-m^2 - t) + m^2 (m^2 + 8 M^2) t + 2 (-m^2 - 2 M^2) t^2 + t^3 
\nonumber \\
 && +{s_u}  (12 m^2 M^2 + (-3 m^2 - 4 M^2) t + t^2))  x  \nonumber \\
&+& F_2^V F_3^A t^2 (-4 m^4 M^2 - {s_u} ^3 + {s_u} ^2 (-m^2 - t) + m^2 (m^2 + 8 M^2) t + 2 (-m^2 - 2 M^2) t^2 + t^3 \nonumber \\
&&+ {s_u}  (12 m^2 M^2 + (-3 m^2 - 4 M^2) t + t^2))  x \left. \right]
\end{eqnarray}
\begin{eqnarray}
 \mathcal{A}^{TT}_{lN'(1)} &=&
\frac{m}{8|\mathbf{q}| |\mathbf{k'}|M^3\mathcal{I}} 
\left[ \right. 
( 4 {F_1^V}^2 M^2 + {F_2^V}^2 t ) t ({s_u} ^2 + {s_u}  (m^2 - t) + 2 (-4 m^2 M^2 + (m^2 + 4 M^2) t - t^2)) \nonumber \\
&+& ( 4 F_A^2 M^2 + F_A F_P t) (-4 m^4 M^2 + m^2 (m^2 + 8 M^2) t + 2 (-m^2 - 2 M^2) t^2 + t^3 + {s_u}  (-4 m^2 M^2 + (m^2 - 4 M^2) t + t^2)) \nonumber \\
&+& F_1^V F_2^V t ({s_u} ^2 (4 M^2 + t) + {s_u}  (4 m^2 M^2 + (m^2 - 4 M^2) t - t^2) + 2 (-16 m^2 M^4 + 16 M^4 t + m^2 t^2 - t^3)) \nonumber \\
&+& F_2^V F_A (-32 m^4 M^4 + 4 m^2 M^2 (3 m^2 + 8 M^2) t + m^2 (-m^2 - 16 M^2) t^2 + 2 (m^2 + 2 M^2) t^3 - t^4 + {s_u}  t^2 (m^2 - 8 M^2 + t) \nonumber \\
&&+ 2 {s_u} ^2 t (-2 M^2 + t))  x  \nonumber \\
&+& 4 F_1^V F_A M^2 (-4 m^4 M^2 + m^4 t + {s_u} ^2 t + 4 M^2 t^2 - t^3 + 2 {s_u}  (-2 m^2 M^2 + (m^2 - 2 M^2) t))  x  \nonumber \\
&+& (F_1^V + F_2^V) F_P t (-({s_u} ^2 t) + {s_u}  (m^2 - t) t + 2 m^2 (-4 m^2 M^2 + (m^2 + 4 M^2) t - t^2))  x 
  \left. \right]
\\
\mathcal{A}^{TT}_{lN'(12)} &=&
\frac{m}{4|\mathbf{q}| |\mathbf{k'}|M^3\mathcal{I}} 
 \left[ \right. 
F_3^A F_A t ({s_u} ^2 (-4 M^2 + t) + {s_u}  (-4 m^2 M^2 + (m^2 + 4 M^2) t - t^2) + 2 (16 m^2 M^4 + 8 M^2 (-m^2 - 2 M^2) t \nonumber \\
&&+ (m^2 + 8 M^2) t^2 - t^3)) \nonumber \\
&+& (F_1^V + F_2^V) F_3^A t (4 m^4 M^2 + m^2 (-m^2 - 8 M^2) t + 2 {s_u} ^2 t + 2 (m^2 + 2 M^2) t^2 - t^3 + {s_u}  (-4 m^2 M^2 + (m^2 - 4 M^2) t + t^2))  x   \left. \right]\nonumber \\
\end{eqnarray}
\begin{eqnarray}
\mathcal{A}^{NN}_{lN'(1)} &=&
\frac{m}{ M\mathcal{I} } \left[ \right.
-4 F_A^2 M^2 {s_u}  
+ 4 {F_1^V}^2 M^2 (-m^2 + t) 
+ {F_2^V}^2 t (-m^2 + t)
+ F_1^V F_2^V (-4 m^2 M^2 + (-m^2 + 4 M^2) t + t^2) \nonumber \\
&+& (F_1^V + F_2^V) F_P t (-m^2 + t)  x - F_A F_P {s_u}  t  + 
 4 F_1^V F_A M^2 (-m^2 - {s_u}  + t)  x  
- F_2^V F_A ({s_u}  t+ 4 M^2 (m^2 - t))  x
 \left. \right] 
\\
\mathcal{A}^{NN}_{lN'(12)} &=&
\frac{2m}{ M\mathcal{I}} \left[ \right. 
(F_1^V +F_2^V) F_3^A x {s_u}  t    
+F_3^A  F_A (4 m^2 M^2 - (m^2 + 4 M^2) t + t^2)  
 \left. \right]
\end{eqnarray}
\begin{eqnarray}
\mathcal{A}^{LT}_{lN'(1)} &=&
\frac{m (s-M^2) \sin(\beta)}{4 M^3|\mathbf{k'}|\mathcal{I}} 
\left[ \right.
8 F_A^2 m^2 M^2 (-4 M^2 + t) + 2 F_A F_P m^2 t (-4 M^2 + t) + 4 {F_1^V}^2 M^2 t (-m^2 + {s_u}  + t)  \nonumber \\
&+&{F_2^V}^2 t^2 (-m^2 + {s_u}  + t) 
+ F_1^V F_2^V t (-4 m^2 M^2 + (-m^2 + 4 M^2) t + t^2 + {s_u}  (4 M^2 + t))  \nonumber \\
&+& 4 F_1^V F_A M^2 (-{s_u} ^2 + {s_u}  (-2 m^2 - t) + m^2 (-m^2 + 8 M^2 - t))  x  
+( F_1^V + F_2^V )F_P m^2 t (-m^2 - {s_u}  + t)  x   \nonumber \\
&+& F_2^V F_A (-({s_u} ^2 t) + {s_u}  (-4 m^2 M^2 - m^2 t - t^2) + 2 m^2 (-2 m^2 M^2 + 6 M^2 t - t^2))  x 
\left. \right]
\end{eqnarray}
\begin{eqnarray}
\mathcal{A}^{LT}_{lN'(12)} &=&
\frac{m (s-M^2) \sin(\beta)}{ M^3|\mathbf{k'}|\mathcal{I}} 
\left[ \right. 
  F_3^A F_A t (4 m^2 M^2 + (-m^2 - 4 M^2) t + t^2 + {s_u}  (-4 M^2 + t))  \nonumber \\
&+& (F_1^V + F_2^V) F_3^A (-{s_u} ^2 + {s_u}  (-m^2 - t) + 2 m^2 (4 M^2 - t)) t  x   \left. \right]
\end{eqnarray}
\begin{eqnarray}
\mathcal{A}^{TL}_{lN'(1)} &=&
\frac{m (s-M^2) \sin(\beta)}{ 2 M^4|\mathbf{k'}|\mathcal{I}} 
\left[ \right.
{F_1^V}^2 M^2 t (-m^2 - {s_u}  + t) 
+ 2 F_1^V F_2^V M^2 t (-m^2 - {s_u}  + t)\nonumber \\
&+& {F_2^V}^2 M^2 t (-m^2 - {s_u}  + t) 
+ F_A^2 M^2 (4 m^2 M^2 + {s_u}  (4 M^2 - t) + (-m^2 - 4 M^2) t + t^2) \nonumber \\
&+& F_1^V F_P M^2 t (-m^2 + {s_u}  + t)  x  
+ 2 F_1^V F_A M^2 (2 m^2 M^2 + {s_u}  (2 M^2 - t) + (-m^2 - 2 M^2) t + t^2)  x \nonumber \\
&+& F_2^V F_A M^2 (8 m^2 M^2 + (-3 m^2 - 8 M^2) t - {s_u}  t + 3 t^2)  x + \frac{F_2^V F_P }{4}t^2 (-m^2 + {s_u}  + t)  x  
 \left. \right]
\\
\mathcal{A}^{TL}_{lN'(12)} &=&
\frac{m (s-M^2) \sin(\beta) x}{ 2 M^4|\mathbf{k'}|\mathcal{I}}
\left[ \right.
F_3^A (4 M^2 - {s_u}  - t) t (4 F_1^V M^2 + F_2^V t) 
 \left. \right]
\end{eqnarray}
\end{footnotesize}

\subsubsection{Target-recoil asymmetry}

\begin{footnotesize}
\begin{eqnarray}
 \mathcal{C}^{LL}_{NN'(1)} &=&\frac{1}{16M^3(s-M^2)|\textbf{p}'|\mathcal{I}}
\left[ \right.
8 F_A F_P m^2 M^2 {s_u}  (m^2 + {s_u}  - t) t \nonumber \\
&+& F_P^2 m^2 t (-4 m^4 M^2 + m^2 (m^2 + 8 M^2) t + {s_u}  (m^2 - t) t + 2 (-m^2 - 2 M^2) t^2 + t^3) \nonumber \\
&+& {F_2^V}^2 (16 m^6 M^4 - 4 m^4 M^2 (m^2 + 4 M^2) t - 16 m^2 M^4 t^2 + {s_u} ^3 t^2 + (m^4 + 12 m^2 M^2 + 16 M^4) t^3 - 2 (m^2 + 4 M^2) t^4 \nonumber \\
&&+ t^5 + {s_u}^2 t (4 m^2 M^2 + (m^2 - 4 M^2) t - t^2) + {s_u}  t (4 m^4 M^2 - 16 m^2 M^2 t + (m^2 + 12 M^2) t^2 - t^3)) \nonumber \\
&+& 4 F_A^2 M^2 (4 m^4 M^2 (m^2 - 4 M^2) + m^2 (-m^4 + 32 M^4) t + {s_u} ^3 t + (m^4 - 12 m^2 M^2 - 16 M^4) t^2 - t^4 \nonumber \\
&&+ (m^2 + 8 M^2) t^3 + {s_u} ^2 (4 m^2 M^2 + (m^2 + 4 M^2) t - t^2) + {s_u}  (8 m^4 M^2 - m^2 (m^2 + 4 M^2) t - 4 M^2 t^2 + t^3)) \nonumber \\
&+& 4 {F_1^V}^2 M^2 (4 m^4 M^2 (m^2 - 4 M^2) + m^2 (-m^4 + 32 M^4) t + {s_u} ^3 t + (m^4 - 12 m^2 M^2 - 16 M^4) t^2 - t^4 + (m^2 + 8 M^2) t^3 
\nonumber \\
&&+ {s_u} ^2 (4 m^2 M^2 + (m^2 - 4 M^2) t - t^2) + {s_u}  (8 m^4 M^2 - m^2 (m^2 + 12 M^2) t + 4 M^2 t^2 + t^3)) \nonumber \\
&+& 8 F_1^V F_2^V M^2 ({s_u} ^3 t + 2 {s_u} ^2 (m^2 - t) t + m^2 (4 m^4 M^2 + m^2 (-m^2 - 8 M^2) t + 2 (m^2 + 2 M^2) t^2 - t^3) \nonumber \\
&&+ {s_u}  (4 m^4 M^2 - 8 m^2 M^2 t + (-m^2 + 4 M^2) t^2 + t^3)) \nonumber \\
&+& 8 F_2^V F_A M^2 t (4 m^4 M^2 - m^2 (m^2 + 8 M^2) t - {s_u} ^2 t + 2 (m^2 + 2 M^2) t^2 - t^3 + 2 {s_u}  (4 m^2 M^2 - (m^2 + 4 M^2) t + t^2))  x  \nonumber \\
&+& 16 F_1^V F_A M^2 ({s_u} ^2 (2 M^2 - t) t + {s_u}  t (4 m^2 M^2 - (m^2 + 4 M^2) t + t^2) \nonumber \\
&&+ 2 M^2 (4 m^4 M^2 + m^2 (-m^2 - 8 M^2) t + 2 (m^2 + 2 M^2) t^2 - t^3))  x 
 \left. \right]
\end{eqnarray}
\begin{eqnarray}
 \mathcal{C}^{LL}_{NN'(2)} &=&\frac{1}{4M^3(s-M^2)|\textbf{p}'|\mathcal{I}}
\left[\right. 
 {F_3^A}^2 t (16 m^4 M^4 + 8 m^2 M^2 (-m^2 - 4 M^2) t + {s_u} ^3 t + (m^4 + 16 m^2 M^2 + 16 M^4) t^2 \nonumber \\
&&+ 2 (-m^2 - 4 M^2) t^3 + t^4 + {s_u} ^2 (-4 m^2 M^2 + (m^2 + 4 M^2) t - t^2) + {s_u}  t (-4 m^2 M^2 + (m^2 + 4 M^2) t - t^2))
\left.\right]
\\
\mathcal{C}^{LL}_{NN'(12)} &=&\frac{1}{4M^3(s-M^2)|\textbf{p}'|\mathcal{I}}
\left[\right. 
4 F_3^A F_A M^2 {s_u}  t (-4 m^2 M^2 + m^2 {s_u}  + {s_u} ^2 + (m^2 + 4 M^2) t - t^2) \nonumber \\
&+& F_3^A F_P m^2 {s_u}  t (-4 m^2 M^2 + (m^2 + 4 M^2) t + {s_u}  t - t^2) \nonumber \\
&+& 4 (F_1^V + F_2^V) F_3^A M^2 t (-4 m^4 M^2 + m^2 (m^2 + 8 M^2) t - {s_u} ^2 t + 2 (-m^2 - 2 M^2) t^2 + t^3)  x \left.\right]
\end{eqnarray}
\begin{eqnarray}
\mathcal{C}^{TT}_{NN'(1)} &=&\frac{1}{4M^3(s-M^2)|\textbf{p}'|}
\left[\right. 
2 F_A F_P m^2 M^2 {s_u}  (m^2 + {s_u}  - t) t \nonumber \\
&+& \frac{F_P^2}{4} m^2 t (-4 m^4 M^2 + m^2 (m^2 + 8 M^2) t + {s_u}  (m^2 - t) t + 2 (-m^2 - 2 M^2) t^2 + t^3)  \nonumber \\
&+& F_A^2 M^2 (4 m^4 M^2 (m^2 + 4 M^2) + m^2 (-m^4 - 16 m^2 M^2 - 32 M^4) t + {s_u} ^3 t + (3 m^4 + 20 m^2 M^2 + 16 M^4) t^2 + t^4  \nonumber \\
&&- (3 m^2 + 8 M^2) t^3  + {s_u} ^2 (4 m^2 M^2 + (m^2 + 4 M^2) t - t^2)  + {s_u}  (8 m^4 M^2 - m^2 (m^2 + 12 M^2) t + 2 (m^2 + 2 M^2) t^2 - t^3))  \nonumber \\
&+& 2 F_1^V F_2^V M^2 (-4 m^6 M^2 + m^4 (m^2 + 16 M^2) t - {s_u} ^3 t + 4 m^2 (-m^2 - 5 M^2) t^2 + (5 m^2 + 8 M^2) t^3 - 2 t^4 + 2 {s_u} ^2 t (-m^2 + t)  \nonumber \\
&&+ {s_u}  (-4 m^4 M^2 + 8 m^2 M^2 t + (-m^2 - 4 M^2) t^2 + t^3)) \nonumber \\
&+& \frac{{F_2^V}^2}{4} (-16 m^6 M^4 + 4 m^4 M^2 (m^2 + 12 M^2) t - 8 m^2 M^2 (m^2 + 6 M^2) t^2 - {s_u} ^3 t^2 + (-m^4 + 4 m^2 M^2 + 16 M^4) t^3 \nonumber \\
&&+ 2 m^2 t^4 - t^5 - {s_u} ^2 t (4 m^2 M^2+ (m^2 - 4 M^2) t - t^2) + {s_u}  t (-4 m^4 M^2 + 8 m^2 M^2 t - (m^2 + 4 M^2) t^2 + t^3))  \nonumber \\
&+& {F_1^V}^2 M^2 (4 m^4 M^2 (-m^2 + 4 M^2) + m^2 (m^4 + 8 m^2 M^2 - 32 M^4) t - {s_u} ^3 t + (-3 m^4 - 4 m^2 M^2 + 16 M^4) t^2 - t^4  \nonumber \\
&&+ 3 m^2 t^3  - {s_u} ^2 (4 m^2 M^2 + (m^2 - 4 M^2) t - t^2) + {s_u}  (-8 m^4 M^2 + m^2 (m^2 + 12 M^2) t - 2 (m^2 + 2 M^2) t^2 + t^3))  \nonumber \\
&+& 8 F_1^V F_A M^4 (-4 m^4 M^2 + m^2 (m^2 + 8 M^2) t - {s_u} ^2 t + 2 (-m^2 - 2 M^2) t^2 + t^3)  x   \nonumber \\
&+& 2 F_2^V F_A M^2 t (-4 m^4 M^2 + m^2 (m^2 + 8 M^2) t - {s_u} ^2 t + 2 (-m^2 - 2 M^2) t^2 + t^3)  x 
\left.\right] 
\end{eqnarray}
\begin{eqnarray}
\mathcal{C}^{TT}_{NN'(2)}  &=&\frac{1}{4M^3(s-M^2)|\textbf{p}'|}
\left[\right. 
{F_3^A}^2 t (16 m^4 M^4 + 8 m^2 M^2 (-m^2 - 4 M^2) t + {s_u} ^3 t + (m^4 + 16 m^2 M^2 + 16 M^4) t^2   \nonumber \\
&&+ 2 (-m^2 - 4 M^2) t^3 + t^4 + {s_u} ^2 (-4 m^2 M^2 + (m^2 + 4 M^2) t - t^2) + {s_u}  t (-4 m^2 M^2 + (m^2 + 4 M^2) t - t^2)) 
\left.\right] 
\end{eqnarray}
\begin{eqnarray}
\mathcal{C}^{TT}_{NN'(12)} &=&\frac{1}{4M^3(s-M^2)|\textbf{p}'|}
\left[\right. 
4 F_3^A F_A M^2 {s_u}  t (-4 m^2 M^2 + m^2 {s_u}  + {s_u} ^2 + (m^2 + 4 M^2) t - t^2)   \nonumber \\
&+& F_3^A F_P m^2 {s_u}  t (-4 m^2 M^2 + (m^2 + 4 M^2) t + {s_u}  t - t^2)   \nonumber \\
&+& 4 (F_1^V + F_2^V) F_3^A M^2 t (-4 m^4 M^2 + m^2 (m^2 + 8 M^2) t - {s_u} ^2 t + 2 (-m^2 - 2 M^2) t^2 + t^3)  x \left. \right]
\end{eqnarray}
\begin{eqnarray}
\mathcal{C}^{NN}_{NN'(1)}  &=& \frac{1}{4M^2\mathcal{I}}
\left[\right. 
8 F_A F_P m^2 M^2 (m^2 - t) 
+ F_P^2 m^2 (m^2 - t) t 
+ 8 F_1^V F_2^V m^2 M^2 (-m^2 + t) \nonumber \\
&+& 4 {F_1^V}^2 M^2 (m^2 (-m^2 - 4 M^2) + {s_u} ^2 + 2 (m^2 + 2 M^2) t - t^2) + 4 F_A^2 M^2 (m^2 (m^2 + 4 M^2) - {s_u} ^2 + 2 (-m^2 - 2 M^2) t + t^2) \nonumber \\
&+&{F_2^V}^2 (-4 m^4 M^2 + 8 m^2 M^2 t - {s_u} ^2 t + (-m^2 - 4 M^2) t^2 + t^3)
\left.\right] 
\\
 \mathcal{C}^{NN}_{NN'(2)}  &=& \frac{1}{M^2\mathcal{I}}
\left[\right.
{F_3^A}^2 t (-4 m^2 M^2 + {s_u} ^2 + (m^2 + 4 M^2) t - t^2)
\left.\right] 
\\
\mathcal{C}^{NN}_{NN'(12)}&=&  \frac{m^2  {s_u}}{M^2}
\left[ 4 F_3^A F_A  M^2  
+  F_3^A F_P   t
\right] 
\end{eqnarray}
\begin{eqnarray}
\mathcal{C}^{TL}_{NN'(1)}&=&\frac{\sin(\beta)}{4M^2\mathcal{I}}
\left[\right.  4 {F_1^V}^2 M^2 (m^2 (m^2 - 4 M^2) + {s_u} ^2 + 2 {s_u}  (m^2 - t) + 4 M^2 t - t^2)  \nonumber \\
&+& 4 F_A^2 M^2 (m^2 (-m^2 - 4 M^2) - 2 m^2 {s_u}  - {s_u} ^2 + 2 (m^2 + 2 M^2) t - t^2)   \nonumber \\
&+& {F_2^V}^2 (4 m^4 M^2 + {s_u} ^2 t + 2 {s_u}  (m^2 - t) t - (m^2 + 4 M^2) t^2 + t^3) 
+ 2 F_A F_P m^2 t (-m^2 - {s_u}  + t)  \nonumber \\
&+&  2 F_1^V F_2^V (4 m^2 M^2 (m^2 - t) + {s_u} ^2 t + {s_u}  (4 m^2 M^2 + (m^2 - 4 M^2) t - t^2))  \nonumber \\
&+& 2 F_2^V F_A t (4 m^2 M^2 + {s_u}  (8 M^2 - t) - (m^2 + 4 M^2) t + t^2)  x   \nonumber \\
&+& 8 F_1^V F_A M^2 (4 m^2 M^2 - (m^2 + 4 M^2) t + {s_u}  t + t^2)  x +
F_P^2 m^2 t (-m^2 + t)
\left.\right]  
\\
 \mathcal{C}^{TL}_{NN'(2)} &=&\frac{\sin(\beta)}{M^2\mathcal{I}}
\left[\right. 
  {F_3^A}^2 t (4 m^2 M^2 - {s_u} ^2 + (-m^2 - 4 M^2) t + t^2) 
\left.\right] 
\\
\mathcal{C}^{TL}_{NN'(12)}  &=&\frac{\sin(\beta)}{M^2\mathcal{I}}
\left[\right. 
- F_3^A F_P m^2 {s_u}  t 
+  F_3^A F_A {s_u}  t (-m^2 - 4 M^2 - {s_u}  + t)  
+  (F_1^V + F_2^V) F_3^A t (-4 m^2 M^2 + (m^2 + 4 M^2) t + {s_u}  t - t^2)  x  
\left.\right] \nonumber\\
\end{eqnarray}
\begin{eqnarray}
  \mathcal{C}^{LT}_{NN'(1)} &=&\frac{\sin(\beta)}{4M^2\mathcal{I}}
\left[\right. 
4 {F_1^V}^2 M^2 (m^2 (m^2 - 4 M^2) + {s_u} ^2 + 2 {s_u}  (m^2 - t) + 2 (-m^2 + 2 M^2) t + t^2)  \nonumber \\
&+& 
F_P^2 m^2 (m^2 - t) t 
+ 2 F_A F_P m^2 (m^2 + {s_u}  - t) t 
+ 4 F_A^2 M^2 (m^2 (m^2 - 4 M^2) + 2 m^2 {s_u}  + {s_u} ^2 + 4 M^2 t - t^2) \nonumber \\
&+& {F_2^V}^2 (4 m^4 M^2 - 8 m^2 M^2 t + {s_u} ^2 t + 2 {s_u}  (m^2 - t) t + (-m^2 + 4 M^2) t^2 + t^3)  \nonumber \\
&+& 2 F_1^V F_2^V ({s_u} ^2 t + {s_u}  (4 m^2 M^2 + (m^2 - 4 M^2) t - t^2) + 4 M^2 (m^4 - 3 m^2 t + 2 t^2))  \nonumber \\
&+& 8 F_1^V F_A M^2 (4 m^2 M^2 + (-m^2 - 4 M^2) t - {s_u}  t + t^2)  x  
+ 2 F_2^V F_A t (4 m^2 M^2 + (-m^2 - 4 M^2) t - {s_u}  t + t^2)  x
\left.\right] \\
  \mathcal{C}^{LT}_{NN'(2)} &=&\frac{ \sin(\beta)}{M^2\mathcal{I}}
\left[\right.
 {F_3^A}^2 t (-4 m^2 M^2 + {s_u} ^2 + (m^2 + 4 M^2) t - t^2) 
\left.\right] 
\\
\mathcal{C}^{LT}_{NN'(12)} &=&\frac{ \sin(\beta)}{M^2\mathcal{I}}
\left[\right. 
 F_3^A F_P m^2 {s_u}  t 
+  F_3^A F_A {s_u}  (m^2 + 4 M^2 + {s_u}  - t) t 
+ (F_1^V +F_2^V) F_3^A t (4 m^2 M^2 - (m^2 + 4 M^2) t - {s_u}  t + t^2)  x
\left.\right] 
\end{eqnarray}
\end{footnotesize}

\subsubsection{Target-lepton-recoil asymmetry}
\begin{footnotesize}
\begin{eqnarray}
\mathcal{D}^{LLL}_{lNN'(1)} &=& \frac{1}{64 |\textbf{k}'| |\textbf{p}'| M^4 ( s-M^2)\mathcal{I}}\left[\right.
 {F_2^V}^2   (16 m^8 M^4 + 4 m^6 M^2 (-m^2 - 32 M^2) t + 4 m^4 M^2 (9 m^2 + 56 M^2) t^2 + {s_u} ^4 t^2 \nonumber \\
&&+ m^2 (-m^4 - 68 m^2 M^2 - 128 M^4) t^3 + (3 m^4 + 44 m^2 M^2 + 16 M^4) t^4 + (-3 m^2 - 8 M^2) t^5 + t^6\nonumber \\
&&+ 2 {s_u} ^3 t (2 m^2 M^2 + (m^2 - 2 M^2) t) + {s_u} ^2 t (8 m^4 M^2 + m^2 (m^2 - 16 M^2) t + (3 m^2 + 8 M^2) t^2 - 2 t^3) \nonumber \\
&&+ 2 {s_u}  (8 m^6 M^4 - 40 m^4 M^4 t + 6 m^2 M^2 (m^2 + 4 M^2) t^2 + (m^4 - 8 m^2 M^2 + 8 M^4) t^3 + (-m^2 + 2 M^2) t^4))  x\nonumber \\
&+& 8 F_2^V F_A M^2 (32 m^6 M^4 - 4 m^4 M^2 (7 m^2 + 16 M^2) t + m^2 (5 m^4 + 60 m^2 M^2 + 32 M^4) t^2 \nonumber \\
&&- {s_u} ^3 t^2 + m^2 (-11 m^2 - 36 M^2) t^3 + (7 m^2 + 4 M^2) t^4 - t^5 + {s_u} ^2 t^2 (-m^2 - 8 M^2 + t) \nonumber \\
&&+ {s_u}  t (-20 m^4 M^2 + m^2 (5 m^2 + 24 M^2) t + 2 (-3 m^2 - 2 M^2) t^2 + t^3)) \nonumber \\
&+& 
16 (F_1^V + F_2^V) F_P m^2 M^2 t (-4 m^4 M^2 + m^2 (m^2 + 8 M^2) t - {s_u} ^2 t + 2 (-m^2 - 2 M^2) t^2 + t^3) \nonumber \\
&+&
16 F_1^V F_A M^2 (2 M^2 {s_u} ^2 (-m^2 - t) t + {s_u} ^3 (2 M^2 - t) t + {s_u}  (8 m^4 M^4 + 2 m^2 M^2 (-7 m^2 - 8 M^2) t\nonumber \\
&&+ (3 m^4 + 20 m^2 M^2 + 8 M^4) t^2 + 2 (-2 m^2 - 3 M^2) t^3 + t^4) + 2 (4 m^6 M^4 + m^4 M^2 (-5 m^2 - 4 M^2) t \nonumber \\
&&+ m^2 (m^4 + 9 m^2 M^2 - 4 M^4) t^2 + (-2 m^4 - 3 m^2 M^2 + 4 M^4) t^3 + (m^2 - M^2) t^4)) \nonumber \\
&+& 8 F_A F_P m^2 M^2 {s_u}  t (m^2 (m^2 - 8 M^2) + 2 m^2 {s_u}  + {s_u} ^2 + 8 M^2 t - t^2)  x \nonumber \\
&+& F_P^2 m^2 M^4 t (-4 m^6 M^2 + m^4 (m^2 + 12 M^2) t + 3 m^2 (-m^2 - 4 M^2) t^2 + (3 m^2 + 4 M^2) t^3 - t^4 + {s_u} ^2 t (m^2 + t) \nonumber \\
&&+ 2 {s_u}  (-2 m^4 M^2 + m^4 t + (-m^2 + 2 M^2) t^2))  x  \nonumber \\
&+& 4 F_A^2 M^2 (4 m^6 M^2 (m^2 + 4 M^2) + m^4 (-m^4 - 20 m^2 M^2 - 48 M^4) t + {s_u} ^4 t + 4 m^2 (m^4 + 9 m^2 M^2 + 12 M^4) t^2 \nonumber \\
&&+ 2 (-3 m^4 - 14 m^2 M^2 - 8 M^4) t^3  +4 (m^2 + 2 M^2) t^4 - t^5 + 2 {s_u} ^3 (2 m^2 M^2 + (m^2 + 2 M^2) t) \nonumber \\
&&+ 4 m^2 {s_u} ^2 (3 m^2 M^2 - 5 M^2 t + t^2) + 2 {s_u}  (2 m^4 M^2 (3 m^2 + 4 M^2) + m^4 (-m^2 - 22 M^2) t + 2 (2 m^4 + 7 m^2 M^2 - 4 M^4) t^2 \nonumber \\
&&+ (-3 m^2 + 2 M^2) t^3))  x  \nonumber \\
&+& 4 {F_1^V}^2 M^2 (4 m^6 M^2 (m^2 - 12 M^2) + m^4 (-m^4 - 4 m^2 M^2 + 80 M^4) t + {s_u} ^4 t  +4 m^2 (m^4 + m^2 M^2 - 4 M^4) t^2 \nonumber \\
&&+ 2 (-3 m^4 - 6 m^2 M^2 - 8 M^4) t^3 + 4 (m^2 + 2 M^2) t^4 - t^5 + 2 {s_u} ^3 (2 m^2 M^2 + (m^2 - 2 M^2) t) \nonumber \\
&&+ 4 m^2 {s_u} ^2 (3 m^2 M^2 - 5 M^2 t + t^2)+ 2 {s_u}  (6 m^4 M^2 (m^2 - 4 M^2) + m^2 (-m^4 - 10 m^2 M^2 + 32 M^4) t \nonumber \\
&&+ 2 (2 m^4 - m^2 M^2 - 4 M^4) t^2 + 3 (-m^2 + 2 M^2) t^3))  x  \nonumber \\
&+& 8 F_1^V F_2^V M^2 ({s_u} ^4 t + {s_u} ^3 (3 m^2 - t) t + {s_u} ^2 (4 m^4 M^2 + 2 m^2 (m^2 - 8 M^2) t + (3 m^2 + 4 M^2) t^2 - t^3)\nonumber \\
&&+ {s_u}  (8 m^6 M^2 + m^4 (-m^2 - 36 M^2) t + 3 m^2 (3 m^2 + 8 M^2) t^2 + (-9 m^2 + 4 M^2) t^3 + t^4) \nonumber \\
&&+ m^2 (4 m^4 M^2 (m^2 - 8 M^2) + m^2 (-m^4 - 12 m^2 M^2 + 64 M^4) t + (5 m^4 + 12 m^2 M^2 - 32 M^4) t^2 \nonumber \\
&&+ (-7 m^2 - 4 M^2) t^3 + 3 t^4))  x  \left. \right]
\end{eqnarray}
\begin{eqnarray}
\mathcal{D}^{LLL}_{lNN'(2)}  &=& \frac{1}{16 |\textbf{k}'| |\textbf{p}'| M^4 ( s-M^2)\mathcal{I}}\left[\right.
{F_3^A}^2   t (-16 m^6 M^4 + 8 m^4 M^2 (m^2 + 6 M^2) t + {s_u} ^4 t \nonumber \\
&&+ m^2 (-m^4 - 24 m^2 M^2 - 48 M^4) t^2 + (3 m^4 + 24 m^2 M^2 + 16 M^4) t^3 + (-3 m^2 - 8 M^2) t^4 + t^5\nonumber \\
&&+ 2 {s_u} ^3 (-2 m^2 M^2 + (m^2 + 2 M^2) t) + {s_u} ^2 (-4 m^4 M^2 + m^2 (m^2 - 12 M^2) t + (3 m^2 + 8 M^2) t^2 - 2 t^3) \nonumber \\
&&+ 2 {s_u}  (24 m^4 M^4 + 2 m^2 M^2 (-5 m^2 - 16 M^2) t + (m^4 + 12 m^2 M^2 + 8 M^4) t^2 + (-m^2 - 2 M^2) t^3))  x 
\left. \right]
\\
\mathcal{D}^{LLL}_{lNN'(12)} &= & \frac{1}{16 |\textbf{k}'| |\textbf{p}'| M^4 ( s-M^2)\mathcal{I}}\left[\right.
 4 (F_1^V + F_2^V) F_3^A M^2 t (4 m^6 M^2 + m^4 (-m^2 - 12 M^2) t - {s_u} ^3 t + {s_u} ^2 (m^2 - t) t \nonumber \\
&&+ 3 m^2 (m^2 + 4 M^2) t^2 - (3 m^2 + 4 M^2) t^3 + t^4 + {s_u}  (-4 m^4 M^2 + m^2 (m^2 + 8 M^2) t - 2 (m^2 + 2 M^2) t^2 + t^3)) \nonumber \\
&+& 4 F_3^A F_A M^2 {s_u}  t (-4 m^4 M^2 + {s_u} ^3 + m^4 t + 4 M^2 t^2 - t^3 + {s_u} ^2 (2 m^2 + t) + {s_u}  (m^2 (m^2 - 12 M^2) \nonumber \\
&&+ 2 (m^2 + 2 M^2) t - t^2))  x  \nonumber \\
&+&  F_3^A F_P m^2   t ({s_u} ^3 t + 2 {s_u} ^2 (-2 m^2 M^2 + (m^2 + 2 M^2) t - t^2) + {s_u}  (-4 m^4 M^2 + m^4 t + 4 M^2 t^2 - t^3)\nonumber \\
&&+ 2 (16 m^4 M^4 + 8 m^2 M^2 (-m^2 - 4 M^2) t + (m^4 + 16 m^2 M^2 + 16 M^4) t^2 + 2 (-m^2 - 4 M^2) t^3 + t^4))  x   \left. \right]
\end{eqnarray}
\begin{eqnarray}
\mathcal{D}^{TTT}_{lNN'(1)} &=& \frac{m \sin(\beta)}{ 4 |\textbf{k}'|  M^4\mathcal{I}}\left[\right.
4 F_1^V F_A M^4 (m^2 (m^2 - 8 M^2) + 2 m^2 {s_u}  + {s_u} ^2 + 8 M^2 t - t^2) \nonumber \\
&+& (F_1^V + F_2^V) F_P M^2 t (m^4 - 2 m^2 t + t^2 + {s_u}  (m^2 + t)) \nonumber \\
&+& F_2^V F_A M^2 ({s_u} ^2 t + {s_u}  (4 m^2 M^2 + (m^2 + 4 M^2) t - t^2) + 2 (2 m^4 M^2 - 8 m^2 M^2 t + (m^2 + 6 M^2) t^2 - t^3)) \nonumber \\
&+& F_A F_P M^2 {s_u}  (-m^2 - {s_u}  - t) t  x  + \frac{F_P^2}{4} m^2 t (4 m^2 M^2 + (-m^2 - 4 M^2) t - {s_u}  t + t^2)  x   \nonumber \\
&+& {F_1^V}^2 M^2 (16 m^2 M^4 + (-m^4 + 4 m^2 M^2 - 16 M^4) t + 2 (-m^2 + 2 M^2) {s_u}  t - {s_u} ^2 t - 4 M^2 t^2 + t^3)  x  \nonumber \\
&+& F_A^2 M^2 (4 m^2 M^2 (-m^2 + 4 M^2) + (m^4 - 4 m^2 M^2 - 16 M^4) t + 8 M^2 t^2 - t^3 + {s_u} ^2 (-4 M^2 + t) \nonumber \\
&&+ 2 {s_u}  (-4 m^2 M^2 + (m^2 - 2 M^2) t))  x  \nonumber \\
&+& \frac{{F_2^V}^2}{4} (16 m^4 M^4 - 8 m^4 M^2 t + 4 M^2 (3 m^2 - 4 M^2) t^2 - {s_u} ^2 t^2 + (-m^2 - 4 M^2) t^3 + t^4  + {s_u}  t (-4 m^2 M^2 + (-m^2 + 4 M^2) t) )  x   \nonumber \\
&+& F_1^V F_2^V M^2 (4 m^4 M^2 - m^2 (3 m^2 - 16 M^2) t - 2 {s_u} ^2 t - 20 M^2 t^2 + 3 t^3 + {s_u}  (4 m^2 M^2 - (5 m^2 - 4 M^2) t + t^2))  x \left. \right] 
\end{eqnarray}
\begin{eqnarray}
\mathcal{D}^{TTT}_{lNN'(12)} &=& \frac{ m \sin(\beta)}{ 2 |\textbf{k}'|  M^4 \mathcal{I}}
\left[\right.
 (F_1^V + F_2^V ) F_3^A M^2 t ({s_u} ^2 + {s_u}  (m^2 - t) + 2 (-4 m^2 M^2 + (m^2 + 4 M^2) t - t^2)) \nonumber \\
&+&  F_3^A F_A M^2 {s_u}  t (m^2 - 8 M^2 + {s_u}  + t)  x  \nonumber \\
&+& \frac{F_3^A F_P}{4}  t (-4 m^4 M^2 + 4 M^2 {s_u}  (m^2 - t) + m^2 (m^2 + 8 M^2) t - {s_u} ^2 t + 2 (-m^2 - 2 M^2) t^2 + t^3)  x 
\left. \right]
\end{eqnarray}
\begin{eqnarray}
\mathcal{D}^{LNN}_{lNN'(1)} &=& \frac{1}{16M^3 |\textbf{k}'|\mathcal{I}} 
\left[ \right.
8 F_1^V F_2^V m^2 M^2 (-m^4 + {s_u}  (-m^2 - t) + 2 m^2 t - t^2)  x  \nonumber \\
&+& 8 F_A F_P m^2 M^2 (m^4 - 2 m^2 t + t^2 + {s_u}  (m^2 + t))  x  
+ F_P^2 m^2 t (m^4 - 2 m^2 t + t^2 + {s_u}  (m^2 + t))  x   \nonumber \\
&+& 4 {F_1^V}^2 M^2 (m^4 (-m^2 + 4 M^2) + {s_u} ^3 + m^2 (m^2 - 8 M^2) t + (m^2 + 4 M^2) t^2 - t^3 + {s_u} ^2 (m^2 + t) \nonumber \\
&&+ {s_u}  (m^2 (-m^2 - 12 M^2) + 2 (m^2 + 2 M^2) t - t^2))  x  \nonumber \\
&+& 4 F_A^2 M^2 (m^4 (m^2 - 4 M^2) - {s_u} ^3 + {s_u} ^2 (-m^2 - t) + m^2 (-m^2 + 8 M^2) t + (-m^2 - 4 M^2) t^2 + t^3\nonumber \\
&&+ {s_u}  (m^2 (m^2 + 12 M^2) + 2 (-m^2 - 2 M^2) t + t^2))  x  \nonumber \\
&+& {F_2^V}^2 (-4 m^6 M^2 + 4 m^4 M^2 t - {s_u} ^3 t + {s_u} ^2 (-m^2 - t) t + m^2 (m^2 + 4 M^2) t^2 + 2 (-m^2 - 2 M^2) t^3 + t^4\nonumber \\
&&+ {s_u}  (-4 m^4 M^2 + 8 m^2 M^2 t + (-3 m^2 - 4 M^2) t^2 + t^3))  x 
 \left. \right]
\\
\mathcal{D}^{LNN}_{lNN'(2)} &=&\frac{1}{4M^3 |\textbf{k}'|\mathcal{I}} 
  \left[\right.  
 {F_3^A}^2 t (4 m^4 M^2 + {s_u} ^3 + m^2 (-m^2 - 8 M^2) t + 2 (m^2 + 2 M^2) t^2 - t^3 \nonumber \\
&&+ {s_u} ^2 (m^2 + t) + {s_u}  (-12 m^2 M^2 + (3 m^2 + 4 M^2) t - t^2))  x 
\left.\right]
\\
\mathcal{D}^{LNN}_{lNN'(12)}&=&\frac{1}{4M^3 |\textbf{k}'|\mathcal{I}} 
  \left[\right. 
-x m^2 (8 M^2 t + ({s_u}  - 2 t) ({s_u}  + t) + m^2 (-8 M^2 + {s_u}  + 2 t))
( - F_3^A (4 F_A M^2 + F_P t))
  \left.\right]
\end{eqnarray}
\begin{eqnarray}
\mathcal{D}^{TNN}_{lNN'(1)} &=&\frac{   m  |\textbf{p}'|   (s-M^2 )\sin(\beta)}{4M^3 |\textbf{k}'|\mathcal{I}} 
\left[ \right.
8 F_1^V F_2^V m^2 M^2  x  - 8 F_A F_P m^2 M^2  x  
+ 4 F_A^2 M^2 (-m^2 + 4 M^2 - {s_u}  - t)  x \nonumber \\
&-& F_P^2 m^2 t  x  
+ 4 {F_1^V}^2 M^2 (m^2 - 4 M^2 + {s_u}  + t)  x  + {F_2^V}^2 (4 m^2 M^2 + 4 M^2 t - {s_u}  t - t^2)  x \left. \right]
\\
\mathcal{D}^{TNN}_{lNN'(2)} &=&\frac{   m  |\textbf{p}'|   (s-M^2 )\sin(\beta)}{M^3 |\textbf{k}'|\mathcal{I}} 
\left[ \right.
 4 {F_3^A}^2 t (-4 M^2 + {s_u}  + t)  x 
\left. \right]
\\
\mathcal{D}^{TNN}_{lNN'(12)}&=&\frac{   m  |\textbf{p}'|   (s-M^2 )\sin(\beta)}{ 2M^3 |\textbf{k}'|\mathcal{I}} 
\left[ \right.
-x(m^2 - {s_u}  - t) ( - F_3^A (4 F_A M^2 + F_P t)) 
  \left.\right]
\end{eqnarray}
\end{footnotesize}
\begin{footnotesize}
\begin{eqnarray}
\mathcal{D}^{NLN}_{lNN'(1)} &=& \frac{ m}{2M(s-M^2)\mathcal{I}}
  \left[\right.
F_2^V F_P t (m^4 + {s_u}  (m^2 - t) - 2 m^2 t + t^2) 
+ 4 {F_1^V}^2 M^2 (m^4 + {s_u}  (m^2 - t) - 2 m^2 t + t^2)  x  \nonumber \\
&+& 4 F_1^V F_A M^2 (m^4 + {s_u} ^2 + 2 {s_u}  (m^2 - t) - 2 m^2 t + t^2) 
+ F_1^V F_P ({s_u} ^2 t + {s_u}  (m^2 - t) t + 4 M^2 (m^4 - 2 m^2 t + t^2)) \nonumber \\
&+& F_2^V F_A ({s_u}  (4 m^2 M^2 + (m^2 - 4 M^2) t - t^2) + t (m^4 - 2 m^2 t + t^2)) 
+ 4 F_A^2 M^2 {s_u}  (m^2 + {s_u}  - t)  x \nonumber \\
&+& F_A F_P (-4 m^4 M^2 + m^2 (m^2 + 8 M^2) t + {s_u}  (m^2 - t) t + 2 (-m^2 - 2 M^2) t^2 + t^3)  x  \nonumber \\
&+& {F_2^V}^2 ({s_u} ^2 t + {s_u}  (m^2 - t) t + 4 M^2 (m^4 - 2 m^2 t + t^2))  x  \nonumber \\
&+&F_1^V F_2^V ({s_u} ^2 t + {s_u}  (4 m^2 M^2 + (m^2 - 4 M^2) t - t^2) + 8 M^2 (m^4 - 2 m^2 t + t^2))  x 
  \left.\right]
\\
\mathcal{D}^{NLN}_{lNN'(12)} &=& \frac{ m}{M(s-M^2)\mathcal{I}}
  \left[\right. 
 x F_3^A F_A {s_u}  (-4 m^2 M^2 + (m^2 + 4 M^2) t + {s_u}  t - t^2) \nonumber \\
&+&  (F_1^V + F_2^V) F_3^A  (-4 m^4 M^2 + m^2 (m^2 + 8 M^2) t + {s_u}  (m^2 - t) t + 2 (-m^2 - 2 M^2) t^2 + t^3)  
\left. \right]
\end{eqnarray}
\begin{eqnarray}
\mathcal{D}^{NTN}_{lNN'(1)} &=& \frac{2 m |\textbf{p}'|\sin(\beta) }{M^2\mathcal{I}} 
   \left[\right. 
F_2^V F_A M^2{s_u}  
+ F_1^V F_P M^2 (m^2 - t) + \frac{F_2^V F_P }{4}(m^2 + {s_u}  - t) t   \nonumber \\
&+& F_1^V F_2^V M^2({s_u}  + 2 (m^2 - t))  x  
+ {F_2^V}^2 M^2 (m^2 - t)  x  
+ {F_1^V}^2 M^2 (m^2 + {s_u}  - t)  x  \nonumber \\
&+& F_A F_P M^2 (-m^2 + t)  x  
+ F_A^2 M^2 (-m^2 - {s_u}  + t)  x 
   \left.\right] 
\end{eqnarray}
\begin{eqnarray}
\mathcal{D}^{NTN}_{lNN'(12)} &=& \frac{4 m |\textbf{p}'|\sin(\beta) }{\mathcal{I}} 
\left[\right.
- ( x  F_3^A)F_A {s_u}  
+(F_1^V +F_2^V) F_3^A (-m^2 + t) 
 \left.\right]
\end{eqnarray}
\begin{eqnarray}
\mathcal{D}^{NNL}_{lNN'(1)} &=&
  \frac{m }{2M^2|\textbf{p}'|\mathcal{I} }
  \left[ \right.
-  (F_1^V +  F_2^V) F_P   {s_u}  t^2 
+    F_2^V F_A   t (-4 m^2 M^2 - 4 M^2 {s_u}  + (m^2 + 4 M^2) t - t^2)  \nonumber \\
&+& 4 F_1^V F_A M^2 (-4 m^2 M^2 + (m^2 + 4 M^2) t - {s_u}  t - t^2) 
+ 4 {F_1^V}^2 M^2 {s_u}  t  x  
+   {F_2^V}^2   {s_u}  t^2  x   \nonumber \\
&+&   F_1^V F_2^V   {s_u}  t (4 M^2 + t)  x  
+ F_A (4 M^2 F_A  + t F_P   )(4 m^2 M^2 - (m^2 + 4 M^2) t + t^2)  x   
\left. \right]
\\
\mathcal{D}^{NNL}_{lNN'(12)} &=&
  \frac{ m }{M^2|\textbf{p}'| \mathcal{I}}\left[ \right.
  - x F_3^A   F_A M^2 {s_u}  t (-4 M^2 + t) 
+  (F_1^V + F_2^V) F_3^A  M^2 t (4 m^2 M^2 + (-m^2 - 4 M^2) t + t^2) 
  \left.\right] 
\end{eqnarray}
\begin{eqnarray}
\mathcal{D}^{NNT}_{lNN'(1)}  &=& \frac{ m  (s-M^2) \sin(\beta)}{2M^3  \mathcal{I}}
  \left[ \right. 
  -16 F_1^V F_A M^4 
- 4 F_2^V F_A M^2 t 
- 4 F_1^V F_P t M^2 -  F_2^V F_P t^2   
+ 4 F_A^2 M^2  (4 M^2 - t)  x  \nonumber \\
&+& 4 ({F_1^V}^2  + {F_2^V}^2) t M^2  x 
+8 F_1^V F_2^V t M^2  x  
  \left.\right]
\end{eqnarray}
\begin{eqnarray}
\mathcal{D}^{TLL}_{lNN'(1)} &=& 
\frac{m \sin(\beta)}{16 |\textbf{k}'|M^4 \mathcal{I}} 
 \left[ \right.
4 F_1^V F_P M^2 t (m^4 - 2 m^2 t + t^2 + {s_u}  (m^2 + t)) 
+ 4 F_2^V F_P M^2 t (m^4 - 2 m^2 t + t^2 + {s_u}  (m^2 + t)) \nonumber \\
&+& 8 F_1^V F_A M^2 (2 m^2 M^2 (-m^2 + 8 M^2) + (m^4 - 16 M^4) t + 2 (-m^2 + M^2) t^2 + t^3 + {s_u} ^2 (-2 M^2 + t)\nonumber \\
&&+ 2 {s_u}  (-2 m^2 M^2 + (m^2 + 4 M^2) t - t^2)) \nonumber \\
&+& 4 F_2^V F_A M^2 ({s_u} ^2 t + {s_u}  (-4 m^2 M^2 + 3 (m^2 + 4 M^2) t - 3 t^2) + 2 (-2 m^4 M^2 + m^2 (m^2 + 8 M^2) t + 3 (-m^2 - 2 M^2) t^2 + 2 t^3)) \nonumber \\
&+& 4 F_A F_P M^2 {s_u}  (-m^2 - {s_u}  - t) t  x  
+F_P^2 m^2 t (4 m^2 M^2 + (-m^2 - 4 M^2) t - {s_u}  t + t^2)  x  \nonumber \\
&+& 4 {F_1^V}^2 M^2 (-16 m^2 M^4 + (m^4 + 4 m^2 M^2 + 16 M^4) t + {s_u} ^2 t + 2 {s_u}  (m^2 - 2 M^2 - t) t - 2 (m^2 + 2 M^2) t^2 + t^3)  x  \nonumber \\
&+& {F_2^V}^2 (-16 m^4 M^4 + 8 m^2 M^2 (m^2 + 4 M^2) t + 4 M^2 (-5 m^2 - 4 M^2) t^2 + {s_u} ^2 t^2 + (m^2 + 12 M^2) t^3 - t^4 \nonumber \\
&&+ {s_u}  t (4 m^2 M^2 + (m^2 - 12 M^2) t))  x \nonumber \\
&+& 4 F_1^V F_2^V M^2 (-4 m^4 M^2 + 3 m^4 t + 2 {s_u} ^2 t + 4 (-m^2 + M^2) t^2 + t^3 + {s_u}  (-4 m^2 M^2 + (5 m^2 - 4 M^2) t - 5 t^2))  x  \nonumber \\
&+& 4 F_A^2 M^2 (4 m^2 M^2 (-m^2 - 4 M^2) + (m^4 + 12 m^2 M^2 + 16 M^4) t + 2 (-m^2 - 4 M^2) t^2 + t^3 + {s_u} ^2 (-4 M^2 + t)\nonumber \\
&&+ 2 {s_u}  (-4 m^2 M^2 + (m^2 + 2 M^2) t - t^2))  x
  \left.\right]
\\
\mathcal{D}^{TLL}_{lNN'(2)}&=&
\frac{m \sin(\beta)}{4|\textbf{k}'|M^4 \mathcal{I}} 
 \left[ \right.
 {F_3^A}^2 t (16 m^2 M^4 + 8 M^2 (-m^2 - 2 M^2) t + {s_u} ^2 t + (m^2 + 8 M^2) t^2 - t^3 + m^2 {s_u}  (-4 M^2 + t))  x 
\left.\right]\\
\mathcal{D}^{TLL}_{lNN'(12)} &=& 
\frac{m \sin(\beta)}{8 |\textbf{k}'|M^4 \mathcal{I}} 
 \left[ \right.
4 (F_1^V + F_2^V) F_3^A M^2 t ({s_u} ^2 + {s_u}  (m^2 - t) + 2 (-4 m^2 M^2 + (m^2 + 4 M^2) t - t^2)) \nonumber \\
&+& 2 F_3^A F_P t (-4 m^4 M^2 + 4 M^2 {s_u}  (m^2 - t) + m^2 (m^2 + 8 M^2) t - {s_u} ^2 t + 2 (-m^2 - 2 M^2) t^2 + t^3)  x  \nonumber \\
&+& 4 F_3^A F_A M^2 {s_u}  t (m^2 - 8 M^2 + {s_u}  + t)  x
\left.\right]
\end{eqnarray}
\begin{eqnarray}
\mathcal{D}^{LTL}_{lNN'(1)} &=&  \frac{  \sin(\beta)}{16 M^3|\textbf{k}'|\mathcal{I}}
\left[ \right.
4 (F_1^V + F_2^V) F_P m^2  t (-4 m^2 M^2 + (m^2 + 4 M^2) t + {s_u}  t - t^2) \nonumber \\
&+& 2 F_2^V F_A  (32 m^4 M^4 + 4 m^2 M^2 (-3 m^2 - 8 M^2) t + {s_u} ^2 (8 M^2 - t) t + m^2 (m^2 + 16 M^2) t^2 \nonumber \\
&&+ 2 (-m^2 - 2 M^2) t^3 + t^4 + 4 M^2 {s_u}  t (m^2 + t))
 \nonumber \\
&+& 8 F_1^V F_A M^2 (4 m^4 M^2 - m^4 t + {s_u} ^2 t - 4 M^2 t^2 + t^3 + 2 {s_u}  (2 m^2 M^2 - 2 M^2 t + t^2)) \nonumber \\
&+& F_P^2 m^2   t (-m^4 - {s_u}  (m^2 + t) + 2 m^2 t - t^2)  x  + 2 F_A F_P m^2  t (-m^4 - {s_u} ^2 + 2 m^2 t - t^2 - 2 {s_u}  (m^2 + 4 M^2 - t))  x \nonumber \\
&+& 4 {F_1^V}^2 M^2 (m^4 (m^2 - 12 M^2) + {s_u} ^3 + {s_u} ^2 (3 m^2 - t) + m^2 (m^2 + 8 M^2) t + (-m^2 + 4 M^2) t^2 - t^3 \nonumber \\
&&+ {s_u}  (3 m^2 (m^2 - 4 M^2) + 4 M^2 t - 3 t^2))  x  \nonumber \\
&+& 4 F_A^2 M^2 (m^4 (-m^2 + 4 M^2) - {s_u} ^3 + {s_u} ^2 (-3 m^2 - t) + m^2 (m^2 - 8 M^2) t + (m^2 + 4 M^2) t^2 - t^3 \nonumber \\
&&+ {s_u}  (m^2 (-3 m^2 + 4 M^2) + 4 M^2 t - t^2))  x  \nonumber \\
&+& {F_2^V}^2 (4 m^6 M^2 - 28 m^4 M^2 t + {s_u} ^3 t + {s_u} ^2 (3 m^2 - t) t + m^2 (5 m^2 + 28 M^2) t^2 + 2 (-3 m^2 - 2 M^2) t^3 + t^4\nonumber \\
&& + {s_u}  (4 m^4 M^2 + 2 m^2 (m^2 - 8 M^2) t + (m^2 - 4 M^2) t^2 - t^3))  x  \nonumber \\
&+& 2 F_1^V F_2^V  ({s_u} ^3 t + 2 {s_u} ^2 (2 m^2 M^2 + (m^2 - 2 M^2) t) + {s_u}  (8 m^4 M^2 + m^2 (m^2 - 20 M^2) t + 2 (m^2 - 2 M^2) t^2 - t^3) \nonumber \\
&&+ 2 m^2 (2 m^2 M^2 (m^2 - 8 M^2) + 4 M^2 (-m^2 + 4 M^2) t + (m^2 + 2 M^2) t^2 - t^3))  x 
  \left.\right]
\end{eqnarray}
\begin{eqnarray}
\mathcal{D}^{LTL}_{lNN'(2)} &=&  \frac{  \sin(\beta)}{4 M^3|\textbf{k}'|\mathcal{I}}
   \left[ \right.
   {F_3^A}^2 t (-4 m^4 M^2 - {s_u} ^3 + {s_u} ^2 (-m^2 - t) + m^2 (m^2 + 8 M^2) t + 2 (-m^2 - 2 M^2) t^2 + t^3 \nonumber \\
&&+ {s_u}  (12 m^2 M^2 + (-3 m^2 - 4 M^2) t + t^2))  x 
   \left.\right]
\end{eqnarray}
\begin{eqnarray}
\mathcal{D}^{LTL}_{lNN'(12)}  &=&  \frac{  \sin(\beta)}{4 M^3|\textbf{k}'|\mathcal{I}}
\left[ \right.
(F_1^V + F_2^V )F_3^A   t (4 m^4 M^2 + m^2 (-m^2 - 8 M^2) t + {s_u} ^2 t + 2 (m^2 + 2 M^2) t^2 - t^3 + 4 M^2 {s_u}  (-m^2 + t)) \nonumber \\ 
&+&   F_3^A F_P m^2   t (-{s_u} ^2 + {s_u}  (-m^2 + t) + 2 (4 m^2 M^2 + (-m^2 - 4 M^2) t + t^2))  x \nonumber \\
&+&   F_3^A F_A   t (2 (-m^2 - 2 M^2) {s_u} ^2 - {s_u} ^3 + 2 m^2 (4 m^2 M^2 + (-m^2 - 4 M^2) t + t^2) \nonumber \\
&&+ {s_u}  (m^2 (-m^2 + 12 M^2) + 2 (-m^2 - 2 M^2) t + t^2))  x
\left.\right]
\end{eqnarray}
\begin{eqnarray}
\mathcal{D}^{LLT}_{lNN'(1)} &=& \frac{\sin(\beta)}{ 16 |\textbf{k}'|M^3\mathcal{I}}
\left[ \right.   
4 F_1^V F_P m^2 t (4 m^2 M^2 + (-m^2 - 4 M^2) t - {s_u}  t + t^2)  \nonumber \\
&+& 2 F_2^V F_A  (32 m^4 M^4 + 4 m^2 M^2 (-3 m^2 - 8 M^2) t + 4 M^2 {s_u}  (-m^2 - t) t + m^2 (m^2 + 16 M^2) t^2 \nonumber \\
&&- {s_u} ^2 t^2 + 2 (-m^2 - 2 M^2) t^3 + t^4) 
+ 4 F_2^V F_P m^2 t (4 m^2 M^2 + (-m^2 - 4 M^2) t - {s_u}  t + t^2) \nonumber \\
&+& 8 F_1^V F_A M^2 (4 m^4 M^2 - m^4 t - {s_u} ^2 t - 4 M^2 t^2 + t^3 + 2 {s_u}  (2 m^2 M^2 + (-m^2 - 2 M^2) t)) \nonumber \\
&+& 2 F_A F_P m^2   t (m^4 + {s_u} ^2 + 2 {s_u}  (m^2 + 4 M^2 - t) - 2 m^2 t + t^2)  x  + F_P^2 m^2  t (m^4 - 2 m^2 t + t^2 + {s_u}  (m^2 + t))  x  \nonumber \\
&+& 4 F_A^2 M^2 (m^4 (m^2 + 4 M^2) + {s_u} ^3 + m^2 (-3 m^2 - 8 M^2) t + (3 m^2 + 4 M^2) t^2 - t^3 + {s_u} ^2 (3 m^2 + t) \nonumber \\
&&+ {s_u}  (m^2 (3 m^2 + 4 M^2) + 2 (-m^2 + 2 M^2) t - t^2))  x \nonumber \\
&+& 4 {F_1^V}^2 M^2 (m^4 (m^2 - 12 M^2) + {s_u} ^3 + {s_u} ^2 (3 m^2 - t) + m^2 (3 m^2 + 8 M^2) t + (-5 m^2 + 4 M^2) t^2 + t^3 \nonumber \\
&&+ {s_u}  (3 m^2 (m^2 - 4 M^2) + 2 (m^2 + 2 M^2) t - t^2))  x  \nonumber \\
&+& {F_2^V}^2  (4 m^6 M^2 - 20 m^4 M^2 t + {s_u} ^3 t + {s_u} ^2 (3 m^2 - t) t + m^2 (5 m^2 + 12 M^2) t^2 + 2 (-3 m^2 + 2 M^2) t^3 + t^4 \nonumber \\
&&+ {s_u}  (4 m^4 M^2 + 2 m^2 (m^2 - 4 M^2) t + (m^2 + 4 M^2) t^2 - t^3))  x  \nonumber \\
&+& 2 F_1^V F_2^V  ({s_u} ^3 t + 2 {s_u} ^2 (2 m^2 M^2 + (m^2 - 2 M^2) t) + {s_u}  (8 m^4 M^2 + m^2 (m^2 - 12 M^2) t + 2 (m^2 + 2 M^2) t^2 - t^3) \nonumber \\
&&+ 2 (2 m^4 M^2 (m^2 - 8 M^2) + 16 m^2 M^4 t + m^2 (m^2 - 6 M^2) t^2 + (-m^2 + 4 M^2) t^3))  x
  \left.\right]
\end{eqnarray}
\begin{eqnarray}
\mathcal{D}^{LLT}_{lNN'(2)}  &=& \frac{\sin(\beta)}{ 4 |\textbf{k}'|M^3\mathcal{I}}
\left[ \right.  
  {F_3^A}^2  t (4 m^4 M^2 + {s_u} ^3 + m^2 (-m^2 - 8 M^2) t + 2 (m^2 + 2 M^2) t^2 - t^3 + {s_u} ^2 (m^2 + t) \nonumber \\
&&+ {s_u}  (-12 m^2 M^2 + (3 m^2 + 4 M^2) t - t^2))  x 
  \left.\right]
\\
\mathcal{D}^{LLT}_{lNN'(12)}  &=& \frac{\sin(\beta)}{ 4 |\textbf{k}'|M^3\mathcal{I}}
\left[ \right. 
  F_1^V F_3^A   t (-4 m^4 M^2 + 4 M^2 {s_u}  (m^2 - t) + m^2 (m^2 + 8 M^2) t - {s_u} ^2 t + 2 (-m^2 - 2 M^2) t^2 + t^3) \nonumber \\
&+&   F_2^V F_3^A t (-4 m^4 M^2 + 4 M^2 {s_u}  (m^2 - t) + m^2 (m^2 + 8 M^2) t - {s_u} ^2 t + 2 (-m^2 - 2 M^2) t^2 + t^3) \nonumber \\
&+&   F_3^A F_P m^2   t ({s_u} ^2 + {s_u}  (m^2 - t) + 2 (-4 m^2 M^2 + (m^2 + 4 M^2) t - t^2))  x  \nonumber \\
&+&   F_3^A F_A  t (2 (m^2 + 2 M^2) {s_u} ^2 + {s_u} ^3 + {s_u}  (m^2 (m^2 - 12 M^2) + 2 (m^2 + 2 M^2) t - t^2) \nonumber \\
&&+ 2 m^2 (-4 m^2 M^2 + (m^2 + 4 M^2) t - t^2))  x   
\left.\right]
\end{eqnarray}
\begin{eqnarray}
\mathcal{D}^{LTT}_{lNN'(1)} &=&\frac{1}{ 64 M^4 |\textbf{k}'||\textbf{p}'|   (s-M^2)\mathcal{I}}
\left[ \right.  32 F_1^V F_A M^4 (-4 m^6 M^2 + m^4 (m^2 + 4 M^2) t - {s_u} ^3 t + {s_u} ^2 (-m^2 - t) t \nonumber \\
&&+ m^2 (-m^2 + 4 M^2) t^2 - (m^2 + 4 M^2) t^3 + t^4 + {s_u}  (-4 m^4 M^2 + m^2 (m^2 + 8 M^2) t - 2 (m^2 + 2 M^2) t^2 + t^3)) \nonumber \\
&+& 16 (F_1^V + F_2^V) F_P m^2 M^2 t (-4 m^4 M^2 + m^2 (m^2 + 8 M^2) t - {s_u} ^2 t + 2 (-m^2 - 2 M^2) t^2 + t^3) \nonumber \\
&+& 8 F_2^V F_A M^2 (-32 m^6 M^4 + 4 m^4 M^2 (3 m^2 + 16 M^2) t + m^2 (-m^4 - 28 m^2 M^2 - 32 M^4) t^2 - {s_u} ^3 t^2 + m^2 (3 m^2 + 20 M^2) t^3 \nonumber \\
&&+ (-3 m^2 - 4 M^2) t^4 + t^5 + {s_u} ^2 t (-8 m^2 M^2 + m^2 t - t^2) + {s_u}  t (-4 m^4 M^2 + m^2 (m^2 + 8 M^2) t + 2 (-m^2 - 2 M^2) t^2 + t^3)) \nonumber \\
&+& 8 F_A F_P m^2 M^2 {s_u}  t (m^2 (m^2 - 8 M^2) + 2 m^2 {s_u}  + {s_u} ^2 + 8 M^2 t - t^2)  x  \nonumber \\
&+& F_P^2 m^2 t (-4 m^6 M^2 + m^4 (m^2 + 12 M^2) t + 3 m^2 (-m^2 - 4 M^2) t^2 + (3 m^2 + 4 M^2) t^3 - t^4 + {s_u} ^2 t (m^2 + t) \nonumber \\
&&+ 2 {s_u}  (-2 m^4 M^2 + m^4 t + (-m^2 + 2 M^2) t^2))  x  \nonumber \\
&+& 4 F_A^2 M^2 (4 m^6 M^2 (m^2 - 4 M^2) + m^4 (-m^4 - 4 m^2 M^2 + 48 M^4) t  + {s_u} ^4 t + 2 m^2 (m^4 - 6 m^2 M^2 - 24 M^4) t^2 \nonumber \\
&&+ 4 M^2 (5 m^2 + 4 M^2) t^3 + 2 (-m^2 - 4 M^2) t^4 + t^5 + 2 {s_u} ^3 (2 m^2 M^2 + (m^2 + 2 M^2) t) + 2 {s_u} ^2 (6 m^4 M^2 - 6 m^2 M^2 t - t^3 \nonumber \\
&&+ (m^2 + 4 M^2) t^2 ) + 2 {s_u}  (2 m^4 M^2 (3 m^2 - 4 M^2) + m^4 (-m^2 - 10 M^2) t + 2 (m^4 + 3 m^2 M^2 + 4 M^4) t^2 + (-m^2 - 2 M^2) t^3))  x  \nonumber \\
&+& 4 {F_1^V}^2 M^2 (4 m^6 M^2 (-m^2 + 12 M^2) + m^4 (m^4 - 4 m^2 M^2 - 80 M^4) t - {s_u} ^4 t + 2 m^2 (-m^4 + 10 m^2 M^2 + 8 M^4) t^2 \nonumber \\
&&+ 4 M^2 (-3 m^2 + 4 M^2) t^3 + 2 m^2 t^4 - t^5 + 2 {s_u} ^3 (-2 m^2 M^2 + (-m^2 + 2 M^2) t) + 2 {s_u} ^2 (-6 m^4 M^2 + 10 m^2 M^2 t - m^2 t^2 + t^3) \nonumber \\
&&+ 2 {s_u}  (6 m^4 M^2 (-m^2 + 4 M^2) + m^2 (m^4 + 6 m^2 M^2 - 32 M^4) t + 2 (-m^4 + m^2 M^2 + 4 M^4) t^2 + (m^2 - 2 M^2) t^3))  x \nonumber \\
&+& 8 F_1^V F_2^V M^2 (4 m^6 M^2 (-m^2 + 8 M^2) + m^4 (m^4 + 4 m^2 M^2 - 64 M^4) t - {s_u} ^4 t + m^2 (-3 m^4 + 12 m^2 M^2 + 32 M^4) t^2 \nonumber \\
&&+ m^2 (m^2 - 20 M^2) t^3 + (3 m^2 + 8 M^2) t^4 - 2 t^5 - {s_u} ^3 t (3 m^2 - t) + {s_u} ^2 (-4 m^4 M^2 + 2 m^2 (-m^2 + 8 M^2) t  \nonumber \\
&&+ (-m^2 - 4 M^2) t^2 + 3 t^3) + {s_u}  (-8 m^6 M^2 + m^4 (m^2 + 28 M^2) t + m^2 (-5 m^2 - 24 M^2) t^2 + (5 m^2 + 4 M^2) t^3 - t^4))  x  \nonumber \\
&+& {F_2^V}^2 (-16 m^8 M^4 + 4 m^6 M^2 (m^2 + 24 M^2) t + 4 m^4 M^2 (-7 m^2 - 32 M^2) t^2  \nonumber \\
&&- {s_u} ^4 t^2 + m^2 (m^4 + 44 m^2 M^2 + 32 M^4) t^3 + (-3 m^4 - 20 m^2 M^2 + 16 M^4) t^4 + 3 m^2 t^5 - t^6 \nonumber \\
&&+ 2 {s_u} ^3 t (-2 m^2 M^2 + (-m^2 + 2 M^2) t) + {s_u} ^2 t (-8 m^4 M^2 + m^2 (-m^2 + 24 M^2) t - 3 m^2 t^2 + 2 t^3) \nonumber \\
&&+ 2 {s_u}  (-8 m^6 M^4 + 24 m^4 M^4 t + 2 m^2 M^2 (m^2 - 12 M^2) t^2 + (-m^4 + 8 M^4) t^3 + (m^2 - 2 M^2) t^4))  x
\left.\right]
\end{eqnarray}
\begin{eqnarray}
\mathcal{D}^{LTT}_{lNN'(2)} &=&\frac{1}{16 M^4 |\textbf{k}'||\textbf{p}'|   (s-M^2)\mathcal{I}}
\left[ \right. 
 {F_3^A}^2 t (-16 m^6 M^4 + 8 m^4 M^2 (m^2 + 6 M^2) t \nonumber \\
&&+ {s_u} ^4 t + m^2 (-m^4 - 24 m^2 M^2 - 48 M^4) t^2 + (3 m^4 + 24 m^2 M^2 + 16 M^4) t^3 + (-3 m^2 - 8 M^2) t^4 + t^5 \nonumber \\
&&+ 2 {s_u} ^3 (-2 m^2 M^2 + (m^2 + 2 M^2) t) + {s_u} ^2 (-4 m^4 M^2 + m^2 (m^2 - 12 M^2) t + (3 m^2 + 8 M^2) t^2 - 2 t^3) \nonumber \\
&&+ 2 {s_u}  (24 m^4 M^4 + 2 m^2 M^2 (-5 m^2 - 16 M^2) t + (m^4 + 12 m^2 M^2 + 8 M^4) t^2 + (-m^2 - 2 M^2) t^3))  x  
\left.\right]
\end{eqnarray}
\begin{eqnarray}
\mathcal{D}^{LTT}_{lNN'(12)} &=&\frac{1}{16 M^4 |\textbf{k}'||\textbf{p}'|   (s-M^2)\mathcal{I}}
\left[ \right. 
 4 F_1^V F_3^A M^2 t (4 m^6 M^2 + m^4 (-m^2 - 12 M^2) t - {s_u} ^3 t + {s_u} ^2 (m^2 - t) t  \nonumber \\
 &&+ 3 m^2 (m^2 + 4 M^2) t^2 - (3 m^2 + 4 M^2) t^3 + t^4 + {s_u}  (-4 m^4 M^2 + m^2 (m^2 + 8 M^2) t - 2 (m^2 + 2 M^2) t^2 + t^3)) \nonumber \\
&+& 4 F_2^V F_3^A M^2 t (4 m^6 M^2 + m^4 (-m^2 - 12 M^2) t - {s_u} ^3 t + {s_u}^2 (m^2 - t) t + 3 m^2 (m^2 + 4 M^2) t^2  \nonumber \\
&&+ (-3 m^2 - 4 M^2) t^3 + t^4 + {s_u}  (-4 m^4 M^2 + m^2 (m^2 + 8 M^2) t + 2 (-m^2 - 2 M^2) t^2 + t^3))  \nonumber \\
&+& 4 F_3^A F_A M^2 {s_u}  t (-4 m^4 M^2 + {s_u} ^3 + m^4 t + 4 M^2 t^2 - t^3 + {s_u} ^2 (2 m^2 + t) + {s_u}  (m^2 (m^2 - 12 M^2) + 2 (m^2 + 2 M^2) t - t^2))  x   \nonumber \\
&+&   F_3^A F_P m^2 t ({s_u} ^3 t + 2 {s_u} ^2 (-2 m^2 M^2 + (m^2 + 2 M^2) t - t^2) + {s_u}  (-4 m^4 M^2 + m^4 t + 4 M^2 t^2 - t^3)  \nonumber \\
&&+ 2 (16 m^4 M^4 + 8 m^2 M^2 (-m^2 - 4 M^2) t + (m^4 + 16 m^2 M^2 + 16 M^4) t^2 + 2 (-m^2 - 4 M^2) t^3 + t^4))  x   \nonumber \\
&&+ {s_u}  (16 m^4 M^4 - 32 m^2 M^4 t + (-m^4 + 8 m^2 M^2 + 16 M^4) t^2 - 8 M^2 t^3 + t^4) + 14 m^2 M^2 (m^2 + 4 M^2) t^2   \nonumber \\
&&+ 2 (8 m^6 M^4 + 2 m^4 M^2 (-m^2 - 20 M^2) t
+ (-m^4 - 22 m^2 M^2 - 24 M^4) t^3 + 2 (m^2 + 5 M^2) t^4 - t^5))  x 
\left.\right]
\end{eqnarray}
\begin{eqnarray}
\mathcal{D}^{TLT}_{lNN'(1)} &=&\frac{m}{ 16 |\textbf{k}'||\textbf{p}'|   M^3 (s-M^2)\mathcal{I}}
\left[ \right.
(F_1^V + F_2^V) F_P t (4 m^6 M^2 + m^4 (-m^2 - 12 M^2) t + {s_u} ^2 (-m^2 - t) t  \nonumber \\
&&+ 3 m^2 (m^2 + 4 M^2) t^2 + (-3 m^2 - 4 M^2) t^3 + t^4 + 2 {s_u}  (2 m^4 M^2 - m^4 t + (m^2 - 2 M^2) t^2))  \nonumber \\
&+& 4 F_1^V F_A M^2 (4 m^4 M^2 (m^2 - 8 M^2) + m^2 (-m^4 + 4 m^2 M^2 + 64 M^4) t - {s_u} ^3 t + (m^4 - 20 m^2 M^2 - 32 M^4) t^2  \nonumber \\
&&+ (m^2 + 12 M^2) t^3 - t^4 + {s_u} ^2 (4 m^2 M^2 + (-3 m^2 - 4 M^2) t + t^2) + {s_u}  (8 m^4 M^2 - 3 m^4 t + 2 (m^2 - 4 M^2) t^2 + t^3))  \nonumber \\
&+& F_2^V F_A (-({s_u} ^3 t^2) + 2 {s_u} ^2 t^2 (-m^2 - 4 M^2 + t) + {s_u}  (16 m^4 M^4 - 4 m^4 M^2 t - (m^2 - 4 M^2)^2 t^2 - 4 M^2 t^3 + t^4)  \nonumber \\
&&+ 2 (8 m^6 M^4 + 2 m^4 M^2 (-m^2 - 20 M^2) t + 14 m^2 M^2 (m^2 + 4 M^2) t^2  \nonumber \\
&&+ (-m^4 - 22 m^2 M^2 - 24 M^4) t^3 + 2 (m^2 + 5 M^2) t^4 - t^5))  \nonumber \\
&+& F_P^2 m^2 t (4 m^4 M^2 + m^2 (-m^2 - 8 M^2) t + {s_u} ^2 t + 2 (m^2 + 2 M^2) t^2 - t^3)  x  \nonumber \\
&+& 4 F_A^2 M^2 (4 m^4 M^2 (m^2 + 4 M^2) + m^2 (-m^4 - 16 m^2 M^2 - 32 M^4) t +
 (3 m^4 + 20 m^2 M^2 + 16 M^4) t^2  \nonumber \\
 &&- (3 m^2 + 8 M^2) t^3 + t^4 + {s_u} ^2 (4 m^2 M^2 + (-m^2 + 8 M^2) t - t^2) + 2 m^2 {s_u}  (4 m^2 M^2 - (m^2 + 4 M^2) t + t^2))  x  \nonumber \\
&+& {F_2^V}^2 (16 m^6 M^4 + 8 m^4 M^2 (-m^2 - 2 M^2) t + m^2 (m^4 + 20 m^2 M^2 - 16 M^4) t^2 - {s_u} ^3 t^2  + 4 (-m^4 - 4 m^2 M^2 + 4 M^4) t^3 \nonumber \\
&&+ (5 m^2 + 4 M^2) t^4 - 2 t^5 + {s_u}  t^2 (m^2 (-m^2 + 8 M^2) - 8 M^2 t + t^2)  + {s_u} ^2 t (4 m^2 M^2 + (-3 m^2 + 4 M^2) t + 2 t^2))  x   \nonumber \\
&+& F_A F_P t ({s_u} ^3 t + 4 M^2 {s_u} ^2 (m^2 + t) + {s_u}  (12 m^4 M^2 - m^2 (3 m^2 + 16 M^2) t + 4 (m^2 + M^2) t^2 - t^3)  \nonumber \\
&&+ 2 m^2 (4 m^4 M^2 + m^2 (-m^2 - 8 M^2) t + 2 (m^2 + 2 M^2) t^2 - t^3))  x   \nonumber \\
&+& 4 {F_1^V}^2 M^2 (-({s_u} ^3 t) + 2 {s_u} ^2 t (-m^2 + 2 M^2 + t) 
+ {s_u}  t (m^2 (-m^2 + 8 M^2) - 8 M^2 t + t^2)  \nonumber \\
&&+ 2 (8 m^4 M^4 + 2 m^2 M^2 (m^2 - 8 M^2) t + (-m^4 - 4 m^2 M^2 + 8 M^4) t^2 + 2 (m^2 + M^2) t^3 - t^4))  x   \nonumber \\
&+& F_1^V F_2^V (16 m^6 M^4 + 8 m^4 M^2 (-m^2 + 6 M^2) t + m^2 (m^4 - 144 M^4) t^2 - 2 {s_u} ^3 t^2  \nonumber \\
&&+ (-3 m^4 + 24 m^2 M^2 + 80 M^4) t^3 + (3 m^2 - 16 M^2) t^4 - t^5 + {s_u} ^2 t (-4 m^2 M^2 + (-3 m^2 + 20 M^2) t + t^2)  \nonumber \\
&&+ 2 {s_u}  (8 m^4 M^4 - 6 m^4 M^2 t + 4 M^2 (3 m^2 - 2 M^2) t^2 + (-m^2 - 6 M^2) t^3 + t^4))  x 
\left.\right]
\end{eqnarray}
\begin{eqnarray}
\mathcal{D}^{TLT}_{lNN'(2)} &=&\frac{m}{ 4 |\textbf{k}'||\textbf{p}'|   M^3 (s-M^2)\mathcal{I}}
\left[ \right. 
 {F_3^A}^2 t (16 m^4 M^4 + 8 m^2 M^2 (-m^2 - 4 M^2) t - {s_u} ^3 t 
+ {s_u} ^2 (4 M^2 - t) t \nonumber \\
&&+ (m^4 + 16 m^2 M^2 + 16 M^4) t^2 + 2 (-m^2 - 4 M^2) t^3 + t^4  \nonumber \\
&&+ {s_u}  (-4 m^4 M^2 + m^2 (m^2 + 8 M^2) t + 2 (-m^2 - 2 M^2) t^2 + t^3))  x  
\left.\right]
\end{eqnarray}
\begin{eqnarray}
\mathcal{D}^{TLT}_{lNN'(12)} &=&\frac{m}{ 8 |\textbf{k}'||\textbf{p}'|   M^3 (s-M^2)\mathcal{I}}
\left[ \right.
 (F_1^V + F_2^V) F_3^A t (-({s_u} ^3 t) + 2 {s_u} ^2 (2 m^2 M^2 + (-m^2 - 2 M^2) t + t^2)  \nonumber \\
&&+ {s_u}  (4 m^4 M^2 - m^4 t - 4 M^2 t^2 + t^3) + 2 (-16 m^4 M^4 + 8 m^2 M^2 (m^2 + 4 M^2) t  + (-m^4 - 16 m^2 M^2 - 16 M^4) t^2 \nonumber \\
&&
+ 2 (m^2 + 4 M^2) t^3 - t^4))  \nonumber \\
&+& F_3^A F_P t (-4 m^6 M^2 + m^4 (m^2 + 12 M^2) t + {s_u} ^3 t + 3 m^2 (-m^2 - 4 M^2) t^2  \nonumber \\
&&+ (3 m^2 + 4 M^2) t^3 - t^4 + {s_u} ^2 t (-m^2 + t) + {s_u}  (4 m^4 M^2 + m^2 (-m^2 - 8 M^2) t + 2 (m^2 + 2 M^2) t^2 - t^3))  x   \nonumber \\
&+& F_3^A F_A t (-4 m^6 M^2 + 2 {s_u} ^3 (2 M^2 - t) + m^4 (m^2 + 12 M^2) t + 3 m^2 (-m^2 - 4 M^2) t^2 + (3 m^2 + 4 M^2) t^3 - t^4  \nonumber \\
&&+ {s_u} ^2 (8 m^2 M^2 - 3 m^2 t + t^2) + 2 {s_u}  (-16 m^2 M^4 + 8 M^2 (m^2 + 2 M^2) t + (-m^2 - 8 M^2) t^2 + t^3))  x 
\left. \right]
\end{eqnarray}
\begin{eqnarray}
\mathcal{D}^{TTL}_{lNN'(1)} &=&\frac{m}{ 16 |\textbf{k}'||\textbf{p}'| M^3  (s-M^2)\mathcal{I}} 
\left[ \right.
(F_1^V + F_2^V) F_P t (-4 m^6 M^2 + m^4 (m^2 + 12 M^2) t + 3 m^2 (-m^2 - 4 M^2) t^2  \nonumber \\
&&+ (3 m^2 + 4 M^2) t^3 - t^4 + {s_u} ^2 t (m^2 + t) + 2 {s_u}  (-2 m^4 M^2 + m^4 t + (-m^2 + 2 M^2) t^2))  \nonumber \\
&+& 4 F_1^V F_A M^2 (4 m^4 M^2 (m^2 - 8 M^2) + m^2 (-m^4 + 4 m^2 M^2 + 64 M^4) t + {s_u} ^3 t + (m^4 - 20 m^2 M^2 - 32 M^4) t^2  \nonumber \\
&&+ (m^2 + 12 M^2) t^3 - t^4 + {s_u} ^2 (4 m^2 M^2 + (m^2 - 4 M^2) t + t^2)  \nonumber \\
&&+ {s_u}  (8 m^4 M^2 + m^2 (-m^2 - 16 M^2) t + 2 (m^2 + 4 M^2) t^2 - t^3))  \nonumber \\
&+& F_2^V F_A ({s_u} ^3 (8 M^2 - t) t 
+ 2 {s_u} ^2 t (8 m^2 M^2 + (-m^2 - 4 M^2) t + t^2) + {s_u}  (16 m^4 M^4 + 4 m^2 M^2 (m^2 - 16 M^2) t  \nonumber \\
&&+ (-m^4 + 8 m^2 M^2 + 48 M^4) t^2 - 12 M^2 t^3 + t^4) + 2 (8 m^6 M^4 + 2 m^4 M^2 (-m^2 - 20 M^2) t  \nonumber \\
&&+ 14 m^2 M^2 (m^2 + 4 M^2) t^2 + (-m^4 - 22 m^2 M^2 - 24 M^4) t^3 + 2 (m^2 + 5 M^2) t^4 - t^5))  \nonumber \\
&+& F_P^2 m^2 t (-4 m^4 M^2 + m^2 (m^2 + 8 M^2) t - {s_u} ^2 t + 2 (-m^2 - 2 M^2) t^2 + t^3)  x  \nonumber \\
&+& 4 F_A^2 M^2 (4 m^4 M^2 (-m^2 + 4 M^2) + m^2 (m^4 - 32 M^4) t + (-m^4 + 12 m^2 M^2 + 16 M^4) t^2 - (m^2 + 8 M^2) t^3  \nonumber \\
&&+ t^4 + {s_u} ^2 (-4 m^2 M^2 + m^2 t - t^2) + 2 m^2 {s_u}  (-4 m^2 M^2 + (m^2 + 4 M^2) t - t^2))  x  \nonumber \\
&+& {F_2^V}^2 (16 m^6 M^4 - 8 m^4 M^2 (m^2 + 6 M^2) t + m^2 (m^4 + 28 m^2 M^2 + 48 M^4) t^2 - {s_u} ^3 t^2  - 4 (m^4 + 8 m^2 M^2 + 4 M^4) t^3 \nonumber \\
&&+ (5 m^2 + 12 M^2) t^4 - 2 t^5 + {s_u}  t^2 (m^2 (-m^2 + 8 M^2) - 8 M^2 t + t^2)  + {s_u} ^2 t (4 m^2 M^2 - (3 m^2 + 4 M^2) t + 2 t^2))  x   \nonumber \\
&+& 4 {F_1^V}^2 M^2 (2 (-m^2 + 2 M^2) {s_u} ^2 t - {s_u} ^3 t + {s_u}  t (m^2 (-m^2 + 8 M^2) - 8 M^2 t + t^2)  \nonumber \\
&&+ 4 M^2 (4 m^4 M^2 + m^2 (-m^2 - 8 M^2) t + 2 (m^2 + 2 M^2) t^2 - t^3))  x   \nonumber \\
&+& F_A F_P t (4 M^2 {s_u} ^2 (-m^2 - t) - {s_u} ^3 t + 2 m^2 (-4 m^4 M^2 + m^2 (m^2 + 8 M^2) t + 2 (-m^2 - 2 M^2) t^2 + t^3)  \nonumber \\
&&+ {s_u}  (-12 m^4 M^2 + m^2 (3 m^2 + 16 M^2) t + 4 (-m^2 - M^2) t^2 + t^3))  x   \nonumber \\
&+& F_1^V F_2^V (16 m^6 M^4 + 8 m^4 M^2 (-m^2 - 2 M^2) t + m^2 (m^4 + 16 m^2 M^2 - 16 M^4) t^2  \nonumber \\
&&- 2 {s_u} ^3 t^2 + (-3 m^4 - 8 m^2 M^2 + 16 M^4) t^3 + 3 m^2 t^4 - t^5 + {s_u} ^2 t (-4 m^2 M^2 + (-3 m^2 + 4 M^2) t + t^2)  \nonumber \\
&&+ 2 {s_u}  (8 m^4 M^4 - 6 m^4 M^2 t + 4 M^2 (3 m^2 - 2 M^2) t^2 + (-m^2 - 6 M^2) t^3 + t^4))  x 
\left.\right]
\end{eqnarray}

\begin{eqnarray}
\mathcal{D}^{TTL}_{lNN'(2)} &=&\frac{m x}{4 |\textbf{k}'||\textbf{p}'| M^3  (s-M^2)\mathcal{I}} 
\left[ \right.
{F_3^A}^2 t (-4 M^2 + {s_u}  + t)  (m^4 (4 M^2 - t) - 2 m^2 t (4 M^2 - t) + t ({s_u} ^2 + (4 M^2 - t) t))  
\left.\right]  \\
\mathcal{D}^{TTL}_{lNN'(12)} &=&\frac{m}{8 |\textbf{k}'||\textbf{p}'| M^3  (s-M^2)\mathcal{I}} 
\left[ \right.
 (F_1^V + F_2^V) F_3^A t ({s_u} ^3 t + 2 {s_u} ^2 (-2 m^2 M^2 + (m^2 + 2 M^2) t - t^2)  \nonumber \\
&&+ {s_u}  (-4 m^4 M^2 + m^4 t + 4 M^2 t^2 - t^3) + 2 (16 m^4 M^4 + 8 m^2 M^2 (-m^2 - 4 M^2) t + (m^4 + 16 m^2 M^2 + 16 M^4) t^2  \nonumber \\
&&+ 2 (-m^2 - 4 M^2) t^3 + t^4))  \nonumber \\
&+& F_3^A F_A t (4 m^6 M^2 + m^4 (-m^2 - 12 M^2) t + 3 m^2 (m^2 + 4 M^2) t^2 + (-3 m^2 - 4 M^2) t^3 + t^4 + 2 {s_u} ^3 (-2 M^2 + t)  \nonumber \\
&&+ {s_u} ^2 (-8 m^2 M^2 + 3 m^2 t - t^2) 
+ 2 {s_u}  (16 m^2 M^4 + 8 M^2 (-m^2 - 2 M^2) t + (m^2 + 8 M^2) t^2 - t^3))  x   \nonumber \\
&+& F_3^A F_P t (4 m^6 M^2 + m^4 (-m^2 - 12 M^2) t - {s_u} ^3 t 
+ {s_u} ^2 (m^2 - t) t + 3 m^2 (m^2 + 4 M^2) t^2 + (-3 m^2 - 4 M^2) t^3  \nonumber \\
&&+ t^4 + {s_u}  (-4 m^4 M^2 + m^2 (m^2 + 8 M^2) t + 2 (-m^2 - 2 M^2) t^2 + t^3))  x 
\left.\right]
\end{eqnarray}
\end{footnotesize}
\normalem
\bibliographystyle{apsrev4-1}
\bibliography{bibdrat,bibmoje,bibdratbook}

\end{document}